\documentclass[twocolumn]{emulateapj}

\usepackage{subfigure}

\usepackage{bm}

\usepackage{amsmath, amssymb}
\usepackage{times}

\usepackage[usenames]{xcolor}
\definecolor{mpl_red}{HTML}{D62728}

\usepackage[backref, breaklinks, plainpages=false, colorlinks=true, anchorcolor=blue!50!black, citecolor=blue!50!black, linkcolor=blue!50!black, urlcolor=mpl_red, bookmarks=false]{hyperref}
\usepackage{natbib}

\usepackage[strict]{changepage}

\begin{document}

\renewcommand*{\backref}[1]{[#1]}

\newcommand{\Pearlman}{Aaron~B.~Pearlman}
\newcommand{\Corbet}{Robin~H.~D.~Corbet}
\newcommand{\Pottschmidt}{Katja~Pottschmidt}
\newcommand{\Coley}{Joel~B.~Coley}

\newcommand{\CaltechPhysics}{Division of Physics, Mathematics, and Astronomy, California Institute of Technology, Pasadena, CA 91125, USA; \textcolor{blue}{aaron.b.pearlman@caltech.edu}}
\newcommand{\Howard}{Department of Physics and Astronomy, Howard University, Washington, DC 20059, USA}
\newcommand{\CRESSTAstroPart}{CRESST/Mail Code 661, Astroparticle Physics Laboratory, NASA Goddard Space Flight Center, Greenbelt, MD 20771, USA}
\newcommand{\CRESSTXRay}{CRESST/Mail Code 662, \text{X-ray} Astrophysics Laboratory, NASA Goddard Space Flight Center, Greenbelt, MD 20771, USA}
\newcommand{\UMBC}{University of Maryland, Baltimore County, Baltimore, MD 21250, USA}
\newcommand{\NDSEG}{$^{\text{6}}$~NDSEG Research Fellow.}
\newcommand{\NSF}{$^{\text{7}}$~NSF Graduate Research Fellow.}


\journalinfo{\textsc{Accepted for publication in The Astrophysical Journal}\textup{, 2018~November~9}}
\submitted{Accepted for publication in The Astrophysical Journal on 2018~November~9}

\shorttitle{ORBITAL PARAMETERS OF \text{IGR~J16493--4348}}
\shortauthors{PEARLMAN ET AL.}

\title{The Orbital Parameters of the Eclipsing \text{High-Mass} \text{X-ray} Binary Pulsar \text{IGR~J16493--4348} from Pulsar Timing}

\author{\Pearlman\altaffilmark{1,6,7}, \Coley\altaffilmark{2,3}, \Corbet\altaffilmark{4,5}, and~\Pottschmidt\altaffilmark{3,5}}

\address{
$^{\text{1}}$~\CaltechPhysics \\
$^{\text{2}}$~\Howard \\
$^{\text{3}}$~\CRESSTAstroPart \\
$^{\text{4}}$~\CRESSTXRay \\
$^{\text{5}}$~\UMBC}

\thanks{\NDSEG}
\thanks{\NSF}


\begin{abstract}
\label{Section:Abstract}

\text{IGR~J16493--4348} is an eclipsing supergiant \text{high-mass} \text{X-ray} binary~(sgHMXB), where accretion onto the compact object occurs via the radially outflowing stellar wind of its early \text{B-type} companion. We present an analysis of the system's \text{X-ray} variability and periodic modulation using pointed observations~\text{(2.5--25\,keV)} and Galactic bulge scans~\text{(2--10\,keV)} from the \textit{Rossi \text{X-ray} Timing Explorer~(RXTE)} Proportional Counter Array~(PCA), along with \textit{Swift}~Burst Alert Telescope~(BAT) \text{70-month} snapshot~\text{(14--195\,keV)} and transient monitor~\text{(15--50\,keV)} observations. The orbital eclipse profiles in the PCA~bulge scans and BAT~light curves are modeled using asymmetric and symmetric step and ramp functions. We obtain an improved orbital period measurement of 6.7828\,$\pm$\,0.0004\,days from an observed minus calculated~\text{(O--C)} analysis of \text{mid-eclipse} times derived from the BAT~transient monitor and PCA~scan data. No evidence is found for the presence of a strong photoionization or accretion wake. We refine the superorbital period to 20.067\,$\pm$\,0.009\,days from the discrete Fourier transform~(DFT) of the BAT~transient monitor light curve. A pulse period of 1093.1036\,$\pm$\,0.0004\,s is measured from a pulsar timing analysis using pointed PCA~observations spanning $\sim$1.4~binary orbits. We present pulse times of arrival~(ToAs), circular and eccentric timing models, and calculations of the system's Keplerian binary orbital parameters. We derive an \text{X-ray} mass function of $f_{x}(M)$\,$=$\,13.2$^{\text{+2.4}}_{\text{--2.5}}$\,$M_{\odot}$ and find a spectral type of B0.5~Ia for the supergiant companion through constraints on the mass and radius of the donor. Measurements of the eclipse \text{half-angle} and additional parameters describing the system geometry are provided.

\end{abstract}




\section{Introduction}
\label{Section:Introduction}

\setcounter{footnote}{7}

\text{IGR~J16493--4348} is a persistently accreting supergiant \text{high-mass} \text{X-ray} binary~(sgHMXB), where mass transfer onto the neutron star is driven by the stellar wind from its early \text{B-type} companion. It was first detected during a survey of the Galactic plane~\citep{Bird2004} using the \textit{INTErnational \text{Gamma-Ray} Astrophysics Laboratory}~(\textit{INTEGRAL};~\citealt{Winkler2003}) with the \textit{INTEGRAL} Soft \text{Gamma-Ray} Imager~(ISGRI;~\citealt{Lebrun2003}) camera of the Imager on Board the \textit{INTEGRAL} Satellite~(IBIS;~\citealt{Ubertini2003}) telescope. The source was also detected with \textit{INTEGRAL} during a deep scan of the Norma Arm region~\citep{Grebenev2005} and in subsequent \text{IBIS/ISGRI}~\citep{Bird2006, Bird2007, Bird2010, Bird2016} and \textit{Swift}~Burst Alert Telescope~(BAT)~surveys~\citep{Baumgartner2013, Oh2018}.

Pointed \textit{Rossi \text{X-ray} Timing Explorer~(RXTE)} Proportional Counter Array~(PCA) observations of \text{IGR~J16493--4348} by~\citet{Markwardt2005a} revealed that the mean spectrum was consistent with a heavily absorbed power law, with a photon index of $\Gamma$\,$=$\,1.4 and $N_{\text{H}}$\,$\approx$\,10$^{\text{23}}$\,cm$^{\text{--2}}$. The measured fluxes in the \text{2--10}, \text{10--20}, and \text{20--40\,keV}~energy bands were 1.0\,$\times$\,10$^{\text{--11}}$, 1.3\,$\times$\,10$^{\text{--11}}$, and 2.1\,$\times$\,10$^{\text{--11}}$\,erg\,cm$^{\text{--2}}$\,s$^{\text{--1}}$, respectively. \textit{Chandra} imaging of the field of \text{IGR~J16493--4348} was performed by~\citet{Kuiper2005} using the High Resolution Camera~\text{(HRC-I)} instrument, which identified \text{2MASS~J16492695--4349090} as the infrared counterpart. A $K_{\text{S}}$~magnitude of 12.0\,$\pm$\,0.1 was found using the Son~of~ISAAC~(SOFI) infrared camera at the European Southern Observatory~(ESO) 3.5\,m~New Technology Telescope~(NTT), which is consistent with the 11.94\,$\pm$\,0.04 $K_{\text{S}}$~magnitude reported in the 2MASS~catalog and suggests that the source is not highly variable in this band. No optical counterpart was found in the Digital Sky Survey~(DSS) maps due to strong absorption along the line of sight.

A spectral type of \text{B0.5-1~Ia-Ib} was estimated by \citet{Nespoli2008, Nespoli2010} from \text{$K_{\text{S}}$-band} spectroscopy of \text{IGR~J16493--4348's} infrared counterpart using observations from the Infrared Spectrometer and Array Camera~(ISAAC) spectrograph on~UT1 of the ESO~Very Large Telescope~(VLT). The spectrum showed a strong He\,I~(20581\,\AA) emission line with He\,I~(21126\,\AA) in absorption, along with a strong Br$\gamma$~(21661\,\AA) absorption line, which are all characteristic features in OB stellar spectra. This led \citet{Nespoli2008, Nespoli2010} to classify the system as an~sgHMXB. They also provided an estimate of the interstellar extinction and calculated a hydrogen column density of $N_{\text{H}}$\,$=$\,(2.92\,$\pm$\,1.96)\,$\times$\,10$^{\text{22}}$\,cm$^{\text{--2}}$, which they attributed to a significant absorbing envelope surrounding the neutron star. The distance to the source was estimated to be between \text{6--26}\,kpc. \citet{Romano2015} found that the cumulative luminosity distribution~(CLD) and small dynamic range in \text{X-ray} flux from \textit{Swift}\footnote{The \textit{Swift} \text{Gamma-Ray} Burst Explorer was renamed the ``Neil~Gehrels \textit{Swift} Observatory'' in honor of Neil~Gehrels, \textit{Swift's} principal investigator.}~X-ray Telescope~(XRT) observations were also typical of a classical~sgHMXB system, rather than a Supergiant Fast~\text{X-ray} Transient~(SFXT).

\citet{Hill2008} carried out a spectral analysis of the source using \text{22--100\,keV}~\textit{INTEGRAL} \text{IBIS/ISGRI} and \text{1--9\,keV}~\textit{Swift}~XRT data. They found that the source was best modeled by a highly absorbed power law, with $\Gamma$\,$=$\,0.6\,$\pm$\,0.3 and $N_{\text{H}}$\,$=$\,5.4$^{\text{+1.3}}_{\text{--1.0}}$\,$\times$\,10$^{\text{22}}$\,cm$^{\text{--2}}$, and a high energy \text{cut-off} at $E_{\text{cut}}$\,$=$\,17$^{\text{+5}}_{\text{--3}}$\,keV. An average source flux of 1.1\,$\times$\,10$^{\text{--10}}$\,erg\,cm$^{\text{--2}}$\,s$^{\text{--1}}$ was measured in the \text{1--100\,keV}~energy band. No coherent periodicities were found in their \textit{INTEGRAL}~or \textit{Swift}~data.

\citet{Morris2009} analyzed the \text{0.2--150.0\,keV}~spectrum of \text{IGR~J16493--4348} obtained from \textit{Suzaku} observations with the Hard \text{X-ray} Detector~(HXD) and \text{X-ray} Imaging Spectrometer~(XIS) instruments. The spectrum was fit with a power law modified by a fully and partially covering absorber. The partially covered and fully covered neutral hydrogen column densities were 26$^{\text{+9.4}}_{\text{--7.9}}$\,$\times$\,10$^{\text{22}}$ and 8.6$^{\text{+0.9}}_{\text{--1.0}}$\,$\times$\,10$^{\text{22}}$\,cm$^{\text{--2}}$, respectively, with a partial covering fraction of 0.62$^{\text{+0.06}}_{\text{--0.07}}$ and photon index of $\Gamma$\,$=$\,2.4\,$\pm$\,0.2. A 6.4\,keV~Fe~emission line, with an equivalent width less than 84\,eV, was also included in their spectral model, and a flux of 13.5$^{\text{+0.3}}_{\text{--2.0}}$\,$\times$\,10$^{\text{--12}}$\,erg\,cm$^{\text{--2}}$\,s$^{\text{--1}}$ was measured between 0.2~and~10\,keV.

Spectral analysis in the hard \text{X-ray} band was also performed by~\citet{DAi2011} using \text{15--150\,keV} \textit{Swift}~BAT and \text{\textit{INTEGRAL}/ISGRI} data, together with pointed soft \text{X-ray} observations from \textit{Suzaku} and the \textit{Swift}~XRT. They found that a \text{negative-positive} exponential power law model, with a broad~(10\,keV~width) absorption line at 33\,$\pm$\,4\,keV, yielded the best fit to the broadband spectrum. This absorption feature was interpreted as evidence of a cyclotron resonance scattering feature~(CRSF). Assuming cyclotron absorption occurs above the magnetic poles of the neutron star,~\citet{DAi2011} inferred a surface magnetic field of $B_{\text{surf}}$\,$=$\,(3.7\,$\pm$\,0.4)\,$\times$\,10$^{\text{12}}$\,G from the energy of the cyclotron line for a canonical neutron star with a mass of 1.4\,$M_{\odot}$ and a radius of 10\,km.

The 6.8\,day binary orbital period was independently discovered by~\citet{Corbet2010a} and~\citet{Cusumano2010}. \citet{Cusumano2010} also found evidence of an eclipse in the folded \textit{Swift}~BAT survey light curve lasting approximately 0.8\,days. Assuming a neutron star mass of 1.4\,$M_{\odot}$ and a B0.5~Ib companion with a mass and radius of 47\,$M_{\odot}$ and 32.2\,$R_{\odot}$, respectively, \citet{Cusumano2010} estimated a \text{semi-major} axis of $a$\,$\approx$\,55\,$R_{\odot}$ for the binary orbit and derived an upper limit of $e$\,$\le$\,0.15 on the eccentricity.

A 20\,day superorbital period was first detected by~\citet{Corbet2010a} in the power spectra of the \textit{Swift}~BAT survey and \textit{RXTE}~PCA~scan light curves. The superorbital period was refined to 20.07\,$\pm$\,0.01\,days using data from the \textit{Swift}~BAT~transient monitor~\text{(15--50\,keV)}, and a monotonic relationship between the superorbital and orbital modulation was suggested~\citep{Corbet2013}. Recently,~\citet{Coley2018} analyzed \textit{Nuclear Spectroscopic Telescope Array~(NuSTAR)} and \textit{Swift}~XRT observations near the maximum and minimum of one cycle of the 20\,day superorbital modulation. They found that the \text{3--40\,keV} spectra were well modeled by an absorbed power law, with $N_{\text{H}}$\,$\approx$\,10$^{\text{23}}$\,cm$^{\text{--2}}$, and a high energy \text{cut-off}. Evidence of an Fe~K$\alpha$ emission line was also found at superorbital maximum near 6.4\,keV. A comprehensive discussion of possible mechanisms responsible for the superorbital variability is presented in~\citet{Coley2018}, along with a timing analysis characterizing its long term behavior.

A 1069\,s period was detected in the power spectrum of the light curve from pointed \textit{RXTE}~PCA~observations, which was suggested to be linked to the neutron star's rotational period~\citep{Corbet2010b}. We find strong evidence for a 1093\,s pulse period from pulsar timing~measurements using additional archival pointed \textit{RXTE}~PCA~observations. Pulse phase resolved spectroscopy near the maximum and minimum of the superorbital cycle has recently been carried out by~\citet{Coley2018} using these pulsar timing results.

In this paper, we present improved measurements of \text{IGR~J16493--4348's} superorbital, orbital, and pulse periods using \textit{Swift}~BAT and \textit{RXTE}~PCA~observations. We also measure the system's Keplerian binary orbital parameters and study the nature of the supergiant donor and the geometry of the binary. The observations and our data reduction procedure are described in Section~\ref{Section:Observations}. In Section~\ref{Section:Period_Measurements}, we provide refined period measurements and model the system's eclipse profile. A pulsar timing analysis is presented in \text{Sections~\ref{Section:Pulsar_Timing_Analysis}--\ref{Section:Pulsar_Timing_Models}}, along with pulse times of arrival~(ToAs) and orbital timing models. New constraints on the spectral type of the supergiant donor and possible system geometries are given in Section~\ref{Section:Supergiant_Companion}. A discussion of the spectral type of the supergiant companion, eclipse asymmetries, orbital eccentricity, and possible superorbital mechanisms is provided in Section~\ref{Section:Discussion}. We summarize our results and conclusions in Section~\ref{Section:Conclusion}.



\section{Observations}
\label{Section:Observations}


\subsection{RXTE~PCA~Pointed Observations}
\label{Section:RXTE_PCA_Pointed_Observations}

The PCA~\citep{Jahoda1996, Jahoda2006} was the primary instrument on board the \textit{RXTE}~satellite. The detector was comprised of five nearly identical Proportional Counter Units~(PCUs), with a total effective area of $\sim$6500\,cm$^{\text{2}}$, and was sensitive to \text{X-rays} with energies between 2~and~60\,keV. The mechanically collimated array had a 1$^{\circ}$ field of view~(FoV) at full width at half maximum~(FWHM). Each~PCU had a \text{multi-anode} main volume filled with xenon and methane and a front \text{propane-filled} ``veto'' volume that was used primarily for background rejection.

We analyzed 24~pointed \textit{RXTE}~PCA~observations collected at $\sim$0.4\,day intervals during 2011~October, which spanned $\sim$9.5\,days. The total exposure time was $\sim$160.5\,ks, and the exposure times of individual observations ranged between 1.7~and~16.0\,ks. A catalog of these observations is provided in Table~\ref{Table:RXTEObservationLog}, and they are accessible through the High Energy Astrophysics Science Archive Research Center~(HEASARC) archive\footnote{See \href{https://heasarc.gsfc.nasa.gov/cgi-bin/W3Browse/w3browse.pl}{https://heasarc.gsfc.nasa.gov/cgi-bin/W3Browse/w3browse.pl.}}.

Background-subtracted PCA~light curves were created using the Standard~2 mode and FTOOLS~v6.22\footnote{See \href{https://heasarc.gsfc.nasa.gov/ftools}{https://heasarc.gsfc.nasa.gov/ftools.}}~\citep{Blackburn1995}. The time resolution of the light curves was~16\,s, and we used data obtained from the top xenon layers~(1L~and~1R) of the~PCUs to maximize the \text{signal-to-noise} ratio. The Faint background model\footnote{See \href{https://heasarc.nasa.gov/docs/xte/pca\_news.html}{https://heasarc.nasa.gov/docs/xte/pca\_news.html.}} was used to background subtract all of the PCA~data since the net count rate did not exceed 40\,counts\,s$^{\text{--1}}$\,PCU$^{\text{--1}}$. To reduce the uncertainty of the PCA~background model, we excluded data: (1)~up to 10\,minutes immediately following the peak of South Atlantic Anomaly~(SAA) passages, (2)~with an elevation angle less than~10$^{\circ}$ above the limb of the Earth, (3)~with an electron ratio larger than~0.1 in at least one of the operating~PCUs, indicating high electron contamination according to the ratio of veto rates in the detectors, (4)~with an offset between the source position and \textit{RXTE}'s~pointing position larger than~0.01$^{\circ}$, and (5)~within~150\,s of the start of a PCU~breakdown event and up to~600\,s following a PCU~breakdown event using the PCA~breakdown history. A detailed discussion of PCA~calibration issues was provided by~\citet{Jahoda2006}. The light curve times were corrected to the solar system barycenter using the \texttt{faxbary} FTOOLS~routine and the Jet Propulsion Laboratory~(JPL) \text{DE-200}~ephemeris~\citep{Standish1990}. Throughout this paper, Modified Julian Dates~(MJDs) refer to the barycentric times.

\begin{figure}[b]
	\centering
	\includegraphics[trim=0cm 0cm 0cm 0cm, clip=false, scale=0.48, angle=0]{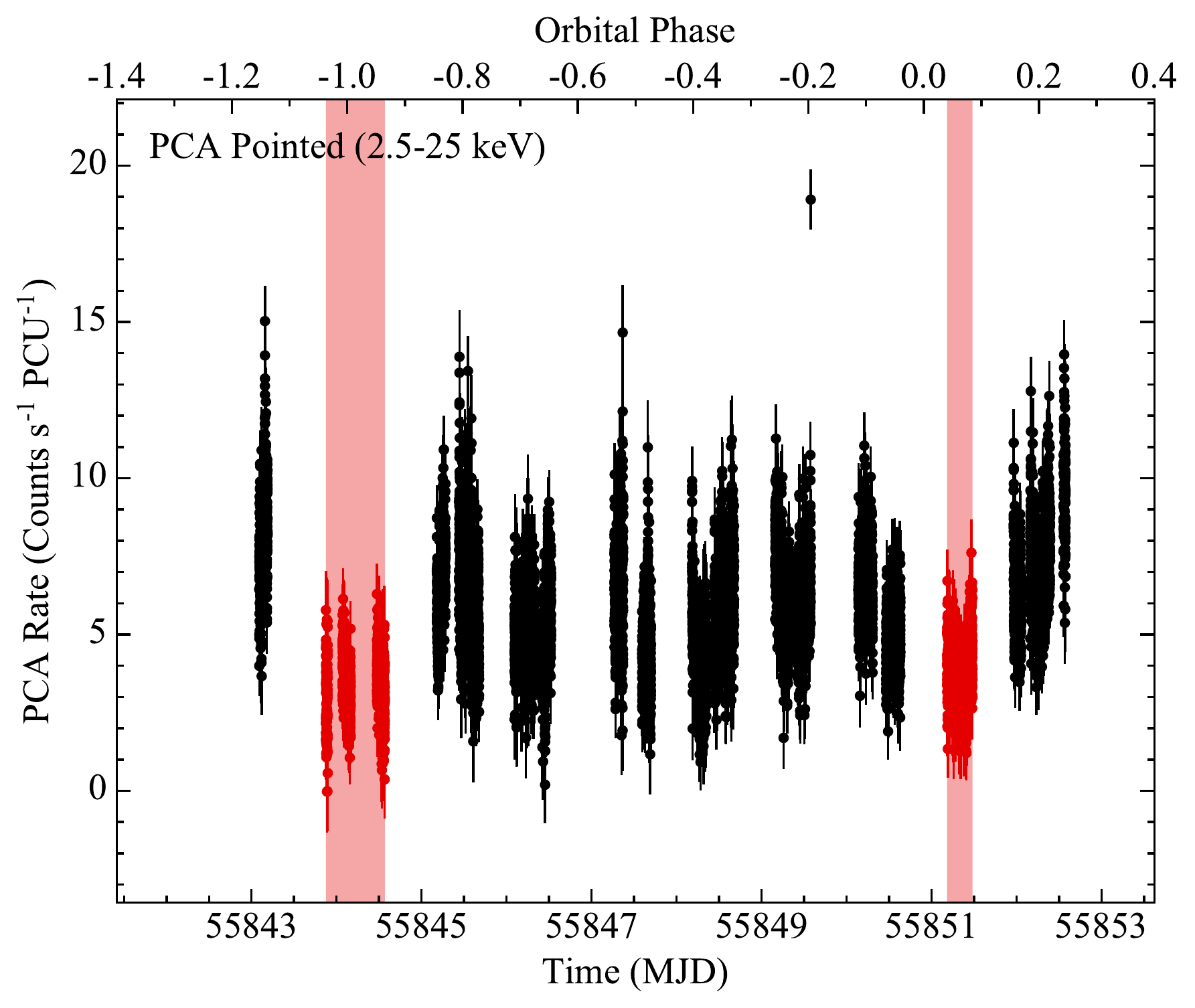}
	\caption{\text{Background-subtracted} pointed \textit{RXTE}~PCA~\text{(2.5--25\,keV)} light curve of \text{IGR~J16493--4348} using all available PCUs with 16\,s~time resolution. Orbital phase~0 corresponds to~$T_{\pi/2}$ from circular solution~1 in Table~\ref{Table:OrbitalParameters}. Data plotted in red were excluded from the pulsar timing analysis since pulsations were not strongly detected.}
	\label{Figure:Figure1}
\end{figure}

Data from all available~PCUs were used during each pointed observation. The \textit{RXTE}~PCA~\text{(2.5--25\,keV)} light curve is shown in Figure~\ref{Figure:Figure1} with count rates normalized by the number of operational~PCUs. Orbital phase~0 corresponds to~$T_{\pi/2}$ from circular solution~1 in Table~\ref{Table:OrbitalParameters}. In Figure~\ref{Figure:Figure2}(d), we show the pointed \textit{RXTE}~PCA~\text{(2--10\,keV)} light curve, produced using the same data filtering criteria and 128\,s time resolution, with nearly simultaneous PCA~scan~\text{(2--10\,keV)} observations overlaid in blue.



\begin{deluxetable*}{ccccccc} 
\tablenum{1}
\tabletypesize{\small}
\tablecolumns{8}
\tablewidth{0pt}
\tablecaption{\textsc{\textit{RXTE}~PCA Pointed Observations of \text{IGR~J16493--4348}}} 
\tablehead{
\colhead{Observation ID} &
\colhead{Date} &
\colhead{Observation Time$^{\mathrm{a}}$} &
\colhead{Time Span} &
\colhead{Exposure Time} &
\colhead{Count Rate$^{\mathrm{b}}$} &
\colhead{Orbital Phase$^{\mathrm{c}}$} \\
\colhead{} & 
\colhead{(UTC)} &
\colhead{(MJD)} &
\colhead{(ks)} &
\colhead{(ks)} &
\colhead{(Counts\,s$^{\text{--1}}$\,PCU$^{\text{--1}}$)} &
\colhead{} 
}
\startdata
96358-01-01-01 & 2011 Oct 09.09--09.19 & 55843.13981 & 8.42 & 5.09 & 8.04 & --1.145 \\
\textbf{96358-01-01-02} & 2011 Oct 09.87--09.90 & 55843.88558 & 1.87 & 1.87 & 2.94 & --1.036 \\
\textbf{96358-01-01-12} & 2011 Oct 10.07--10.17 & 55844.11862 & 8.29 & 4.91 & 3.50 & --1.001 \\
\textbf{96358-01-01-03} & 2011 Oct 10.48--10.57 & 55844.52174 & 8.00 & 5.84 & 3.44 & --0.942 \\
96358-01-01-04 & 2011 Oct 11.18--11.28 & 55845.23159 & 8.85 & 6.05 & 6.73 & --0.837 \\
96358-01-01-05 & 2011 Oct 11.44--11.48 & 55845.46157 & 3.44 & 3.44 & 7.35 & --0.803 \\
96358-01-01-00 & 2011 Oct 11.51--11.61 & 55845.55934 & 9.07 & 6.88 & 6.31 & --0.789 \\
96358-01-01-06 & 2011 Oct 11.64--11.68 & 55845.65720 & 3.42 & 3.42 & 5.57 & --0.774 \\
96358-01-01-07 & 2011 Oct 12.09--12.12 & 55846.10809 & 2.37 & 1.71 & 5.06 & --0.708 \\
96358-01-01-08 & 2011 Oct 12.18--12.33 & 55846.25345 & 13.12 & 7.68 & 5.22 & --0.686 \\
96358-01-01-09 & 2011 Oct 12.42--12.53 & 55846.47288 & 9.02 & 6.78 & 5.10 & --0.654 \\
96358-01-01-10 & 2011 Oct 13.27--13.37 & 55847.32124 & 9.01 & 6.72 & 6.37 & --0.529 \\
96358-01-01-11 & 2011 Oct 13.60--13.70 & 55847.64742 & 8.99 & 6.70 & 4.63 & --0.481 \\
96358-01-02-00 & 2011 Oct 14.18--14.35 & 55848.26709 & 14.61 & 9.34 & 4.37 & --0.390 \\
96358-01-02-01 & 2011 Oct 14.44--14.68 & 55848.56078 & 20.24 & 13.29 & 5.92 & --0.346 \\
96358-01-02-02 & 2011 Oct 15.16--15.33 & 55849.24572 & 14.57 & 9.14 & 6.36 & --0.245 \\
96358-01-02-03 & 2011 Oct 15.42--15.58 & 55849.50098 & 13.57 & 8.90 & 6.03 & --0.208 \\
96358-01-02-04 & 2011 Oct 16.14--16.31 & 55850.22417 & 14.54 & 8.83 & 6.64 & --0.101 \\
96358-01-02-05 & 2011 Oct 16.47--16.63 & 55850.55044 & 14.54 & 9.82 & 4.89 & --0.053 \\
\textbf{96358-01-02-06} & 2011 Oct 17.18--17.48 & 55851.33335 & 25.81 & 15.95 & 3.73 & 0.063 \\
96358-01-02-09 & 2011 Oct 17.97--17.99 & 55851.97617 & 1.66 & 1.66 & 6.56 & 0.157 \\
96358-01-02-10 & 2011 Oct 18.03--18.06 & 55852.04394 & 2.08 & 2.08 & 5.89 & 0.167 \\
96358-01-02-07 & 2011 Oct 18.16--18.40 & 55852.27911 & 20.16 & 12.50 & 7.13 & 0.202 \\
96358-01-02-08 & 2011 Oct 18.55--18.58 & 55852.56585 & 2.10 & 1.89 & 9.55 & 0.244
\enddata
\tablecomments{Observation~IDs in bold were excluded from the pulsar timing analysis since pulsed emission was not strongly detected (see also Figure~\ref{Figure:Figure1}). \\
$^\mathrm{a}$ Mid-time of observation. \\
$^\mathrm{b}$ Average \text{2.5--25\,keV} count rate after background subtraction using all available PCUs. \\
$^\mathrm{c}$ Orbital phase at the observation mid-time using the refined 6.7828\,day orbital period measurement in Section~\ref{Section:O-C_Analysis}. Phase~0 corresponds to~$T_{\pi/2}$ from circular solution~1 in Table~\ref{Table:OrbitalParameters}.}
\label{Table:RXTEObservationLog}
\end{deluxetable*}



\newpage


\subsection{RXTE~PCA~Galactic Bulge Scans}
\label{Section:RXTE_PCA_Galactic_Bulge_Scans}

From 1999~February to 2011~November, \textit{RXTE}~performed raster scans of an approximately 16$^{\circ}$\,$\times$\,16$^{\circ}$~rectangular region near the Galactic Center~(GC) using the~PCA~\citep{Swank2001}. The count rates were modulated by the 1$^{\circ}$ collimators as the source moved into and out of the PCA's~FoV during the Galactic bulge scans. These observations were carried out twice weekly, excluding \text{November--January} and June when the positions of the Sun and \text{anti-Sun} crossed the GC~region. A single scan of a source had a typical exposure time of approximately 20\,s per observation and was sensitive to \text{0.5--1\,mCrab} variations in the source flux. The light curves were generated in the \text{2--10\,keV}~energy band using only the top layer of the~PCA to optimize detections of faint sources. We corrected the light curve times to the solar system barycenter using the tools available through the Ohio~State University~(OSU) Department of Astronomy website\footnote{See \href{http://astroutils.astronomy.ohio-state.edu/time}{http://astroutils.astronomy.ohio-state.edu/time.\label{FootnoteBarycenter}}}~\citep{Eastman2010}.

We present 524~measurements from a series of PCA~Galactic bulge scans between MJDs\,53163.8 and~55863.4~\text{(2004~June~7--2011~October~29)}. The PCA~scan~\text{(2--10\,keV)} weighted average light curve is shown in Figure~\ref{Figure:Figure2}(a). The Galactic bulge scan data are publicly available\footnote{See \href{https://asd.gsfc.nasa.gov/Craig.Markwardt/galscan/main.html}{https://asd.gsfc.nasa.gov/Craig.Markwardt/galscan/main.html.}}.



\subsection{Swift~BAT}
\label{Section:Swift_BAT}

The \textit{Swift}~BAT~\citep{Barthelmy2005, Gehrels2004} is a wide~FoV (1.4\,sr~\text{half-coded}), hard \text{X-ray} telescope that uses a 2.7\,m$^{\text{2}}$~\text{coded-aperture} mask and is sensitive to \text{X-rays} in the \text{14--195\,keV}~band. Although the~BAT is primarily designed for studying \text{gamma-ray} bursts and their afterglows, it also serves as a hard \text{X-ray} transient monitor~\citep{Krimm2013} and surveys the hard \text{X-ray} sky with $\sim$0.4\,mCrab sensitivity~\citep{Baumgartner2013}. Thus, BAT~observations of \text{X-ray} sources are usually performed in an unpredictable and serendipitous manner. Due to the nonuniform nature of the BAT~sky survey, the~\text{signal-to-noise} ratio of a source during an observation depends strongly on the location of the source within the BAT's~FoV. The~BAT typically covers \text{50--80\%} of the sky each day. The data reduction procedures are described in detail in~\citet{Krimm2013} and~\citet{Baumgartner2013}.

\begin{figure*}[t]
	\centering
	\begin{tabular}{cc}
	
	\subfigure
	{
		\includegraphics[trim=0cm 0cm 0cm 0cm, clip=false, scale=0.475, angle=0]{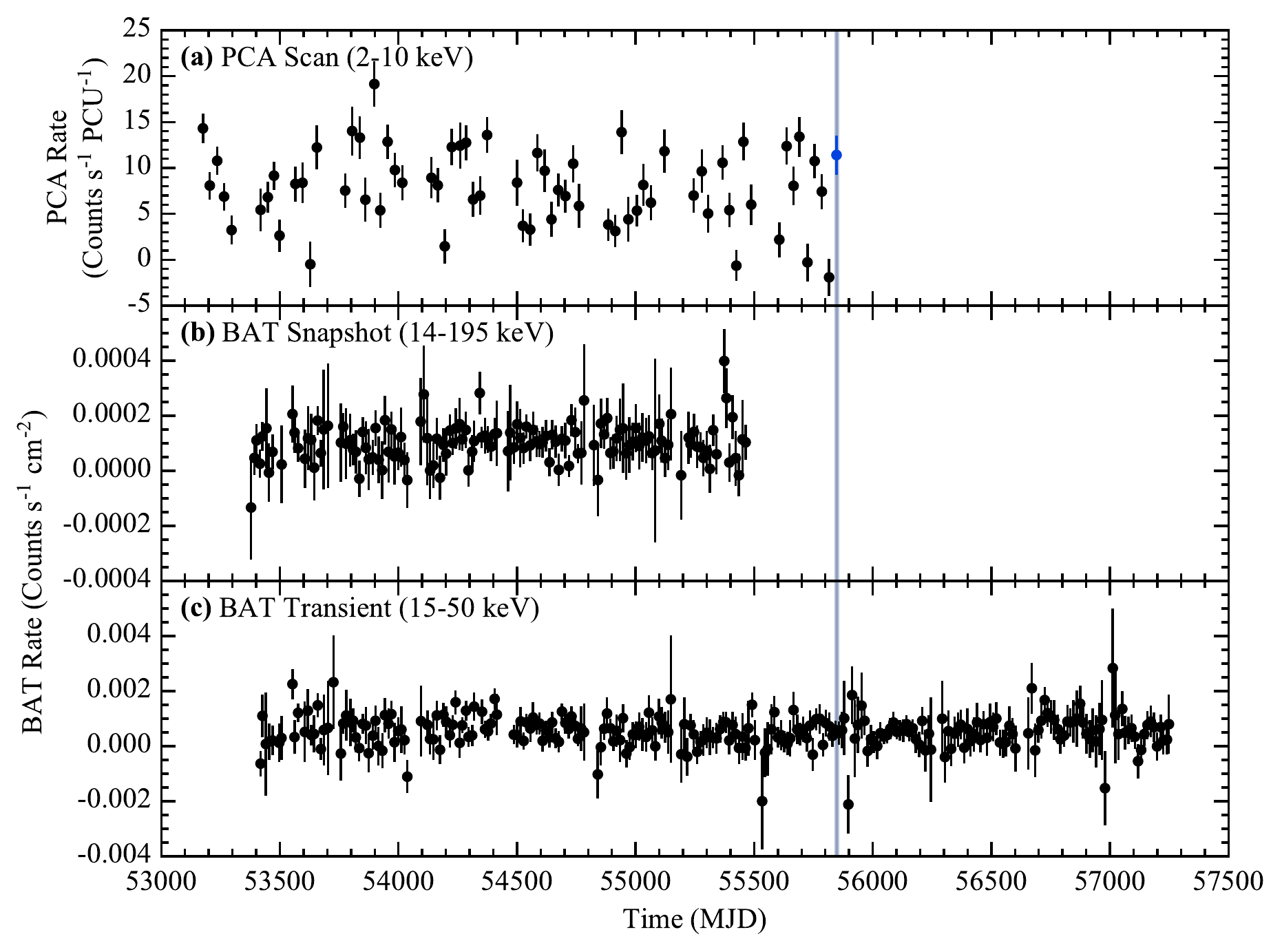}
		\label{Figure:Figure2abc}
	}
	
	&
	
	\subfigure
	{
		\includegraphics[trim=0cm 0cm 0cm 0cm, clip=false, scale=0.47, angle=0]{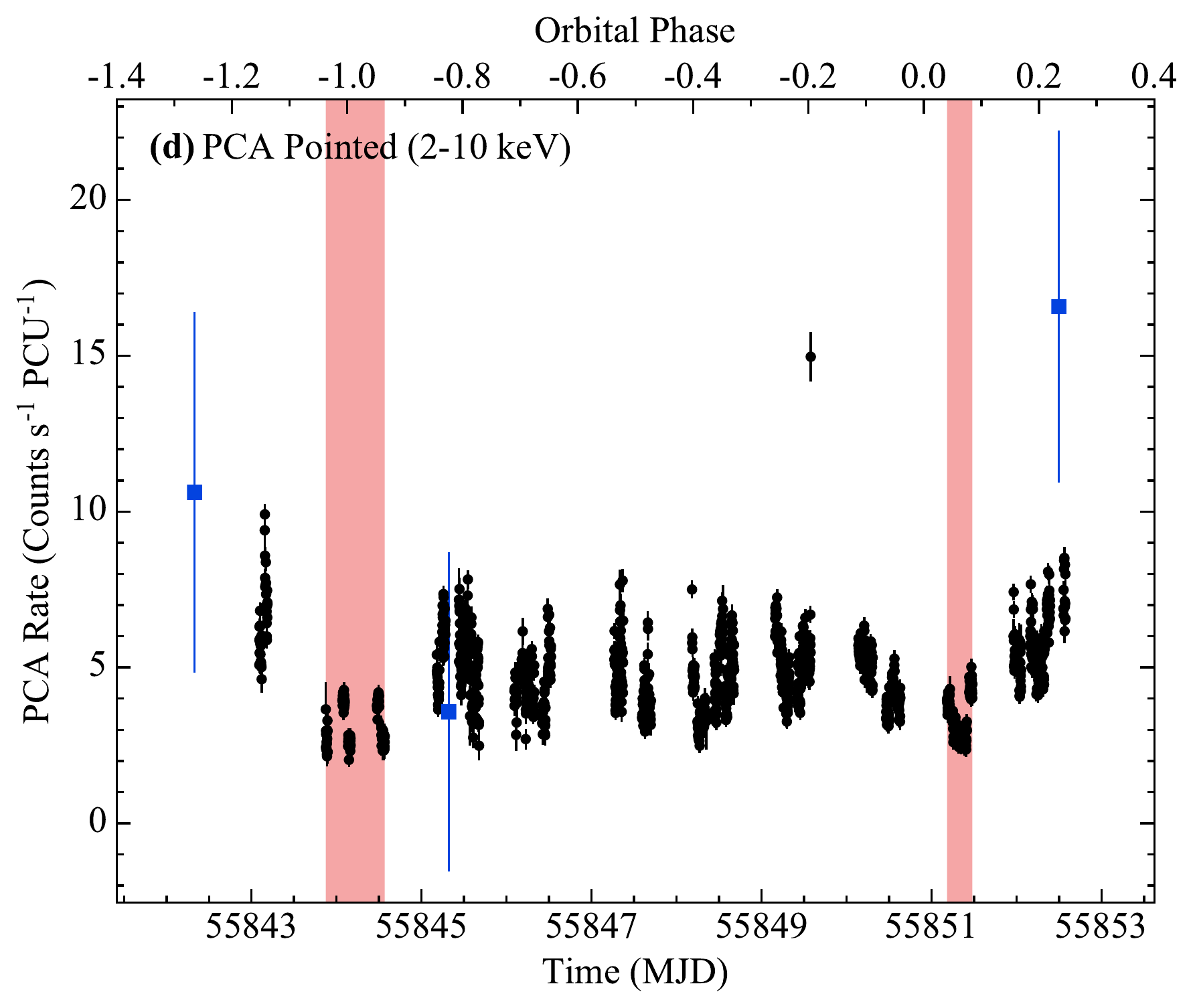}
		\label{Figure:Figure2d}
	}
	
	\end{tabular}
	
	\caption{(a)~\textit{RXTE}~PCA~scan~\text{(2--10\,keV)} weighted average light curve of \text{IGR~J16493--4348} using 30\,day bin widths. (b)~\textit{Swift}~BAT~\text{70-month} snapshot~\text{(14--195\,keV)} and (c)~\textit{Swift}~BAT~transient monitor~\text{(15--50\,keV)} weighted average light curves of \text{IGR~J16493--4348} using bin widths equal to twice the 6.7828\,day orbital period. The horizontal uncertainties in \text{Figures~\ref{Figure:Figure2}(a)--(c)} correspond to the half bin widths in the light curves, and the vertical uncertainties are obtained from the standard error. The pointed PCA~observation times are indicated by the blue shaded regions (smaller than the symbol size) in \text{Figures~\ref{Figure:Figure2}(a)--(c).} (d)~\text{Background-subtracted} pointed \textit{RXTE}~PCA~\text{(2--10\,keV)} light curve of \text{IGR~J16493--4348} using all operational~PCUs with 128\,s time resolution. The red shaded regions correspond to observation times with weak pulsed emission, and nearly simultaneous \textit{RXTE}~PCA~scan~\text{(2--10\,keV)} observations are overlaid as blue squares. Orbital phase~0 corresponds to~$T_{\pi/2}$ from circular solution~1 in Table~\ref{Table:OrbitalParameters}.}
	\label{Figure:Figure2}
\end{figure*}

We analyzed the BAT~\text{70-month} snapshot\footnote{See \href{https://swift.gsfc.nasa.gov/results/bs70mon}{https://swift.gsfc.nasa.gov/results/bs70mon.}}~\text{(14--195\,keV)} light curve, which is shown in Figure~\ref{Figure:Figure2}(b) and spans MJDs~53360.0 through~55470.0 \text{(2004~December~21--2010~October~1)}. The light curve is comprised of continuous, individual observations pointed at the same sky location. Exposure times ranged from 150~to~2679\,s, and the mean exposure time was 783\,s. The time resolution is determined by the sampling of the individual observations.

We also analyzed the BAT transient monitor~\text{(15--50\,keV)} light curve between MJDs~53416.0 and~57249.8 \text{(2005~February~15--2015~August~15)}, which is shown in Figure~\ref{Figure:Figure2}(c). Orbital and daily averaged light curves are available through the~\textit{Swift}~NASA~Goddard Space Flight Center~(GSFC) website\footnote{See \href{https://swift.gsfc.nasa.gov/results/transients}{https://swift.gsfc.nasa.gov/results/transients.}} after they have been processed with the data reduction procedures described in~\citet{Krimm2013}. We used the orbital light curve in our analysis, which had typical exposure times ranging from 64~to~2640\,s and a mean exposure time of 720\,s. We excluded ``bad''~times from the light curve, which were indicated by nonzero data quality flag~(\texttt{DATA\_FLAG}) values. A small number of data points with very low fluxes and unusually small uncertainties were also identified and removed, even though they were flagged as ``good''~\citep{Corbet2013}. The BAT~\text{70-month} snapshot and transient monitor light curve times were corrected to the solar system barycenter using the tools available through the~OSU~Department of Astronomy website\textsuperscript{\ref{FootnoteBarycenter}}.



\section{Period Measurements}
\label{Section:Period_Measurements}

The \textit{RXTE}~PCA~scan and \textit{Swift}~BAT light curves were used to search for orbital and superorbital modulation since they were longer in duration. A refined superorbital period measurement was obtained from the \text{semi-weighted}~discrete Fourier transform~(DFT) of the BAT~transient monitor light curve~(see Section~\ref{Section:Superorbital_Period}). In this paper, uncertainties on period measurements obtained from~DFTs were determined according to~\citet{Horne1986}. We report an improved orbital period from an observed minus calculated~\text{(O--C)} analysis of \text{mid-eclipse} times derived from the BAT~transient monitor and PCA~scan light curves using a Bayesian Markov chain Monte Carlo~(MCMC) fitting procedure~(see Section \ref{Section:O-C_Analysis}). Pulsations were detected in the unweighted power spectrum of the pointed~PCA light curve after the removal of low frequency noise~(see Section~\ref{Section:Pulse_Period}). We refine the pulse period using the~ToAs derived from the pulsar timing analysis described in \text{Sections~\ref{Section:Pulsar_Timing_Analysis}--\ref{Section:Pulsar_Timing_Models}}.


\subsection{Superorbital Period}
\label{Section:Superorbital_Period}

\begin{figure}[t]
	\centering
	\includegraphics[trim=0cm 0cm 0cm 0cm, clip=false, scale=0.459, angle=0]{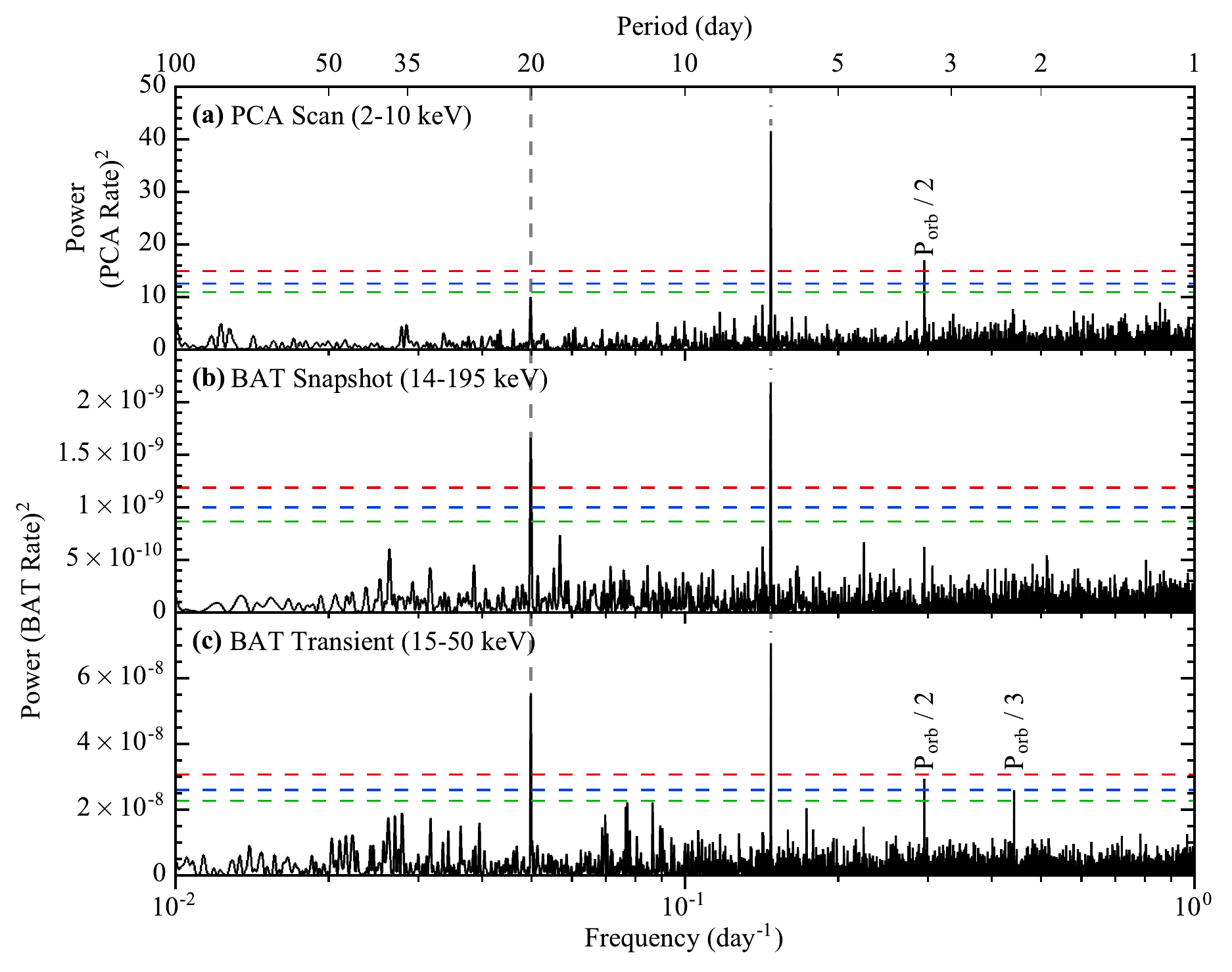}
	\caption{Semi-weighted power spectra of \text{IGR~J16493--4348} using the (a)~\textit{RXTE}~PCA~scan~\text{(2--10\,keV)}, (b)~\textit{Swift}~BAT~\text{70-month} snapshot~\text{(14-195\,keV)}, and (c)~\textit{Swift}~BAT~transient monitor~\text{(15--50\,keV)} light curves. The horizontal dashed lines indicate 95\%~(shown in green), 99\%~(shown in blue), and 99.9\%~(shown in red) significance levels. The grey vertical dashed line corresponds to the 20.067\,day superorbital period from the \text{semi-weighted}~DFT of the BAT~transient monitor data. The 6.7821\,day orbital period from the \text{semi-weighted}~DFT of the BAT~transient monitor light curve is indicated by the grey vertical \text{dot-dashed} line. Significant harmonics of the orbital period are labeled in each power spectrum.}
	\label{Figure:Figure3}
\end{figure}

We searched for orbital and superorbital modulation between 1~and 100\,days in the power spectra of the~PCA~scan,~BAT~\text{70-month} snapshot, and~BAT~transient monitor light curves shown in Figure~\ref{Figure:Figure3}. Each of these power spectra was oversampled by a factor of five compared to their nominal frequency resolution, given by the length of the light curve.

The measurements from the PCA~scan and BAT~light curves showed a wide range of nonuniform error bar sizes. It can be advantageous to use these errors to weight the contribution of each data point when calculating the power spectrum~\citep{Scargle1989}. The \text{semi-weighting} technique uses the error bar of each data point and the excess light curve variability to weight the data points in the power spectrum, which is analogous to a \text{semi-weighted} mean~\citep{Cochran1937, Cochran1954}. \citet{Corbet2007a, Corbet2007b} showed that \text{semi-weighting} can be very beneficial for faint sources, such as \text{IGR~J16493-4348}. We used \text{semi-weighting} in all of the power spectra in Figure~\ref{Figure:Figure3} since it yielded more significant period detections with smaller uncertainties. 

Strong evidence of superorbital modulation near 20.07\,days was found in the power spectra of the PCA~scan, BAT~\text{70-month} snapshot, and BAT~transient monitor light curves~(see~Figure~\ref{Figure:Figure3}). The superorbital period was most significantly detected in the power spectrum of the BAT~transient monitor light curve in~Figure~\ref{Figure:Figure3}(c). We refine the superorbital period to 20.067\,$\pm$\,0.009\,days from a semi-weighted~DFT of the BAT~transient monitor data, which covered an additional 798\,days compared to the data in~\citet{Corbet2013}. A coherent signal well above the 99.9\%~significance level, with a false alarm probability~(FAP;~\citealt{Scargle1982}) of~7\,$\times$\,10$^{\text{--6}}$, was found at this period. The power spectrum of the BAT~\text{70-month} snapshot data in Figure~\ref{Figure:Figure3}(b) showed a peak at 20.07\,$\pm$\,0.02\,days, with a significance level above~99.9\% and a~FAP of~1\,$\times$\,10$^{\text{--4}}$. Evidence of a peak at 20.08\,$\pm$\,0.02\,days, with a~FAP of~5\,$\times$\,10$^{\text{--3}}$, was also found in the power spectrum of the PCA~scan data in~Figure~\ref{Figure:Figure3}(a), but it was less significant than the corresponding peaks in the BAT~power spectra. We note that these superorbital period measurements are all consistent with each other to within~1$\sigma$.

\begin{figure}[t]
	\centering
	\includegraphics[trim=0cm 0cm 0cm 0cm, clip=false, scale=0.459, angle=0]{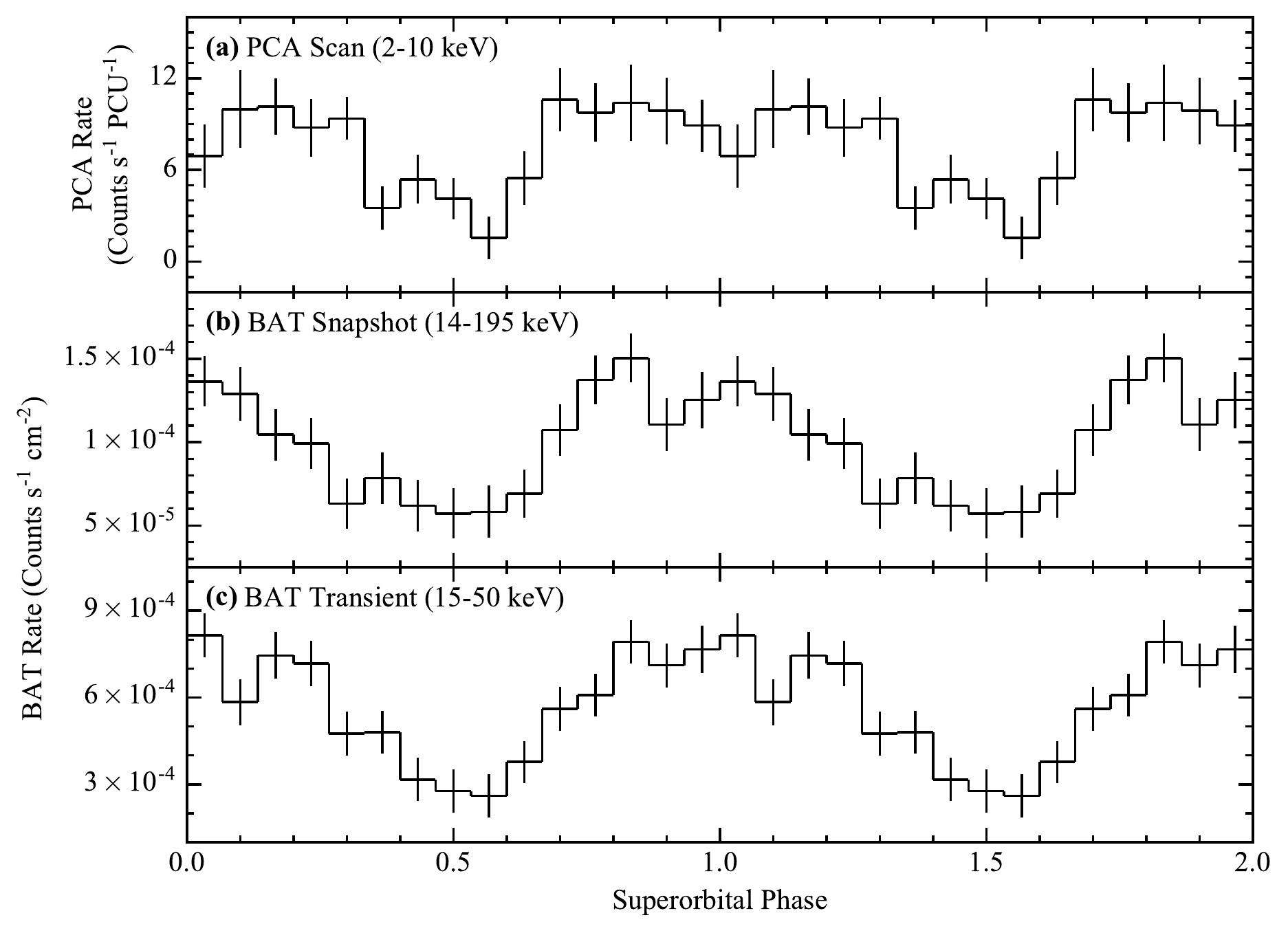}
	\caption{(a)~\textit{RXTE}~PCA~scan~\text{(2--10\,keV)}, (b)~\textit{Swift}~BAT~\text{70-month} snapshot~\text{(14--195\,keV)}, and (c)~\textit{Swift}~BAT~transient monitor~\text{(15--50\,keV)} light curves of \text{IGR~J16493--4348} folded using 15~bins on the 20.067\,day superorbital period measurement from the \text{semi-weighted}~DFT of the BAT~transient monitor data. Superorbital phase~0 corresponds to the time of maximum flux in the BAT~transient monitor data~(MJD\,55329.65647), which was determined from a sine wave fit to the light curve.}
	\label{Figure:Figure4}
\end{figure}

In Figure~\ref{Figure:Figure4}, we show the PCA~scan, BAT~\text{70-month} snapshot, and BAT~transient monitor light curves folded on our refined superorbital period measurement using 15~bins. Superorbital phase~0 in all of the folded light curves is defined to be the time of maximum flux in the BAT~transient monitor data~(MJD\,55329.65647), which was determined from a sine wave fit to the light curve. The folded~PCA and BAT~profiles show \text{quasi-sinusoidal} variability over many superorbital cycles.



Next, we investigated whether there was an energy dependent phase shift between the superorbital modulation detected in the \text{semi-weighted}~DFTs of the~PCA and BAT~light curves by \text{cross-correlating} the folded, binned light curves against each other using Equation~(\ref{Equation:CrossCorrelation}), after applying phase offsets to one of the light curves. Linear interpolation was used to determine the count rates of the phase shifted light curve at the phase bins of the unshifted light curve. These linearly interpolated count rates and the count rates from the unshifted light curve were used to compute the \text{cross-correlation} statistics. The normalized, linear \text{cross-correlation} coefficient between two vectors~$\mathbf{u}$ and~$\mathbf{v}$ of length~$N$ is defined by:
\begin{equation}
r=\mathrm{Re}\frac{\left\langle \mathbf{u,}\mathbf{v}\right\rangle }{\left\Vert \mathbf{u}\right\Vert \left\Vert \mathbf{v}\right\Vert },
\label{Equation:CrossCorrelation}
\end{equation}
where~$\left\Vert\mathbf{u}\right\Vert$ and~$\left\Vert\mathbf{v}\right\Vert$ are the magnitudes of vectors~$\mathbf{u}$ and~$\mathbf{v}$, and~$\mathbf{u}$ and~$\mathbf{v}$ are normalized vectors given by~\mbox{$\mathbf{u}=\mathbf{U}-\bar{U}$} and~\mbox{$\mathbf{v}=\mathbf{V}-\bar{V}$}, with~$\bar{U}$ and~$\bar{V}$ denoting the mean of data vectors~$\mathbf{U}$ and~$\mathbf{V}$, respectively. The inner product between vectors~$\mathbf{u}$ and~$\mathbf{v}$ is given by:
\begin{equation}
\left\langle \mathbf{u,}\mathbf{v}\right\rangle =\sum_{i=1}^{N}u_{i}v_{i}^{*}
\label{Equation:InnerProduct}
\end{equation}

Each pair of light curves was folded on the superorbital period using~$N$\,$=$\,20, 40, 50, 60, and~80~bins. Phase shifts, in steps of $N^{\text{--1}}$, 0.5$N^{\text{--1}}$, 0.1$N^{\text{--1}}$, and~0.01$N^{\text{--1}}$, were applied to the shifted light curve at each iteration during separate analyses using each of these binnings. A total of 20~analyses were performed for each set of light curves, and the superorbital phase corresponding to the maximum \text{cross-correlation} value is given by the average of the superorbital phase bins with maximum \text{cross-correlation} values from all 20~analyses.

Uncertainties on these measurements were derived from a total of 2,000,000~Monte~Carlo simulations, where 100,000~simulations were performed for each binning and phase shift value. At the beginning of each Monte~Carlo iteration, we replaced each count rate in both of the unfolded light curves with a value selected randomly from a Gaussian distribution, whose mean was equal to the measured count rate from the unmodified light curves and standard deviation was given by its associated uncertainty. The resultant light curves were then folded on the superorbital period, binned, and \text{cross-correlated} using Equation~(\ref{Equation:CrossCorrelation}). The error for each set of Monte~Carlo analyses was calculated from the standard deviation of the superorbital phase bins with maximum \text{cross-correlation} values, and we quote an uncertainty given by the average of all the standard deviations produced by the Monte~Carlo procedure.

We find a maximum \text{cross-correlation} value between the folded PCA~scan and BAT~transient monitor light curves at phase~0.02\,$\pm$\,0.04 of the superorbital period. We repeated this analysis for the PCA~scan and BAT~\text{70-month} snapshot light curves, and also pairing the BAT~\text{70-month} snapshot and transient monitor light curves, and found that the maximum \text{cross-correlation} value occured at superorbital phases~\text{--0.02}\,$\pm$\,0.04 and~0.00\,$\pm$\,0.05, respectively. No significantly detected phase offset is observed between the folded~PCA and BAT~superorbital profiles, which indicates that an energy dependent phase shift is not present.


\subsection{Orbital Period}
\label{Section:Orbital_Period}

Highly significant peaks were detected in the power spectra shown in Figure~\ref{Figure:Figure3} at the previously reported 6.8\,day orbital period~\citep{Corbet2010a, Cusumano2010}. We measured orbital periods of~6.784\,$\pm$\,0.001, 6.788\,$\pm$\,0.002, and 6.7821\,$\pm$\,0.0008\,days from \text{semi-weighted}~DFTs of the PCA~scan, BAT~\text{70-month} snapshot, and BAT~transient monitor light curves. The~FAPs were 3\,$\times$\,10$^{\text{--15}}$, 5\,$\times$\,10$^{\text{--9}}$, and~2\,$\times$\,10$^{\text{--12}}$, respectively. The orbital period was most significantly detected in the power spectrum of the PCA~scan light curve, but the BAT~transient monitor data yielded the most precise orbital period measurement due to the longer light curve duration. Harmonics of the orbital period were also significantly detected in the power spectra of the PCA~scan and BAT~transient monitor light curves and are labeled in Figures~\ref{Figure:Figure3}(a) and~\ref{Figure:Figure3}(c).


\newpage

\subsubsection{Observed Minus Calculated Analysis}
\label{Section:O-C_Analysis}

We carried out an \text{O--C}~analysis to obtain improved measurements of \text{IGR~J16493--4348's} orbital period and orbital period derivative using observed \text{mid-eclipse} times from the BAT~transient monitor and PCA~scan light curves. The BAT~light curve was divided into six 638\,day time intervals, and the PCA~light curve was split into two 1348\,day segments. At the beginning of the first iteration, we folded each of these divided light curves on the 6.7821\,day orbital period from the semi-weighted~DFT of the BAT~transient monitor light curve. Each divided BAT~light curve was folded on the orbital period using 200~bins. Since the PCA~light curves were sampled every $\sim$5\,days on average, they were not binned to prevent \text{cycle-to-cycle} source brightness variations from affecting the folded orbital profiles.

\begin{figure}[t]
	\centering
	\includegraphics[trim=0cm 0cm 0cm 0cm, clip=false, scale=0.37, angle=0]{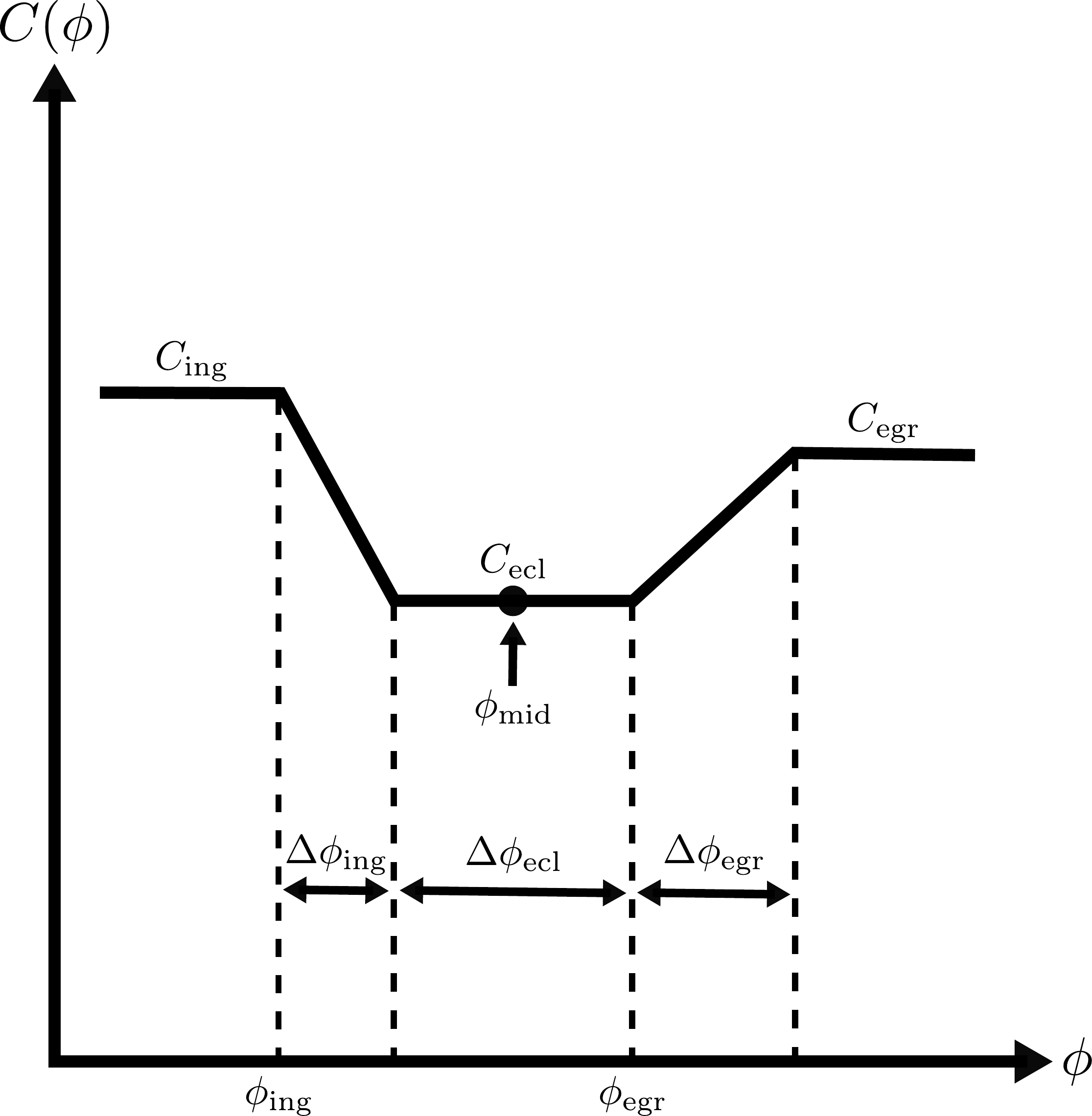}
	\caption{Schematic of the asymmetric step and ramp function in Equation~(\ref{Equation:StepAndRamp}), which was used to model the eclipses in the folded \textit{RXTE}~PCA~scan~\text{(2--10\,keV)}, \textit{Swift}~BAT \text{70-month} snapshot~\text{(14--195\,keV)}, and \textit{Swift}~BAT transient monitor~\text{(15--50\,keV)} light curves.}
	\label{Figure:Figure5}
\end{figure}

Eclipses were only visible in the~BAT and PCA~scan light curves after folding the data on the orbital period. We modeled the eclipses in each folded light curve using asymmetric and symmetric step and ramp functions defined in Equation~(\ref{Equation:StepAndRamp}), where the intensities before ingress, during eclipse, and after egress were assumed to remain constant and change linearly during the ingress and egress transitions~\citep{Coley2015}. The symmetric model imposes constraints requiring that both the ingress and egress durations and \text{pre-ingress} and \text{post-egress} count rates be equal. In the asymmetric model, these constraints were removed and the ingress duration, egress duration, and count rates before ingress and after egress were independent free parameters in the model. The adjustable parameters in these models were the: phases corresponding to the start of ingress and start of egress,~$\phi_{\text{ing}}$ and~$\phi_{\text{egr}}$, ingress duration,~$\Delta\phi_{\text{ing}}$, egress duration,~$\Delta\phi_{\text{egr}}$, \text{pre-ingress} count rate,~$C_{\text{ing}}$, \text{post-egress} count rate,~$C_{\text{egr}}$, and eclipse count rate,~$C_{\text{ecl}}$. $C_{\text{ing}}$~was fit from orbital phase \mbox{$\phi$\,$=$\,--0.2} to the start of ingress, $C_{\text{ecl}}$~was fit during the eclipse, and $C_{\text{egr}}$~was fit from the end of egress to orbital phase $\phi$\,$=$\,0.2. A schematic of the asymmetric eclipse model is shown in Figure~\ref{Figure:Figure5}. The eclipse duration was calculated using Equation~(\ref{Equation:EclipseDuration}), and the \text{mid-eclipse} phase was found using Equation~(\ref{Equation:MidEclipsePhase}). The eclipse \text{half-angle} is defined by Equation~(\ref{Equation:EclipseHalfAngle}).


\begin{widetext}

\begin{equation}
C(\phi)=\left\{ \begin{array}{lc}
C_{\text{ing}}, & -0.2\le\phi\le\phi_{\text{ing}}\\
C_{\text{ing}}+\left(\frac{C_{\text{ecl}}-C_{\text{ing}}}{\Delta\phi_{\text{ing}}}\right)(\phi-\phi_{\text{ing}}), & \phi_{\text{ing}}\le\phi\le\phi_{\text{ing}}+\Delta\phi_{\text{ing}}\\
C_{\text{ecl}}, & \phi_{\text{ing}}+\Delta\phi_{\text{ing}}\le\phi\le\phi_{\text{egr}}\\
C_{\text{ecl}}+\left(\frac{C_{\text{egr}}-C_{\text{ecl}}}{\Delta\phi_{\text{egr}}}\right)(\phi-\phi_{\text{egr}}), & \phi_{\text{egr}}\le\phi\le\phi_{\text{egr}}+\Delta\phi_{\text{egr}}\\
C_{\text{egr}}, & \phi_{\text{egr}}+\Delta\phi_{\text{egr}}\le\phi\le0.2
\end{array}\right.
\label{Equation:StepAndRamp}
\end{equation}

\begin{equation}
\Delta\phi_{\text{ecl}}=\phi_{\text{egr}}-\left(\phi_{\text{ing}}+\Delta\phi_{\text{ing}}\right)
\label{Equation:EclipseDuration}
\end{equation}

\begin{equation}
\phi_{\text{mid}}=\frac{1}{2}\left[\phi_{\text{egr}}+\left(\phi_{\text{ing}}+\Delta\phi_{\text{ing}}\right)\right]
\label{Equation:MidEclipsePhase}
\end{equation}

\begin{equation}
\Theta_{e}=\Delta\phi_{\text{ecl}}\times180^{\circ}
\label{Equation:EclipseHalfAngle}
\end{equation}

\end{widetext}


Flares were excluded when fitting the eclipse models to the folded PCA~scan light curves, which were identified by data points with count rates above 20\,counts\,s$^{\text{--1}}$\,PCU$^{\text{--1}}$ at orbital phases near the start of ingress or end of egress. This resulted in the removal of approximately~6\% of the data from the fitted PCA~scan light curves. We chose to remove these data points with high count rates since they increased the fitted $\chi^{\mathrm{2}}$ values, but did not significantly affect the \text{best-fit} parameters or their uncertainties.

Observed \text{mid-eclipse} times from each folded light curve were determined using Equation~(\ref{Equation:MidEclipsePhase}). We fit the observed \text{mid-eclipse} times using the orbital change function:
\begin{equation}
T_{n}=T_{\pi/2}+nP_{\text{orb}}+\frac{1}{2}n^{2}P_{\text{orb}}\dot{P}_{\text{orb}},
\label{Equation:OrbitalChangeFunction}
\end{equation}
where~$T_{n}$ is the \text{mid-eclipse} time in days, $n$~is the nearest integer number of elapsed binary orbits, $P_{\text{orb}}$~is the orbital period in days, and $\dot{P}_{\text{orb}}$~is the orbital period derivative at~$T_{\pi/2}$. Each \text{mid-eclipse} time was weighted by its maximum asymmetric error in Table~\ref{Table:MidEclipseTimesOC} during the fitting procedure. After each iteration, the orbital period and $T_{\pi/2}$~were updated with the values obtained from fitting the \text{mid-eclipse} times with the orbital change function in Equation~(\ref{Equation:OrbitalChangeFunction}), and these values were used to refold the~BAT and PCA~light curves in the next iteration. The \text{O--C}~procedure was repeated until there were no significant changes in the orbital period and~$T_{\pi/2}$ between successive iterations.

We observed that many of the parameters in these models were highly covariant from projections of their posterior distributions. A Bayesian~MCMC fitting procedure was used to incorporate these covariances into the model parameter uncertainties by marginalizing over \text{multi-dimensional} joint posterior distributions. From Bayes' theorem, the posterior probability of a set of model parameters,~$\bm{\theta}$, given the observed data,~$\bm{D}$, and any prior information,~$\bm{I}$, is defined by:
\begin{equation}
p(\bm{\theta}\vert \bm{D},\bm{I})=\frac{p(\bm{D}\vert\bm{\theta},\bm{I})p(\bm{\theta}\vert \bm{I})}{p(\bm{D}\vert \bm{I})}
\label{Equation:BayesTheorem}
\end{equation}
Here, \mbox{$p(\bm{D}\vert\bm{\theta},\bm{I})=\mathcal{L}(\bm{\theta}\vert\bm{D},\bm{I})$} is the likelihood function, \mbox{$p(\bm{\theta}\vert\bm{I})=\pi(\bm{\theta}\vert\bm{I})$} is the prior probability distribution for the model parameters, and $p(\bm{D}\vert\bm{I})$ is the marginal likelihood function. The marginal likelihood function can be thought of as a normalization constant, determined by requiring the posterior probability integrate to unity when integrating over all of the parameters in the model. Marginalized single parameter posterior distributions were obtained by integrating the joint posterior distribution over the remaining parameters:
\begin{equation}
p(\theta_{i}\vert \bm{D},\bm{I})\propto\int_{\mathcal{V}}d^{n}\bm{\theta'}\mathcal{L}(\bm{\theta}\vert\bm{D},\bm{I})\pi(\bm{\theta}\vert \bm{I}),
\label{Equation:MarginalizedPostDist}
\end{equation}
where $\bm{\theta'}$ is a parameter vector equal to $\bm{\theta}$ excluding $\theta_{i}$ and $\mathcal{V}$ is the integration volume of the parameter space. We assumed uninformed, flat priors on all of our model parameters and used a Gaussian likelihood function, such that \mbox{$\mathcal{L}(\bm{\theta}\vert\bm{D},\bm{I})\propto\text{exp}(-\chi^{2}/2)$}.

An \text{affine-invariant}~MCMC ensemble sampler~\citep{Goodman2010}, implemented in \texttt{emcee}\footnote{See \href{http://dfm.io/emcee/current}{http://dfm.io/emcee/current.}} by~\citet{ForemanMackey2013}, was used to sample the posterior probability density functions~(PDFs) of the model parameters in Equations~(\ref{Equation:StepAndRamp}) and~(\ref{Equation:OrbitalChangeFunction}). The parameter spaces were explored using 200~walkers and a chain length of 1,500~steps per walker. The first 500~steps in each chain were treated as the initial \text{burn-in} phase and were removed from the analysis. The position of each walker was updated using the current positions of all of the other walkers in the ensemble~\citep{Goodman2010}. We initialized the walkers to start from a small Gaussian ball centered around the parameter values obtained from maximizing the likelihood function subject to the constraints given by the priors. The posterior distributions of the model parameters were calculated using the remaining 1,000~steps in each chain. \text{Best-fit} values for the model parameters were derived from the median of the marginalized posterior distributions, and we quote 1$\sigma$ uncertainties using Bayesian credible intervals.

\begin{figure*}[t]
	\centering
	\begin{tabular}{ccc}
		
		\subfigure
		{
			\includegraphics[trim=0cm 0cm 0cm 0cm, clip=false, scale=0.453, angle=0]{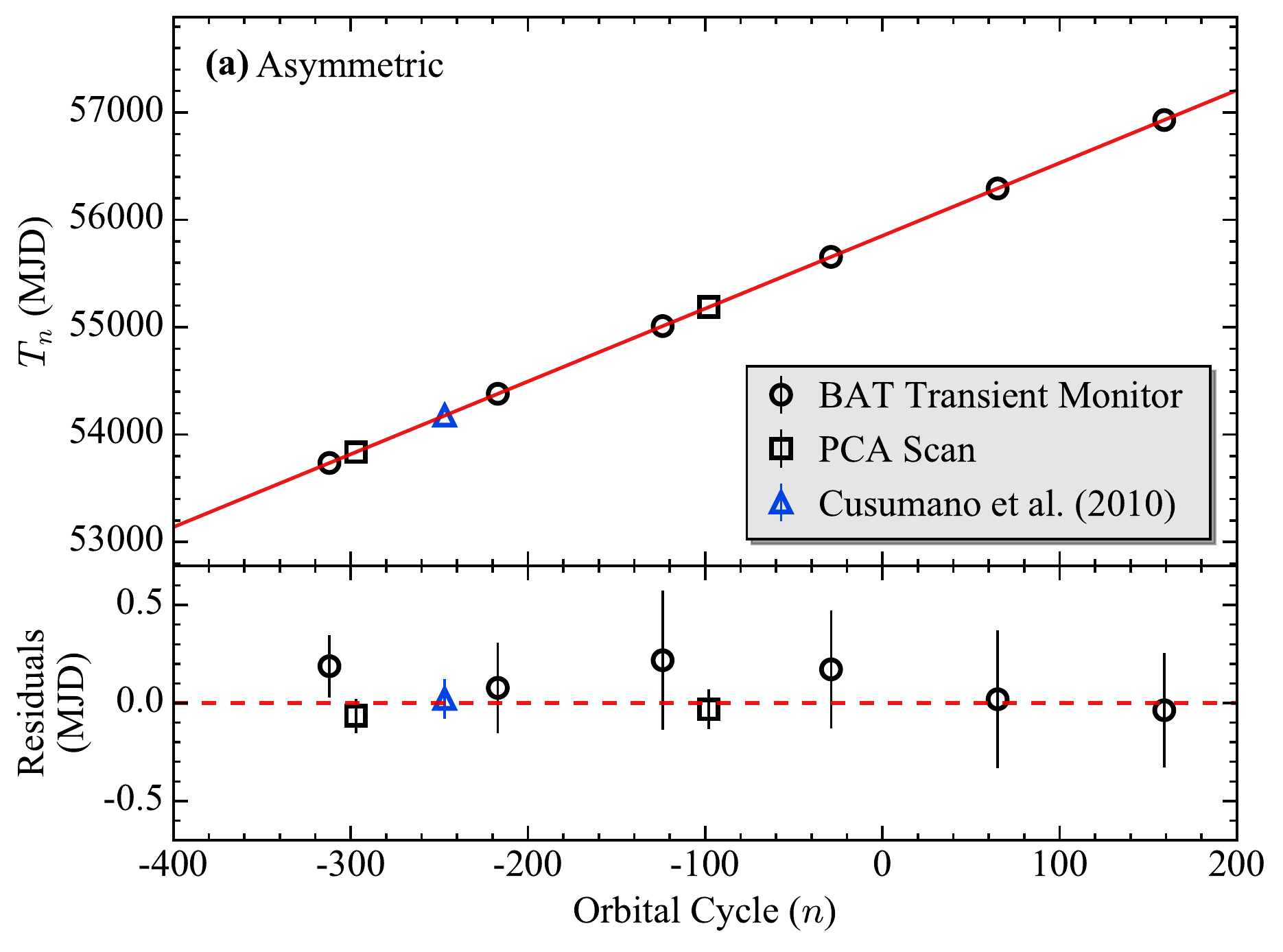}
			\label{Figure:Figure6a}
		}
		
		& &
		
		\subfigure
		{
			\includegraphics[trim=0cm 0cm 0cm 0cm, clip=false, scale=0.453, angle=0]{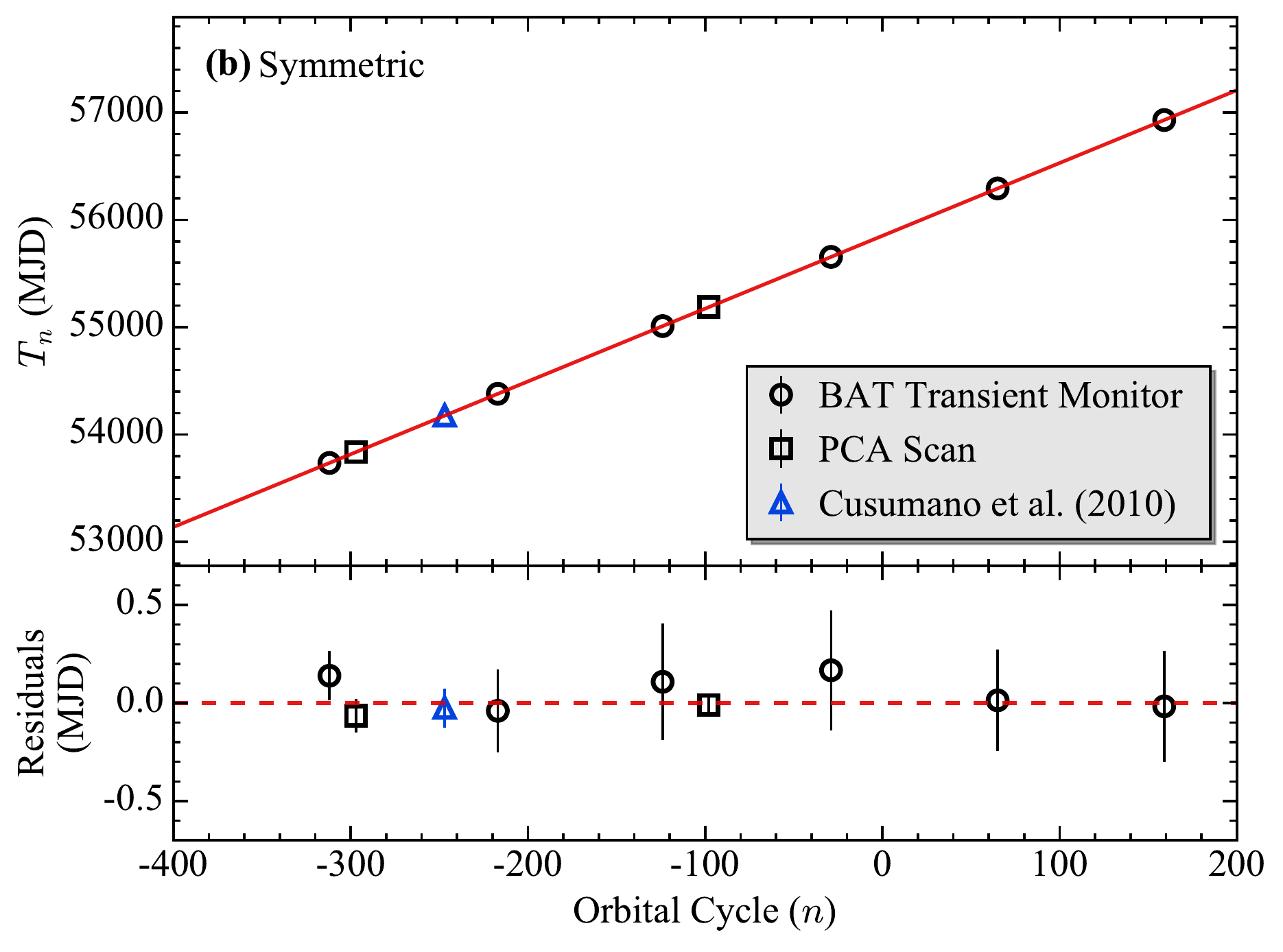}
			\label{Figure:Figure6b}
		}
		
	\end{tabular}
	
	\caption{Top panels:~Observed \text{mid-eclipse} times of \text{IGR~J16493--4348} obtained from an \text{O--C}~analysis using the \textit{RXTE}~PCA~scan~\text{(2--10\,keV)} and \textit{Swift}~BAT~transient monitor~\text{(15--50\,keV)} light curves and (a)~asymmetric and (b)~symmetric eclipse models. The solid red line corresponds to the \text{best-fit} orbital change function using Equation~(\ref{Equation:OrbitalChangeFunction}). Each \text{mid-eclipse} time was weighted by its maximum asymmetric error in Table~\ref{Table:MidEclipseTimesOC} during the fitting procedure. Bottom panels:~Residuals determined by subtracting the best fit from the \text{mid-eclipse} times. \text{Mid-eclipse} times derived from the BAT~transient monitor and PCA~scan light curves are represented by black open circles and black open squares, respectively. The \text{mid-eclipse} time measurement reported by~\citet{Cusumano2010} is indicated with blue open triangles.}
	\label{Figure:Figure6}
\end{figure*}

In Table~\ref{Table:MidEclipseTimesOC}, we list the observed \text{mid-eclipse} times obtained from the \text{O--C}~analysis using asymmetric and symmetric eclipse models. These measurements are also plotted in the top panels of Figures~\ref{Figure:Figure6a} and~\ref{Figure:Figure6b}, along with the \text{best-fit} orbital change functions. The residuals were derived by subtracting the fits from the \text{mid-eclipse} times and are shown in the bottom panels of these figures. We note that our \text{mid-eclipse} times are consistent with the \text{mid-eclipse} time reported by~\citet{Cusumano2010} using \textit{Swift}~BAT survey~\text{(15--50\,keV)} data, which we indicate with blue triangles in these plots.

We refine the orbital period to 6.7828\,$\pm$\,0.0004\,days using an asymmetric eclipse model and a fiducial \text{mid-eclipse} time of~$T_{\pi/2}$\,$=$\,MJD\,55851.3\,$\pm$\,0.1 in our \text{O--C}~analysis. A consistent orbital period of 6.7825\,$\pm$\,0.0004\,days was found using a symmetric eclipse model with~$T_{\pi/2}$\,$=$\,MJD\,55851.21\,$\pm$\,0.07. Orbital period derivatives of 0.01$^{\text{+1.74}}_{\text{--1.77}}$\,$\times$\,10$^{\text{--7}}$\,d\,d$^{\text{--1}}$ and 0.09$^{\text{+1.69}}_{\text{--1.73}}$\,$\times$\,10$^{\text{--7}}$\,d\,d$^{\text{--1}}$ were measured by fitting the asymmetric and symmetric \text{mid-eclipse} times with the orbital change function in Equation~(\ref{Equation:OrbitalChangeFunction}), respectively. These values indicate that there was no significant change in the orbital period over approximately 500~orbital cycles.

We selected the period obtained from using an asymmetric eclipse model in the \text{O--C}~analysis as our preferred orbital period measurement. Since most eclipsing~sgHMXBs show evidence of asymmetry in their \text{X-ray} eclipses~(e.g.,~\citealt{Falanga2015}), we argue that an asymmetric eclipse model is more representative of the eclipse behavior in these systems. Additionally, constraints on the eclipse transition durations and count rates outside of the eclipses could introduce systematic errors when fitting the folded light curves with a symmetric eclipse model.



\addtocounter{table}{1}

\begin{deluxetable*}{cccc} 
	\tablenum{2}
	\tabletypesize{\small}
	\tablecolumns{4}
	\tablewidth{0pt}
	\tablecaption{\textsc{\text{Mid-Eclipse} Times of \text{IGR~J16493--4348} from \text{O--C}~Analysis}}
	\tablehead{
		\colhead{Observation} &
		\colhead{Orbital Cycle} &
		\colhead{Mid-Eclipse Time$^\mathrm{a}$} &
		\colhead{Mid-Eclipse Time$^\mathrm{b}$} \\
		\colhead{} &
		\colhead{($n$)} &
		\colhead{(MJD)} &
		\colhead{(MJD)}
	}
	\startdata
	BAT~Transient Monitor & --312 & 53735.2$^{\text{+0.2}}_{\text{--0.1}}$ & 53735.2\,$\pm$\,0.1  \\
	PCA~Scan & --297 & 53836.69$^{\text{+0.09}}_{\text{--0.08}}$ & 53836.76$^{\text{+0.09}}_{\text{--0.07}}$  \\
	BAT~Transient Monitor & --217 &  54379.5\,$\pm$\,0.2 & 54379.4\,$\pm$\,0.2  \\
	BAT~Transient Monitor & --124 & 55010.4$^{\text{+0.4}}_{\text{--0.3}}$ & 55010.3\,$\pm$\,0.3  \\
	PCA~Scan & --98 & 55186.51$^{\text{+0.10}}_{\text{--0.09}}$ & 55186.52\,$\pm$\,0.06 \\
	BAT~Transient Monitor & --29 & 55654.7$^{\text{+0.2}}_{\text{--0.3}}$ & 55654.7\,$\pm$\,0.3  \\
	BAT~Transient Monitor & 65 & 56292.2$^{\text{+0.4}}_{\text{--0.2}}$ & 56292.1\,$\pm$\,0.3  \\
	BAT~Transient Monitor & 159 & 56929.7$^{\text{+0.3}}_{\text{--0.2}}$ & 56929.6$^{\text{+0.3}}_{\text{--0.2}}$
	\enddata
	\tablecomments{We quote 1$\sigma$ uncertainties using Bayesian credible intervals. \\
		$^\mathrm{a}$ Obtained using an asymmetric eclipse model. \\
		$^\mathrm{b}$ Obtained using a symmetric eclipse model.}
	\label{Table:MidEclipseTimesOC}
\end{deluxetable*}



\subsubsection{Folded Orbital Profiles}
\label{Section:Folded_Orbital_Profiles}

\begin{figure*}[t]
	\centering
	\includegraphics[trim=0cm 0cm 0cm 0cm, clip=false, scale=0.65, angle=0]{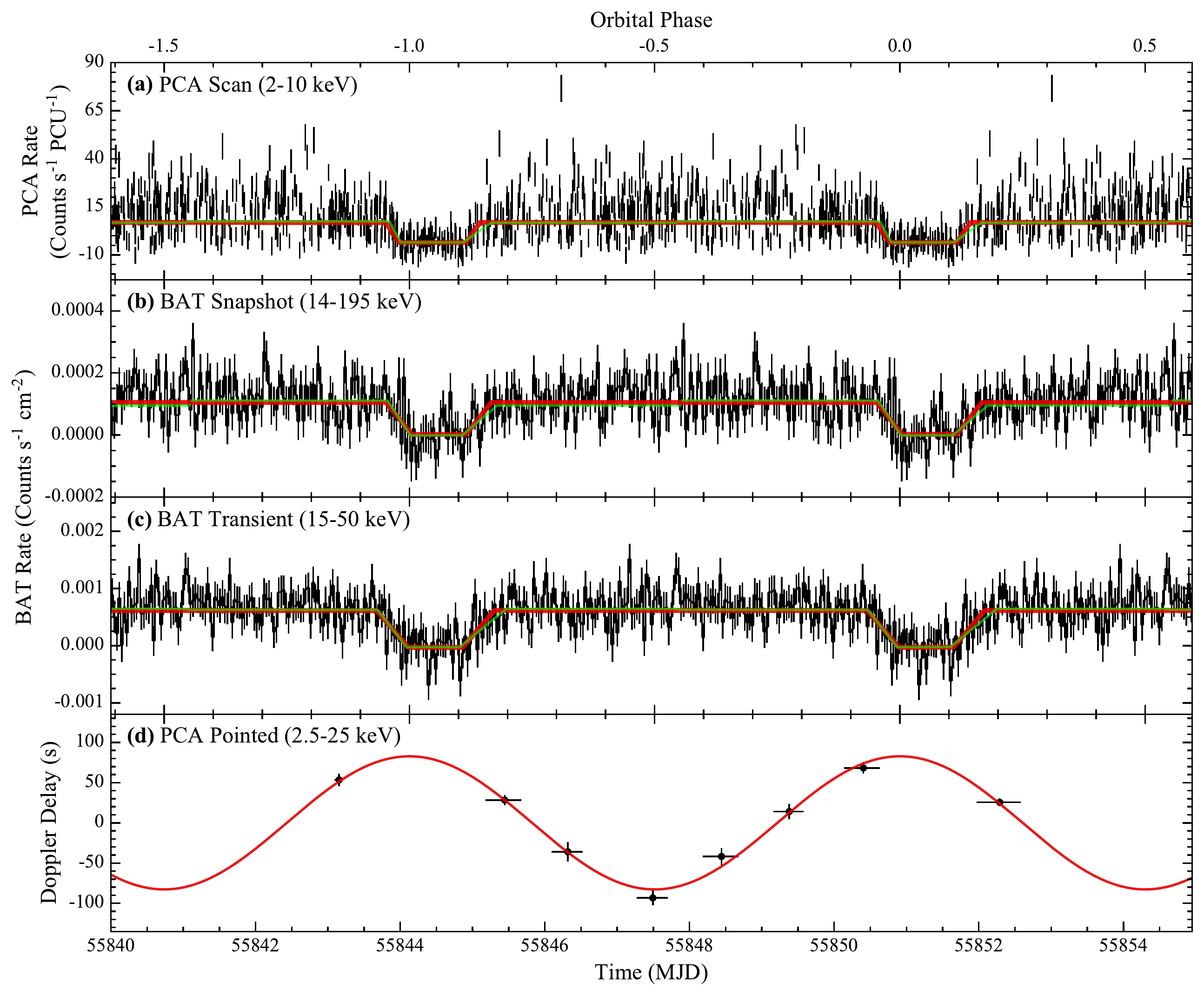}
	\caption{(a)~\textit{RXTE}~PCA~scan~\text{(2--10\,keV),} (b)~\textit{Swift}~BAT~\text{70-month} snapshot~\text{(14--195\,keV),} and (c)~\textit{Swift}~BAT~transient monitor~\text{(15--50\,keV)} light curves of \text{IGR~J16493--4348} folded on the refined 6.7828\,day orbital period from the \text{O--C}~analysis in Section~\ref{Section:O-C_Analysis}. The BAT~light curves were folded using 200~bins. The PCA~scan light curve was not binned to prevent \text{cycle-to-cycle} source brightness variations from affecting the folded orbital profile. We overlay the asymmetric~(shown in green) and symmetric~(shown in red) step and ramp eclipse models from Tables~\ref{Table:AsymmetricEclipseModel} and~\ref{Table:SymmetricEclipseModel}. Discontinuities in the asymmetric eclipse model are included at half orbital cycles from the \text{mid-eclipse} times. (d)~Orbital Doppler delay times measured during the final iteration of the pulsar timing analysis using the pointed \textit{RXTE}~PCA~\text{(2.5--25\,keV)} light curve of \text{IGR~J16493--4348}. The uncertainties on the~ToAs correspond to the statistical errors obtained from Monte Carlo simulations and do not include the additional 3.1\,s systematic uncertainty from circular solution~2 in Table~\ref{Table:OrbitalParameters}. The horizontal error bars indicate the duration of the light curve segments used to derive the~ToAs. The red curve shows the predicted delay times using the fit from circular solution~1 in Table~\ref{Table:OrbitalParameters}, which assumes a constant neutron star rotational period. Orbital phase~0 corresponds to~$T_{\pi/2}$ from circular solution~1 in Table~\ref{Table:OrbitalParameters}.}
	\label{Figure:Figure7}
\end{figure*}

Orbital profiles were produced by folding the PCA~scan, BAT~\text{70-month} snapshot, and BAT~transient monitor light curves on the orbital periods from the \text{O--C}~analysis in Section~\ref{Section:O-C_Analysis}. Asymmetric and symmetric eclipse models, defined in Equation~(\ref{Equation:StepAndRamp}), were fit to each of the folded light curves using the Bayesian~MCMC procedure described in Section~\ref{Section:O-C_Analysis} with the same number of walkers and chain lengths. In Tables~\ref{Table:AsymmetricEclipseModel} and~\ref{Table:SymmetricEclipseModel}, we list the \text{best-fit} eclipse model parameters from the median of the marginalized posterior distributions, along with 1$\sigma$ uncertainties using Bayesian credible intervals. The folded orbital profiles are shown in \text{Figures~\ref{Figure:Figure7}(a)--(c),} and the \text{best-fit} asymmetric and symmetric eclipse models are overlaid in green and red, respectively.

The \text{mid-eclipse} times~($T_{\text{mid}}$), calculated in Tables~\ref{Table:AsymmetricEclipseModel} and~\ref{Table:SymmetricEclipseModel} using Equation~(\ref{Equation:MidEclipsePhase}), are consistent with each other at the 1$\sigma$ level. Fitting an asymmetric eclipse model to the folded PCA~scan, BAT~\text{70-month} snapshot, and BAT~transient monitor light curves yielded eclipse durations of 0.9\,$\pm$\,0.1, 0.7$^{\text{+0.2}}_{\text{--0.3}}$, and 0.7\,$\pm$\,0.3\,days, respectively. Using a symmetric eclipse model, we measured eclipse lengths of 0.92$^{\text{+0.10}}_{\text{--0.09}}$, 0.8\,$\pm$\,0.2, and 0.8\,$\pm$\,0.2\,days, respectively. These eclipse durations are all consistent with each other to within 1$\sigma$ and agree well with the $\sim$0.8\,day eclipse length reported by~\citet{Cusumano2010} from BAT~survey observations. There were no statistically significant differences between the ingress and egress durations or \text{pre-ingress} and \text{post-egress} count rates obtained from fitting the light curves with an asymmetric eclipse model. This suggests that \text{large-scale} structure in the stellar wind from a strong accretion or photoionization wake is unlikely.



\addtocounter{table}{1}

\begin{deluxetable*}{cccc} 
	\tablenum{3}
	\tabletypesize{\small}
	\tablecolumns{4}
	\tablewidth{0pt}
	\tablecaption{\textsc{Asymmetric Eclipse Model Parameters of \text{IGR~J16493--4348}}}
	\tablehead{
		\colhead{Model Parameter} &
		\colhead{\textit{RXTE}~PCA} &
		\colhead{\textit{Swift}~BAT} &
		\colhead{\textit{Swift}~BAT} \\
		\colhead{} & 
		\colhead{Galactic Bulge Scans} &
		\colhead{\text{70-month} Snapshots} &
		\colhead{Transient Monitor} \\
		\colhead{} &
		\colhead{(2--10\,keV)} &
		\colhead{(14--195\,keV)} &
		\colhead{(15--50\,keV)}
	}
	\startdata
	$\phi_{\text{ing}}$ & --0.092$^{\text{+0.005}}_{\text{--0.007}}$ & --0.10$^{\text{+0.02}}_{\text{--0.01}}$ & --0.12$^{\text{+0.01}}_{\text{--0.02}}$ \\
	$\phi_{\text{egr}}$ & 0.059$^{\text{+0.009}}_{\text{--0.011}}$ & 0.06$^{\text{+0.02}}_{\text{--0.03}}$ & 0.05$^{\text{+0.01}}_{\text{--0.03}}$ \\
	$\Delta\phi_{\text{ing}}$ & 0.023$^{\text{+0.011}}_{\text{--0.009}}$ & 0.05$^{\text{+0.03}}_{\text{--0.02}}$ & 0.06$^{\text{+0.05}}_{\text{--0.03}}$ \\
	$\Delta\phi_{\text{egr}}$ & 0.06$^{\text{+0.04}}_{\text{--0.03}}$ & 0.07$^{\text{+0.05}}_{\text{--0.04}}$ & 0.09\,$\pm$\,0.04 \\
	$C_{\text{ing}}$ & 7.5\,$\pm$\,0.8$^{\mathrm{a}}$ & 0.11\,$\pm$\,0.01$^{\mathrm{b}}$ & 0.63\,$\pm$\,0.07$^{\mathrm{b}}$ \\
	$C_{\text{egr}}$ & 6.7\,$\pm$\,1.1$^{\mathrm{a}}$ & 0.09$^{\text{+0.02}}_{\text{--0.01}}$$^{\mathrm{b}}$ & 0.64$^{\text{+0.10}}_{\text{--0.08}}$$^{\mathrm{b}}$ \\
	$C_{\text{ecl}}$ & --3.5\,$\pm$\,0.6$^{\mathrm{a}}$ & --0.001$^{\text{+0.013}}_{\text{--0.014}}$$^{\mathrm{b}}$ & --0.03$^{\text{+0.06}}_{\text{--0.07}}$$^{\mathrm{b}}$ \\
	\tableline
	$\Delta\phi_{\text{ecl}}$ & 0.13$^{\text{+0.01}}_{\text{--0.02}}$ & 0.11\,$\pm$\,0.04 & 0.11$^{\text{+0.05}}_{\text{--0.04}}$ \\
	$P_{\text{orb}}$$^{\mathrm{c}}$ & 6.7828\,$\pm$\,0.0004 & 6.7828\,$\pm$\,0.0004 & 6.7828\,$\pm$\,0.0004 \\
	$\dot{P}_{\text{orb}}$$^{\mathrm{d}}$ & 0.01$^{\text{+1.74}}_{\text{--1.77}}$ & 0.01$^{\text{+1.74}}_{\text{--1.77}}$ & 0.01$^{\text{+1.74}}_{\text{--1.77}}$ \\
	$T_{\text{mid}}$$^{\mathrm{e}}$ & 55851.2\,$\pm$\,0.1 & 55851.3\,$\pm$\,0.2 & 55851.2\,$\pm$\,0.2 \\
	$\Theta_{e}$$^{\mathrm{f}}$ & 22.9$^{\text{+2.7}}_{\text{--2.8}}$ & 19.0$^{\text{+6.4}}_{\text{--7.9}}$ & 19.5$^{\text{+9.2}}_{\text{--7.3}}$ \\
	\tableline
	$\chi_{\nu}^{\mathrm{2}}$ (dof) & 1.22 (197) & 1.10 (73) & 1.16 (73)
	\enddata
	\tablecomments{We quote 1$\sigma$ uncertainties using Bayesian credible intervals. Phase~0 is defined at \text{mid-eclipse}. \\
		$^\mathrm{a}$ Units are counts\,s$^{\text{--1}}$\,PCU$^{\text{--1}}$. \\
		$^\mathrm{b}$ Units are 10$^{\text{--3}}$\,counts\,cm$^{\text{--2}}$\,s$^{\text{--1}}$. \\
		$^\mathrm{c}$ Refined orbital period from \text{O--C}~analysis. Units are days. \\
		$^\mathrm{d}$ Orbital period derivative from orbital change function. Units are~10$^{\text{--7}}$\,d\,d$^{\text{--1}}$. \\
		$^\mathrm{e}$ Units are MJD. \\
		$^\mathrm{f}$ Units are degrees.}
	\label{Table:AsymmetricEclipseModel}
\end{deluxetable*}



\addtocounter{table}{1}

\begin{deluxetable*}{cccc} 
	\tablenum{4}
	\tabletypesize{\small}
	\tablecolumns{4}
	\tablewidth{0pt}
	\tablecaption{\textsc{Symmetric Eclipse Model Parameters of \text{IGR~J16493--4348}}}
	\tablehead{
		\colhead{Model Parameter} &
		\colhead{\textit{RXTE}~PCA} &
		\colhead{\textit{Swift}~BAT} &
		\colhead{\textit{Swift}~BAT} \\
		\colhead{} & 
		\colhead{Galactic Bulge Scans} &
		\colhead{\text{70-month} Snapshots} &
		\colhead{Transient Monitor} \\
		\colhead{} &
		\colhead{(2--10\,keV)} &
		\colhead{(14--195\,keV)} &
		\colhead{(15--50\,keV)}
	}
	\startdata
	$\phi_{\text{ing}}$ & --0.100$^{\text{+0.006}}_{\text{--0.008}}$ & --0.10$^{\text{+0.02}}_{\text{--0.01}}$ & --0.12\,$\pm$\,0.01 \\
	$\phi_{\text{egr}}$ & 0.064\,$\pm$\,0.006 & 0.06$^{\text{+0.02}}_{\text{--0.01}}$ & 0.06\,$\pm$\,0.01 \\
	$\Delta\phi$$^{\mathrm{a}}$ & 0.03\,$\pm$\,0.01 & 0.05$^{\text{+0.02}}_{\text{--0.03}}$ & 0.06$^{\text{+0.03}}_{\text{--0.02}}$ \\
	$C$$^{\mathrm{b}}$ & 6.9\,$\pm$\,0.5$^{\mathrm{c}}$ & 0.105$^{\text{+0.010}}_{\text{--0.009}}$$^{\mathrm{d}}$ & 0.62\,$\pm$\,0.05$^{\mathrm{d}}$ \\
	$C_{\text{ecl}}$ & --3.6\,$\pm$\,0.6$^{\mathrm{c}}$ & 0.002$^{\text{+0.012}}_{\text{--0.013}}$$^{\mathrm{d}}$ & --0.03\,$\pm$\,0.06$^{\mathrm{d}}$ \\
	\tableline
	$\Delta\phi_{\text{ecl}}$ & 0.14$^{\text{+0.02}}_{\text{--0.01}}$ & 0.11$^{\text{+0.04}}_{\text{--0.03}}$ & 0.11\,$\pm$\,0.03 \\
	$P_{\text{orb}}$$^{\mathrm{e}}$ & 6.7825\,$\pm$\,0.0004 & 6.7825\,$\pm$\,0.0004 & 6.7825\,$\pm$\,0.0004 \\
	$\dot{P}_{\text{orb}}$$^{\mathrm{f}}$ & 0.09$^{\text{+1.69}}_{\text{--1.73}}$ & 0.09$^{\text{+1.69}}_{\text{--1.73}}$ & 0.09$^{\text{+1.69}}_{\text{--1.73}}$ \\
	$T_{\text{mid}}$$^{\mathrm{g}}$ & 55851.19\,$\pm$\,0.09 & 55851.3\,$\pm$\,0.1 & 55851.2\,$\pm$\,0.1 \\
	$\Theta_{e}$$^{\mathrm{h}}$ & 24.5$^{\text{+2.8}}_{\text{--2.5}}$ & 20.0$^{\text{+6.6}}_{\text{--6.2}}$ & 19.9$^{\text{+6.1}}_{\text{--5.2}}$ \\
	\tableline
	$\chi_{\nu}^{\mathrm{2}}$ (dof) & 1.17 (198) & 1.25 (75) & 0.97 (75)
	\enddata
	\tablecomments{We quote 1$\sigma$ uncertainties using Bayesian credible intervals. Phase~0 is defined at \text{mid-eclipse}. \\
		$^\mathrm{a}$ $\Delta\phi = \Delta\phi_{\text{ing}} = \Delta\phi_{\text{egr}}$, assuming equal ingress and egress durations. \\
		$^\mathrm{b}$ $C = C_{\text{ing}} = C_{\text{egr}}$, assuming equal \text{pre-ingress} and \text{post-egress} count rates. \\
		$^\mathrm{c}$ Units are counts\,s$^{\text{--1}}$\,PCU$^{\text{--1}}$. \\
		$^\mathrm{d}$ Units are 10$^{\text{--3}}$\,counts\,cm$^{\text{--2}}$\,s$^{\text{--1}}$. \\
		$^\mathrm{e}$ Orbital period from \text{O--C}~analysis. Units are days. \\
		$^\mathrm{f}$ Orbital period derivative from orbital change function. Units are~10$^{\text{--7}}$\,d\,d$^{\text{--1}}$. \\
		$^\mathrm{g}$ Units are MJD. \\
		$^\mathrm{h}$ Units are degrees.}
	\label{Table:SymmetricEclipseModel}
\end{deluxetable*}



\subsection{Pulse Period}
\label{Section:Pulse_Period}

We searched for pulsations with periods between 32\,s and 9.5\,days in the unweighted power spectrum of the entire pointed \textit{RXTE}~PCA~\text{(2.5--25\,keV)} light curve. The power spectrum was oversampled by a factor of five compared to the nominal frequency resolution, which was found to be 1.22\,$\times$\,10$^{\text{--6}}$\,Hz from the length of the light curve. The data in the pointed~PCA power spectrum were not weighted since the observations were performed using the same pointing and the light curve had a uniform time resolution of 16\,s.

\begin{figure}[t]
	\centering
	\includegraphics[trim=0cm 0cm 0cm 0cm, clip=false, scale=0.48, angle=0]{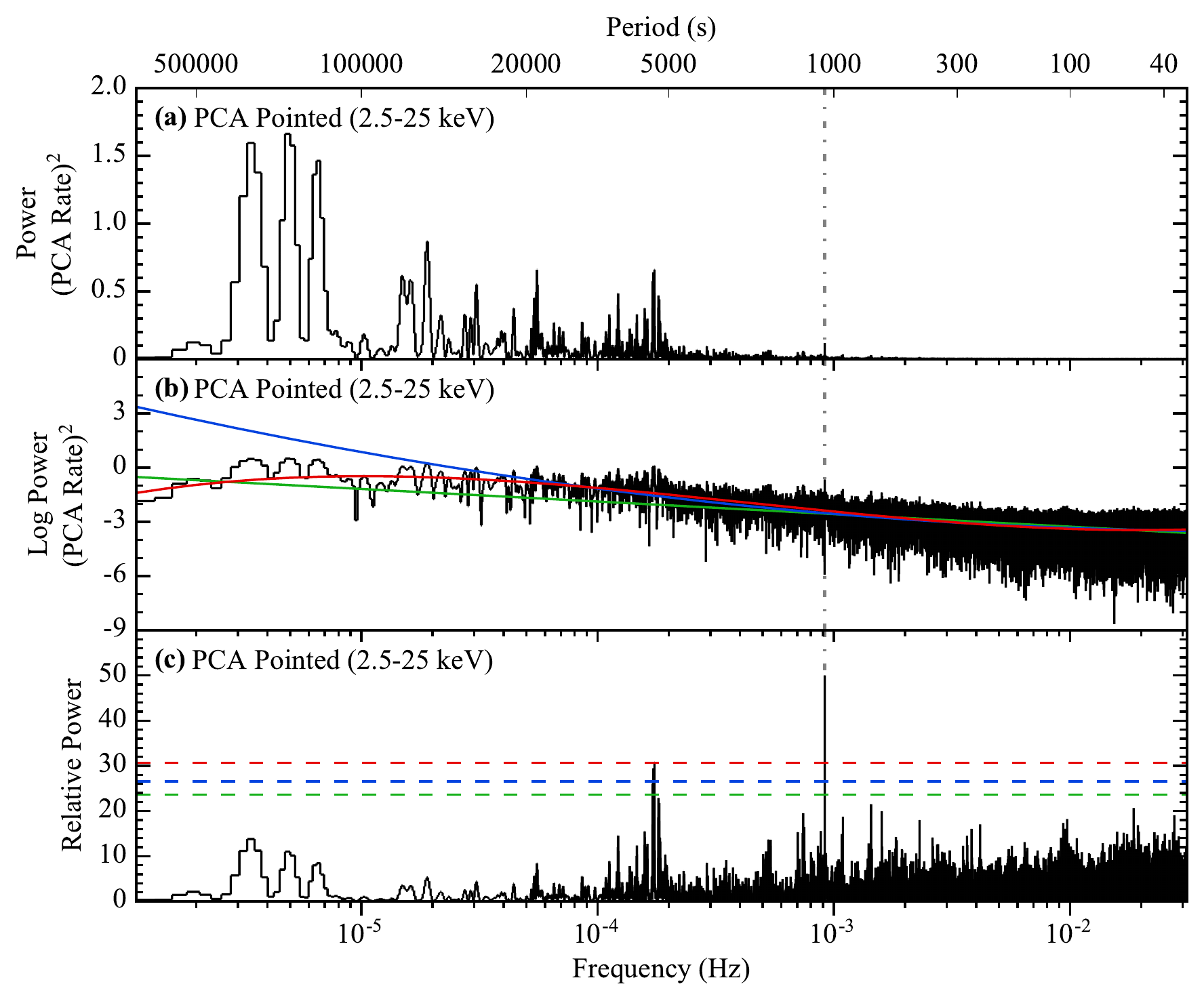}
	\caption{(a)~Unweighted power spectrum of \text{IGR~J16493--4348} using the pointed \textit{RXTE}~PCA~\text{(2.5--25\,keV)} light curve without low frequency noise subtracted from the continuum. (b)~Linear (shown in green), quadratic (shown in blue), and cubic (shown in red) fits to the logarithm of the power spectrum. The cubic fit was used to estimate and remove the continuum noise. (c)~Corrected power spectrum after subtracting the cubic continuum noise model. The horizontal dashed lines indicate the 95\% (shown in green), 99\% (shown in blue), and 99.9\% (shown in red) significance levels. The vertical dot-dashed line corresponds to the 1093\,s pulse period. The statistically significant peaks near $\sim$5800\,s are attributed to the orbital period of \textit{RXTE}.}
	\label{Figure:Figure8}
\end{figure}

A significant amount of low frequency noise was detected in the power spectrum shown in Figure~\ref{Figure:Figure8}(a). We estimated the continuum noise level by fitting polynomials to the logarithm of the power spectrum in Figure~\ref{Figure:Figure8}(b) after adding a constant value of~0.25068 to remove the bias from the $\chi^{\mathrm{2}}$~distribution of the \text{log-spectrum}~\citep{Papadakis1993, Vaughan2005}. We found that fitting a linear function to the log-spectrum, which would indicate a power law relationship in the raw power spectrum, was not optimal for describing the power at low frequencies. A quadratic fit significantly overestimated the low frequency power, and we found that a cubic fit to the log-spectrum sufficiently characterized the continuum noise level. The red noise was removed by subtracting the cubic fit from the logarithm of the power spectrum, which produced the corrected power spectrum in Figure~\ref{Figure:Figure8}(c). We find strong evidence of pulsed emission at a period of~1093.3\,$\pm$\,0.1\,s in the corrected power spectrum, which we associate with the rotational period of the neutron star. This pulse period is labeled by the vertical \text{dot-dashed} line in Figure~\ref{Figure:Figure8}.

At lower frequencies, there are significant peaks near the $\sim$5800\,s~orbital period of~\textit{RXTE}, which exhibit complex structure. The~95\%, 99\%, and~99.9\% significance levels are labeled in Figure~\ref{Figure:Figure8}(c), but do not account for the uncertainty in the model used to fit the continuum or the red noise subtraction. These effects are greatest at low frequencies, and larger power levels would be required to achieve these true levels of statistical significance. We also note that none of the low frequency peaks in the uncorrected power spectrum are statistically significant after the continuum noise was removed.



\section{System Geometry}
\label{Section:System_Geometry}

A pulsar timing algorithm was developed to accurately measure the neutron star rotational period and orbital parameters of \text{IGR~J16493--4348} by fitting circular and eccentric orbital timing models to the~ToAs. These results are presented here, along with a rigorous treatment of the statistical and systematic uncertainties associated with the~ToAs. We also provide measurements of additional parameters constraining the system geometry, such as the eclipse \text{half-angle}, donor star spectral type, and Roche lobe radius, for different possible inclinations and neutron star masses.

\subsection{Pulsar Timing Analysis}
\label{Section:Pulsar_Timing_Analysis}

We carried out a \text{phase-coherent} pulsar timing analysis using the pointed \textit{RXTE}~PCA~\text{(2.5--25\,keV)} light curve, where each rotation of the pulsar was unambiguously accounted for over the time span of the observations. An iterative epoch folding algorithm~\citep{Leahy1983, Schwarzenberg-Czerny1989, Chakrabarty1996} was used to derive the~ToAs. Epoch folding is useful because of its higher sensitivity to \text{non-sinusoidal} pulse shapes and ability to handle gaps in the light curve. The~ToAs were obtained by measuring phase offsets between a pulse template and individual measured profiles, which were created by dividing the light curve into smaller segments.

The measured profiles were created by first dividing the pointed PCA~light curve into individual segments spanning at least one pulse period in duration. Data within~20\,ks of another segment were merged together. Neighboring segments were separated by at least~36\,ks, and the segment durations ranged from 8.4 to 52.8\,ks. This produced a total of eight segments from which~ToAs were derived. We partitioned the data in this manner to optimize both the \text{signal-to-noise} of the pulsations in each segment and the final number of~ToAs.

An initial set of measured profiles were produced by folding each segment on the 1093.3\,s pulse period found in the noise subtracted power spectrum of the pointed PCA~light curve (see~Section~\ref{Section:Pulse_Period}). The segments were folded using 68~bins, which provided a time resolution equal to the 16\,s~sampling rate in the light curve. A preliminary pulse template was created by aligning the measured profiles and averaging the count rates in each bin. The uncertainties on the count rates in the pulse template were calculated by summing the errors in each bin in quadrature and then normalizing by the total number of measured profiles. This method of generating an initial template is beneficial because it incorporates the effects of the binary system's orbital motion, which are neglected when the entire light curve is folded on the pulse period.

Next, the phase offset between the pulse template and each of the measured profiles was determined by \text{cross-correlating} the two profiles in the Fourier frequency domain~\citep{Taylor1992}. It is advantageous to calculate the \text{cross-correlation} in the frequency domain, as opposed to using \text{time-domain} techniques, because it circumvents systematic errors due to binning and allows for the~ToAs to be measured with greater accuracy. Assuming that the measured profile is a scaled and shifted version of the pulse template, a phase shift is equivalent to multiplying the template by a complex exponential. The phase shift between each measured profile, $d(\phi)$, and the pulse template, $p(\phi)$, was found by minimizing~\citep{Taylor1992, Koh1997, Demorest2007}:
\begin{equation}
\chi^{2}(A,\phi)=\sum_{k=1}^{k_{\text{max}}}\frac{\left|d_{k}-Ap_{k}e^{-2\pi ik\phi}\right|^{2}}{\sigma_{k}^{2}},
\label{Equation:FourierCCF}
\end{equation}
where \mbox{$d_{k}=\sum_{j}d(j/N)e^{-2\pi ijk/N}$} is the~DFT of~$d(\phi)$, \mbox{$p_{k}=\sum_{j}p(j/N)e^{-2\pi ijk/N}$} is the~DFT of~$p(\phi)$, $\sigma_{k}^{2}$~is the noise power at each frequency bin of the~DFT, and~$A$ and~$\phi$ are the measured amplitude and phase shift, respectively.

Several of the observations did not exhibit strong pulsed emission~(see~the bold entries in Table~\ref{Table:RXTEObservationLog}) and were excluded from the pulsar timing analysis since their measured profiles yielded phase offsets that could not be well constrained during the \text{cross-correlation} procedure. The omitted data spanned MJDs\,55843.87475 to~55844.56803~(orbital phases~\text{--0.037} to~0.065) and MJDs\,55851.18401 to~55851.48269~(orbital phases~0.041 to~0.085), where orbital phase~0 is defined at~$T_{\pi/2}$ from circular solution~1 in Table~\ref{Table:OrbitalParameters}. The excised data are shown in red in Figure~\ref{Figure:Figure1} and coincide with the eclipse and eclipse transition phases measured from fitting the folded~PCA and BAT~orbital profiles with asymmetric and symmetric eclipse models~(see~Section~\ref{Section:Folded_Orbital_Profiles}).

Pulse time delays were obtained by multiplying the phase differences between each of the measured profiles and the pulse template by the folding period. The~ToAs were derived by adding each time delay to the time nearest to the middle of its corresponding observation interval where the pulsar rotational phase was zero. Referencing each~ToA relative to the middle of the interval reduces systematic effects that can arise from folding with an inaccurate timing model and is a standard convention used in pulsar timing~(e.g.,~\citealt{Levine2004}).

Folding the data with a slightly incorrect pulse period can lead to pulse smearing and produce~ToAs that show a drift in pulse phase over the observation duration. At the end of each iteration, a small correction was applied to the pulse period using the drift rate measured from the~ToAs. This improved pulse period measurement was then used to refold each of the segments to produce corrected measured profiles and a sharper pulse template in the following iteration. A refined pulse template was constructed by averaging the corrected measured profiles together without any alignment, and a new set of~ToAs was produced by \text{cross-correlating} the measured profiles with the updated pulse template in the frequency domain. This algorithm was repeated until there was no statistically significant change in the pulse period between successive iterations.

\begin{figure}[t]
	\centering
	\includegraphics[trim=0cm 0cm 0cm 0cm, clip=false, scale=0.475, angle=0]{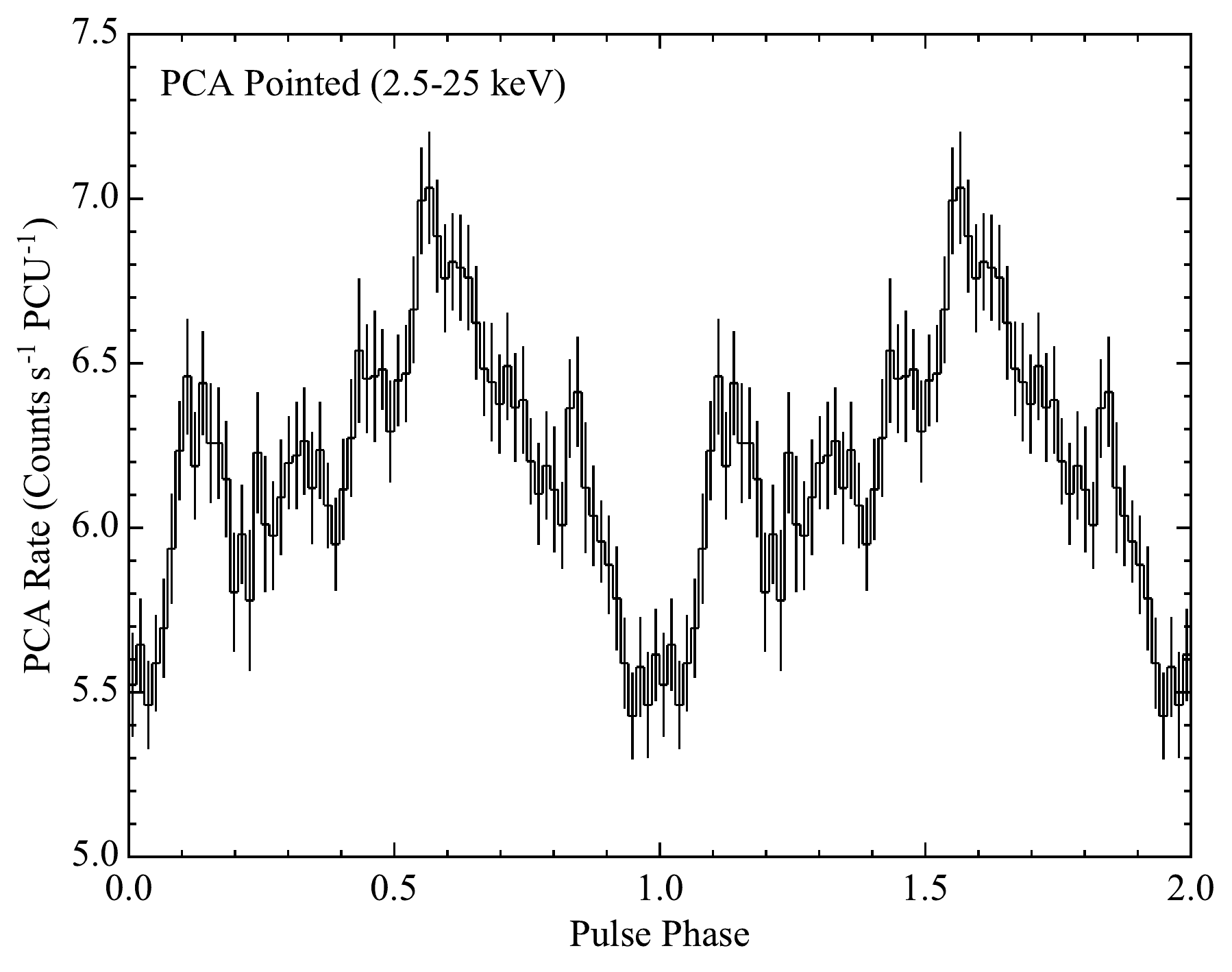}
	\caption{Final pulse template of \text{IGR~J16493--4348} obtained during the last iteration of the pulsar timing analysis using the pointed \textit{RXTE}~PCA~\text{(2.5--25\,keV)} light curve. The count rates in each bin were derived by averaging the count rates in the measured profiles. The uncertainties were calculated by summing the errors on the count rates in each bin of the measured profiles in quadrature and then normalizing by the total number of profiles. Phase~0 corresponds to $T_{\pi/2}$ from circular solution~1 in Table~\ref{Table:OrbitalParameters}.}
	\label{Figure:Figure9}
\end{figure}

\begin{figure}[t]
	\centering
	\includegraphics[trim=0cm 0cm 0cm 0cm, clip=false, scale=0.6, angle=0]{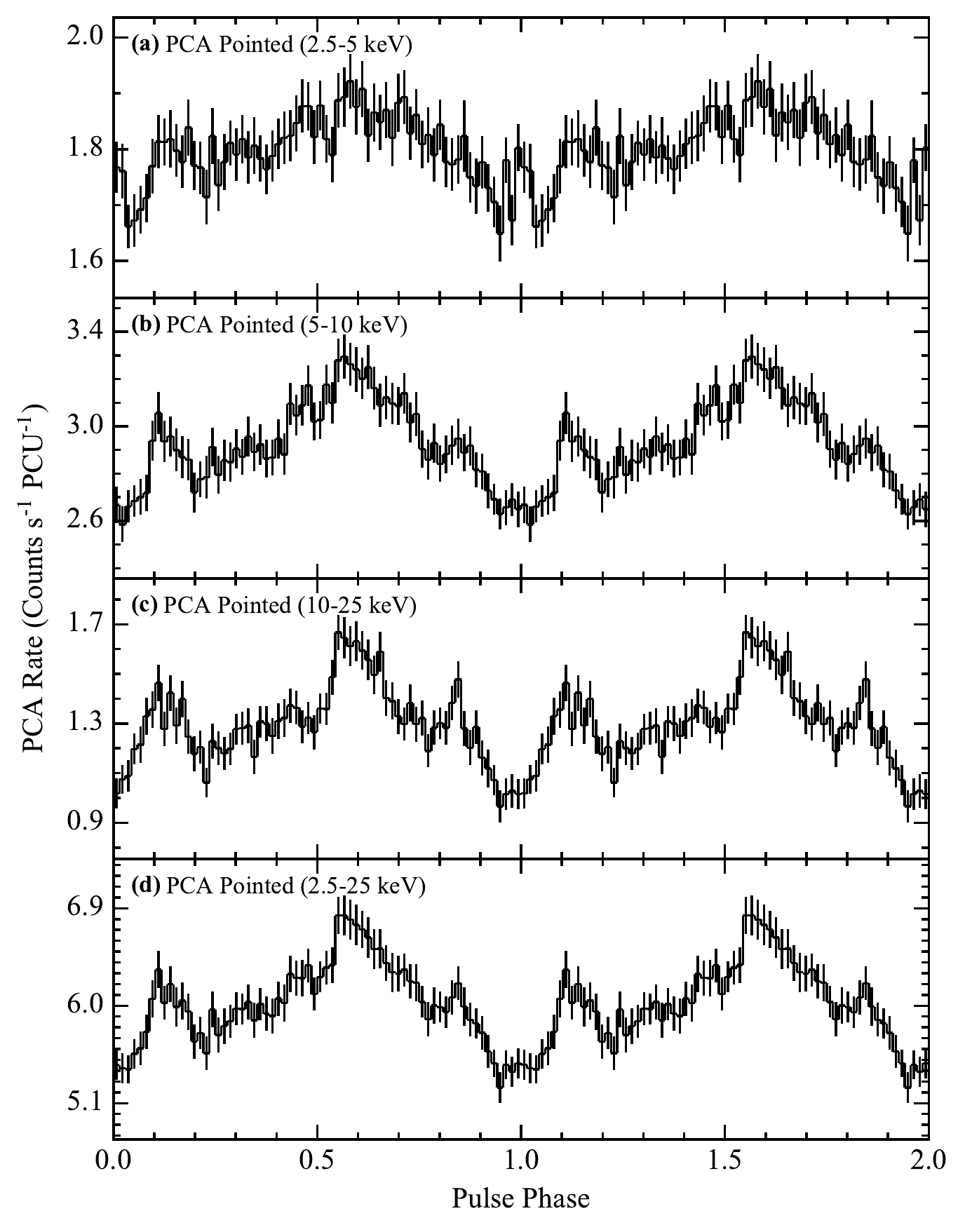}
	\caption{Pulse profiles of \text{IGR~J16493--4348} derived from pointed \textit{RXTE}~PCA observations in the (a)~\text{2.5--5,} (b)~\text{5--10,} (c)~\text{10--25,} and (d)~\text{2.5--25\,keV}~energy bands. The profiles were obtained by folding the light curves on the refined 1093\,s~pulse period measurement from the final iteration of the pulsar timing analysis after correcting for orbital Doppler delays. Phase~0 corresponds to~$T_{\pi/2}$ from circular solution~1 in Table~\ref{Table:OrbitalParameters}.}
	\label{Figure:Figure10}
\end{figure}

We refine the pulse period to 1093.1036\,$\pm$\,0.0004\,s using the~ToAs obtained during the last iteration of the pulsar timing analysis. The final pulse template~\text{(2.5--25\,keV)}, shown in Figure~\ref{Figure:Figure9}, exhibits sharp features on top of a \text{quasi-sinusoidal} template shape. These sharp variations in the pulse template enabled the phase shifts between the measured profiles and the template to be determined more accurately. A list of~ToAs measured during the final iteration of the pulsar timing analysis is provided in Table~\ref{Table:PulseArrivalTimes}. Pulse profiles in the \text{2.5--5,} \text{5--10,} \text{10--25,} and \text{2.5--25\,keV}~energy bands are shown in Figure~\ref{Figure:Figure10} and were obtained by folding the pointed~PCA light curves on the final pulse period measurement after correcting for orbital Doppler delays. We define the \text{peak-to-peak} pulsed fraction as:
\begin{equation}
\mathcal{P}=\frac{(F_{\text{max}}-F_{\text{min}})}{(F_{\text{max}}+F_{\text{min}})},
\label{Equation:PulsedFraction}
\end{equation}
where $F_{\text{max}}$ and $F_{\text{min}}$ are the maximum and minimum count rates in the pulse profile, respectively. Using Equation~(\ref{Equation:PulsedFraction}), the pulsed fractions in the \text{2.5--5,} \text{5--10,} \text{10-25,} and \text{2.5--25\,keV}~energy bands are~0.08\,$\pm$\,0.02, 0.12\,$\pm$\,0.02, 0.27\,$\pm$\,0.04, and~0.13\,$\pm$\,0.02, respectively. These measurements indicate an increase in pulsed fraction with increasing \text{X-ray}~energy.



\addtocounter{table}{1}

\begin{deluxetable*}{ccccc} 
	\tablenum{5}
	\tabletypesize{\small}
	\tablecolumns{5}
	\tablewidth{0pt}
	\tablecaption{\textsc{Pulse Times of Arrival from \text{IGR~J16493--4348}}}
	\tablehead{
		\colhead{ToA} &
		\colhead{Pulse Cycle$^\mathrm{a}$} &
		\colhead{Statistical Uncertainty$^\mathrm{b}$} &
		\colhead{Total Uncertainty$^\mathrm{c}$} & 
		\colhead{Pulse Phase$^\mathrm{d}$} \\
		\colhead{($t_{n}$,~MJD)} &
		\colhead{$(n)$} &
		\colhead{(s)} &
		\colhead{(s)} &
		\colhead{}
	}
	\startdata
	55843.15499 & 5 & 7.7 & 8.3 & 0.09 \\
	55845.44465 & 186 & 6.4 & 7.1 & 0.07 \\
	55846.31687 & 255 & 11.9 & 12.3 & 0.01 \\
	55847.49281 & 348 & 9.1 & 9.6 & 0.96 \\
	55848.44228 & 423 & 10.7 & 11.2 & 0.01 \\
	55849.37915 & 497 & 9.4 & 9.9 & 0.06 \\
	55850.40456 & 578 & 7.2 & 7.9 & 0.11 \\
	55852.28917 & 727 & 5.0 & 5.8 & 0.07
	\enddata
	\tablecomments{\\
		$^\mathrm{a}$ Nearest integer pulse cycle calculated using Equation~(\ref{Equation:PulseCycle}) and $t_{0}$\,$=$\,MJD\,55843.09111. Pulse cycles are referenced with respect to the start of the pointed~PCA observations. \\
		$^\mathrm{b}$ 1$\sigma$ statistical uncertainties derived from Monte Carlo simulations. \\
		$^\mathrm{c}$ Total uncertainties calculated by adding a systematic uncertainty of 3.1\,s from circular solution~2 in Table~\ref{Table:OrbitalParameters} to each statistical uncertainty in quadrature. \\
		$^\mathrm{d}$ Phase~0 corresponds to $T_{\pi/2}$ from circular solution~1 in Table~\ref{Table:OrbitalParameters}.}
	\label{Table:PulseArrivalTimes}
\end{deluxetable*}


\subsection{Pulse Time of Arrival Uncertainties}
\label{Section:ToA_Uncertainties}

Statistical uncertainties on the~ToAs were calculated using 100,000~Monte Carlo simulations~\citep{Thompson2006, Thompson2007}. At the beginning of each simulation, the pointed~PCA light curve was divided into individual segments according to the procedure described in Section~\ref{Section:Pulsar_Timing_Analysis}. Each count rate in the segments was replaced with a value selected randomly from a Gaussian distribution, with mean equal to the original \text{background-subtracted} count rate and standard devation given by its associated uncertainty. Simulated measured profiles were produced by folding these randomly generated segments on the refined 1093\,s~pulse period measurement from the final iteration of the pulsar timing analysis. During each simulation, phase offsets were measured by \text{cross-correlating} each of the simulated measured profiles with the final pulse template in the frequency domain, which were then used to derive an independent set of~ToAs. We quote 1$\sigma$~statistical uncertainties on each~ToA listed in Table~\ref{Table:PulseArrivalTimes} from the median absolute deviation of the distribution of simulated~ToAs obtained for each segment. The median absolute deviation is a robust statistic that is more resilient to outliers than the standard deviation. We used this statistic to reduce the effect that tails in the distribution of the simulated~ToAs had on the statistical errors. The statistical uncertainties ranged from 5.0~to~11.9\,s.

Systematic errors in the~ToAs can arise from changes in the average pulse profile due to varying flux levels, absorption along the line of sight, or contamination from nearby sources~(e.g.,~\citealt{Thompson2006}). We modeled the systematic error as a nuisance parameter when fitting the~ToAs with timing models using the Bayesian~MCMC procedure described in Section~\ref{Section:O-C_Analysis}. The systematic uncertainty was treated as an additional error that was added in quadrature with the statistical uncertainty in the likelihood function. We marginalized over this systematic uncertainty when constructing posterior~PDFs for the model parameters. Due to the limited number of~ToAs and few degrees of freedom in the timing solutions, a systematic error was derived only for circular solution~2, which used the same timing model as in circular solution~1~(see~Section~\ref{Section:Pulsar_Timing_Models} and~Table~\ref{Table:OrbitalParameters}). We find a systematic uncertainty of $\sigma_{\text{sys}}$\,$=$\,3.1$^{\text{+3.0}}_{\text{--2.3}}$\,s from the median of the marginalized posterior distribution, and we report 1$\sigma$ errors on this measurement using Bayesian credible intervals. This systematic uncertainty was added in quadrature with the statistical uncertainty for each~ToA, which yielded the total uncertainties listed in Table~\ref{Table:PulseArrivalTimes}. We note that the inclusion of systematic errors as a nuisance parameter increased the reduced $\chi^{2}_{\nu}$ value from~0.88 in circular solution~1 to~1.04 in circular solution~2 without significantly affecting the \text{best-fit} orbital parameters.


\subsection{Pulsar Timing Models}
\label{Section:Pulsar_Timing_Models}

Assuming the pulsar phase varies smoothly as a function of time, a Taylor expansion can be used to approximate the pulse phase,~$\phi$, at time~$t$:
\begin{equation}
\phi(t)=\phi(t_{0})+\nu_{\text{pulse}}(t-t_{0})+\frac{1}{2}\dot{\nu}_{\text{pulse}}(t-t_{0})^{2}+\cdots,
\label{Equation:TimingPhaseModel}
\end{equation}
where~$\nu_{\text{pulse}}$ and $\dot{\nu}_{\text{pulse}}$~are the neutron star rotational frequency and its time derivative at a reference time~$t_{0}$, typically chosen to be the start time of the observation. Equation~(\ref{Equation:TimingPhaseModel}) can be transformed to give the expected arrival time,~$t_{n}'$, of the $n$th~pulse:
\begin{equation}
t_{n}'=t_{0}+nP_{\text{pulse}}+\frac{1}{2}n^{2}P_{\text{pulse}}\dot{P}_{\text{pulse}}+\cdots,
\label{Equation:PulseArrivalTimeModel}
\end{equation}
where~$P_{\text{pulse}}$\,$=$\,$\nu_{\text{pulse}}^{-1}$ is the pulse period at time~$t_{0}$ and $\dot{P}_{\text{pulse}}$~is the pulse period derivative. The pulse cycle,~$n$, associated with each~ToA is calculated to the nearest integer using:
\begin{equation}
n=\frac{t_{n}'-t_{0}}{P_{\text{pulse}}}-\frac{1}{2}\frac{\dot{P}_{\text{pulse}}}{P_{\text{pulse}}^{2}}(t_{n}'-t_{0})^{2}
\label{Equation:PulseCycle}
\end{equation}

Additional time delays are observed from binary pulsars due to their orbital motion. In these systems, the~ToAs can be described by:
\begin{equation}
t_{n}=t_{n}'+f_{\text{orb}}(t_{n}'),
\label{Equation:ToAsOrbitalMotion}
\end{equation}
where $f_{\text{orb}}(t_{n}')$ is the orbital Doppler delay time associated with $t_{n}'$. For binary pulsars in circular orbits, the orbital Doppler delay times are given by~\citep{Blandford1976, Kelley1980}:
\begin{equation}
f_{\text{orb}}(t_{n}')=a_{x}\sin i\cos\left[\frac{2\pi(t_{n}'-T_{\pi/2})}{P_{\text{orb}}}\right],
\label{Equation:CircularOrbitDelayTime}
\end{equation}
where $a_{x}\sin i$~is the projected \text{semi-major} axis of the orbit, $i$~is the orbital inclination angle relative to the line of sight, and $T_{\pi/2}$ is the time of maximum delay and \text{mid-eclipse}. If the orbit is eccentric, the Doppler delay times are instead given by~\citep{Blandford1976}:
\begin{equation}
\resizebox{0.891\hsize}{!}
{$f_{\text{orb}}(t_{n}')=a_{x}\sin i\left[\sin\omega(\cos E-e)+\sqrt{1-e^{2}}\cos\omega\sin E\right]$,}
\label{Equation:EccentricOrbitDelayTime}
\end{equation}
where $e$~is the eccentricity, $\omega$~is the longitude of periastron, and $E$~is the eccentric anomaly. The eccentric anomaly can be related to the mean anomaly,~$M$, using Kepler's equation:
\begin{equation}
M=E-e\sin E=\frac{2\pi(t_{n}'-T_{\text{peri}})}{P_{\text{orb}}},
\label{Equation:MeanAnomaly}
\end{equation}
where $T_{\text{peri}}$~is the time of periastron passage.

The pulse period behavior and orbital parameters were measured by fitting circular and eccentric orbital timing models to the~ToAs in Table~\ref{Table:PulseArrivalTimes}. The timing models were constructed using Equations~(\ref{Equation:PulseArrivalTimeModel}) and~(\ref{Equation:ToAsOrbitalMotion}), together with Equation~(\ref{Equation:CircularOrbitDelayTime}) for the circular solutions and Equation~(\ref{Equation:EccentricOrbitDelayTime}) for the eccentric solutions. To properly account for covariances between the model parameters in the orbital solutions, we used the Bayesian~MCMC fitting procedure described in Section~\ref{Section:O-C_Analysis}.

Circular solutions~1 and~2 and eccentric solution~1 were fit assuming a constant neutron star rotational period. Pulse period changes were incorporated into the timing models used to derive circular solution~3 and eccentric solution~2. In all of these models, the orbital period was fixed to the refined 6.7828\,day measurement from the \text{O--C}~analysis in Section~\ref{Section:O-C_Analysis}, and the~ToAs were weighted by their corresponding uncertainties during the fitting process.

In Table~\ref{Table:OrbitalParameters}, we list the \text{best-fit} pulse period and orbital parameter measurements for each timing model, along with 1$\sigma$~uncertainties derived from Bayesian credible intervals. Circular solution~1 is our favored orbital timing model since the fit yielded an acceptable reduced $\chi^{2}_{\nu}$ value with the greatest number of degrees of freedom compared to the other timing models in Table~\ref{Table:OrbitalParameters}. The orbital Doppler delay times measured during the final iteration of the pulsar timing analysis are shown in Figure~\ref{Figure:Figure7}(d), along with the predicted delay times from circular solution~1. We find a projected \text{semi-major} axis of~$a_{x}\sin i$\,$=$\,82.8$^{\text{+5.0}}_{\text{--5.2}}$\,\text{lt-s} and a \text{mid-eclipse} time of~$T_{\pi/2}$\,$=$\,MJD\,55850.91\,$\pm$\,0.05 from the fit in circular solution~1. A pulse period derivative of \mbox{$\dot{P}_{\text{pulse}}$\,$=$\,--5.4$^{\text{\text{+7.9}}}_{\text{\text{--9.7}}}$\,$\times$\,10$^{\text{--8}}$\,s\,s$^{\text{--1}}$} was obtained in circular solution~3, which indicates that there was no statistically significant \text{long-term} change in the pulse period during the pointed~PCA observations. In addition, no rapid \text{spin-up} or \text{spin-down} episodes were observed, which suggests that a transient accretion disk is not present in this system~\citep{Koh1997, Jenke2012}.

The \text{X-ray} mass function is given by:
\begin{equation}
f_{x}(M)=\frac{4\pi^{2}(a_{x}\sin i)^{3}}{GP_{\text{orb}}^{2}}=\frac{(M_{c}\sin i)^{3}}{(M_{x}+M_{c})^{2}},
\label{Equation:XrayMassFunction}
\end{equation}
where $M_{x}$ is the mass of the neutron star and $M_{c}$ is the mass of the companion. We find an \text{X-ray} mass function of $f_{x}(M)$\,$=$\,13.2$^{\text{+2.4}}_{\text{--2.5}}$\,$M_{\odot}$ using Equation~(\ref{Equation:XrayMassFunction}) and the values of~$a_{x}\sin i$ and $P_{\text{orb}}$ from circular solution~1. This mass function provides further evidence that \text{IGR~J16493--4348} is an~sgHMXB with an early~\text{B-type} stellar companion. It is also consistent with the mass functions obtained from the other orbital models in Table~\ref{Table:OrbitalParameters} to within~1$\sigma$.

An eccentricity of~$e$\,$=$\,0.17\,$\pm$\,0.09 and a longitude of periastron of~$\omega$\,$=$\,251\,$\pm$\,28$^{\circ}$ were measured from eccentric solution~1, and these values are consistent with the results from eccentric solution~2. A time of periastron passage of $T_{\text{peri}}$\,$=$\,MJD\,55847.1\,$\pm$\,0.5 was obtained using the eccentric timing models. The posterior distributions of the periastron passage time and longitude of periastron were both relatively broad due to the limited number of available~ToAs, which resulted in large uncertainties on these parameters. Therefore, the results from these eccentric solutions should be interpreted with caution since they were obtained from timing model fits with only a few degrees of freedom.



\addtocounter{table}{1}

\begin{deluxetable*}{lccccc} 
	\tablenum{6}
	\tabletypesize{\small}
	\tablecolumns{6}
	\tablewidth{0pt}
	\tablecaption{\textsc{Orbital Parameters of \text{IGR~J16493--4348}}}
	\tablehead{
		\colhead{Parameter} &
		\colhead{\textbf{Circular Solution~1}} &
		\colhead{Circular Solution~2} &
		\colhead{Circular Solution~3} &
		\colhead{Eccentric Solution~1} &
		\colhead{Eccentric Solution~2}
	}
	\startdata
	$P_{\text{pulse}}$$^{\mathrm{a}}$\,(s) & 1093.1036\,$\pm$\,0.0004 & 1093.1036\,$\pm$\,0.0001 & 1093.10\,$\pm$\,0.02 & 1093.1036\,$\pm$\,0.0007 & 1093.10\,$\pm$\,0.01 \\
	$\dot{P}_{\text{pulse}}$\,($\times$\,10$^{\text{--8}}$\,s\,s$^{\text{--1}}$) & \nodata & \nodata & --5.4$^{\text{+7.9}}_{\text{--9.7}}$ & \nodata & --3.0$^{\text{+7.5}}_{\text{--8.9}}$ \\
	$a_{x}\sin i$\,(\text{lt-s}) & 82.8$^{\text{+5.0}}_{\text{--5.2}}$ & 82.4\,$\pm$\,5.5 & 81.3\,$\pm$\,5.6 & 82.9\,$\pm$\,5.2 & 82.3\,$\pm$\,5.5 \\
	$T_{\pi/2}$$^{\mathrm{b}}$\,(MJD) & 55850.91\,$\pm$\,0.05 & 55850.90\,$\pm$\,0.05 & 55850.90\,$\pm$\,0.05 & \nodata & \nodata \\
	$T_{\text{peri}}$$^{\mathrm{c}}$\,(MJD) & \nodata & \nodata & \nodata & 55847.1\,$\pm$\,0.5 & 55847.1\,$\pm$\,0.5 \\
	$e$ & \nodata & \nodata & \nodata & 0.17\,$\pm$\,0.09 & 0.17$^{\text{+0.08}}_{\text{--0.09}}$ \\
	$\omega$\,(deg) & \nodata & \nodata & \nodata & 251\,$\pm$\,28 & 251$^{\text{+28}}_{\text{--27}}$ \\
	$P_{\text{orb}}$$^{\mathrm{d}}$\,(days) & 6.7828 & 6.7828 & 6.7828 & 6.7828 & 6.7828 \\
	$\sigma_{\text{sys}}$$^{\mathrm{e}}$\,(s) & \nodata & 3.1$^{\text{+3.0}}_{\text{--2.3}}$ & \nodata & \nodata & \nodata \\
	\tableline
	$f_{x}(M)$\,$(M_{\odot})$ & 13.2$^{\text{+2.4}}_{\text{--2.5}}$ & 13.0\,$\pm$\,2.6 & 12.5\,$\pm$\,2.6 & 13.3\,$\pm$\,2.5 & 13.0\,$\pm$\,2.6 \\
	\tableline
	$\chi^{2}_{\nu}$ (dof) & 0.88 (4) & 1.04 (3) & 1.06 (3) & 0.89 (2) & 0.98 (1)
	\enddata
	\tablecomments{We quote 1$\sigma$~uncertainties on the model parameters using Bayesian credible intervals. We assumed no change in the neutron star's rotational period in circular solution~1, circular solution~2, and eccentric solution~1. We favor circular solution~1 as our preferred timing model for \text{IGR~J16493--4348}. \\
		$^{\mathrm{a}}$ Pulse period at $t_{0}$\,$=$\,MJD\,55843.09111. \\
		$^{\mathrm{b}}$ Time of maximum delay and \text{mid-eclipse} in the circular orbital models. \\
		$^{\mathrm{c}}$ Time of periastron passage in the eccentric orbital models. \\
		$^{\mathrm{d}}$ Orbital period measurement from the \text{O--C}~analysis using an asymmetric eclipse model. \\
		$^{\mathrm{e}}$ Systematic uncertainty measured from the posterior~PDF in the Bayesian~MCMC fitting procedure.}
	\label{Table:OrbitalParameters}
\end{deluxetable*}


\subsection{Supergiant Companion and System Parameters}
\label{Section:Supergiant_Companion}

We present constraints on the mass and radius of the supergiant donor using the orbital parameters from circular solution~1 and eccentric solution~1~(see~Table~\ref{Table:OrbitalParameters}), together with the asymmetric eclipse model parameters from the BAT~transient monitor orbital profile~(see~Table~\ref{Table:AsymmetricEclipseModel}). The empirical mass distribution of neutron stars is peaked around a canonical value of~1.4\,$M_{\odot}$, and a neutron star mass of~1.9\,$M_{\odot}$ is a reasonable upper limit for~sgHMXB systems with a \text{B-type} companion~\citep{vanKerkwijk1995, Kaper2006}. Therefore, we assumed neutron star masses of~1.4\,$M_{\odot}$ and~1.9\,$M_{\odot}$ in these calculations.

For each neutron star mass, the mass of the supergiant was calculated as a function of inclination angle using Equation~(\ref{Equation:XrayMassFunction}) and a fine grid of inclination angles ranging from~\mbox{$i$\,$=$\,0--90$^{\circ}$}. Assuming a circular orbit~($e$\,$=$\,0), the separation between the center of masses of the two stars in the binary can be found from Kepler's third law:
\begin{equation}
a=\left[\frac{GP_{\text{orb}}^{2}(M_{x}+M_{c})}{4\pi^{2}}\right]^{1/3}
\label{Equation:KeplersThirdLaw}
\end{equation}
For an eccentric orbit, the separation at \text{mid-eclipse} is instead given by:
\begin{equation}
a'=a\frac{1-e^{2}}{1+e\,\mathrm{cos}\,\omega}
\label{Equation:EccentricOrbitalSeparation}
\end{equation}
The radius of the supergiant was determined as a function of inclination angle from~\citep{Joss1984}:
\begin{equation}
R_{c}=a'\sqrt{1-\cos^{2}\Theta_{e}\sin^{2}i},
\label{Equation:CompanionRadius}
\end{equation}
where $\Theta_{e}$ is the eclipse \text{half-angle}. Equation~(\ref{Equation:CompanionRadius}) can be used to derive the relationship between the eclipse \text{half-angle} and the inclination. The Roche lobe radius was calculated using~\citep{Eggleton1983, Goossens2013}:
\begin{equation}
\frac{R_{L}}{a'}=\frac{0.49q^{-2/3}}{0.6q^{-2/3}+\ln\left(1+q^{-1/3}\right)},
\label{Equation:RocheLobeRadius}
\end{equation}
where $q$\,$=$\,$M_{x}/M_{c}$~is the mass ratio. The inclination angle where the donor star would fill its Roche lobe was found by linearly interpolating between inclination angles where $R_{L}$\,$-$\,$R_{c}$ changed sign. We assumed that Roche lobe overflow occurred at periastron in the eccentric orbital models, where the distance between the two stars is $a'$\,$=$\,$a(1-e)$.

In Table~\ref{Table:DonorStarParameters}, we list calculated values for the companion mass, mass ratio, companion radius, Roche lobe radius, and Roche lobe filling factor~($\beta$\,$=$\,$R_{c}/R_{L}$) for neutron star masses of~1.4\,$M_{\odot}$ and~1.9\,$M_{\odot}$ using the orbital parameters from circular solution~1. These values were determined at inclination angles corresponding to Roche lobe overflow and an \text{edge-on} orbit~($i$\,$=$\,90$^{\circ}$). Assuming a canonical neutron star mass of~1.4\,$M_{\odot}$, we find that the supergiant fills its Roche lobe at an inclination angle of~56.0$^{\text{+6.4}}_{\text{--5.8}}$$^{\circ}$. At this inclination, the stellar mass and radius of the supergiant companion are~25.8$^{\text{+6.3}}_{\text{--5.9}}$\,$M_{\odot}$ and~28.3$^{\text{+5.7}}_{\text{--5.1}}$\,$R_{\odot}$, respectively. This yields a mass ratio of~$q$\,$=$\,0.05\,$\pm$\,0.01. If we instead consider an \text{edge-on} orbit, the stellar mass and radius of the supergiant donor are~15.7$^{\text{+2.4}}_{\text{--2.5}}$\,$M_{\odot}$ and~13.0$^{\text{+5.9}}_{\text{--4.7}}$\,$R_{\odot}$, respectively. The Roche lobe radius is~22.8\,$\pm$\,1.2\,$R_{\odot}$ in this case. We find a mass ratio of~$q$\,$=$\,0.09\,$\pm$\,0.01 and a Roche lobe filling factor of~$\beta$\,$=$\,0.57$^{\text{+0.26}}_{\text{--0.21}}$ using these values.

If we now consider a more massive 1.9\,$M_{\odot}$~neutron star, the Roche lobe is filled by the donor star at an inclination angle of~58.1$^{\text{+5.9}}_{\text{--5.1}}$$^{\circ}$. The stellar mass and radius of the supergiant companion are~25.0$^{\text{+5.1}}_{\text{--4.8}}$\,$M_{\odot}$ and~27.1$^{\text{+5.0}}_{\text{--4.4}}$\,$R_{\odot}$, respectively. This gives a mass ratio of~$q$\,$=$\,0.08$^{\text{+0.02}}_{\text{--0.01}}$. For an \text{edge-on} orbit, the stellar mass and radius of the supergiant donor are~16.5\,$\pm$\,2.5\,$M_{\odot}$ and~13.3$^{\text{+6.1}}_{\text{--4.8}}$\,$R_{\odot}$, respectively, and we find a Roche lobe radius of~22.6$^{\text{+1.1}}_{\text{--1.2}}$\,$R_{\odot}$. This yields a mass ratio and Roche lobe filling factor of~$q$\,$=$\,0.12\,$\pm$\,0.02 and~$\beta$\,$=$\,0.59$^{\text{+0.27}}_{\text{--0.22}}$, respectively. These derived masses and radii for the donor star are consistent with a B0.5~Ia~spectral type companion from~\citet{Searle2008}, where the Roche lobe is nearly filled at a moderate inclination angle. A complete list of these parameters is provided in Table~\ref{Table:DonorStarParametersAppendix} in the Appendix for each pulsar timing solution in Table~\ref{Table:OrbitalParameters} using the asymmetric and symmetric eclipse model parameters in Tables~\ref{Table:AsymmetricEclipseModel} and~\ref{Table:SymmetricEclipseModel} from fitting the folded BAT~transient monitor and PCA~scan orbital profiles.

\begin{figure*}[t]
	\centering
	\begin{tabular}{cc}
		
		\subfigure
		{
			\includegraphics[trim=0cm 0cm 0cm 0cm, clip=false, scale=0.46, angle=0]{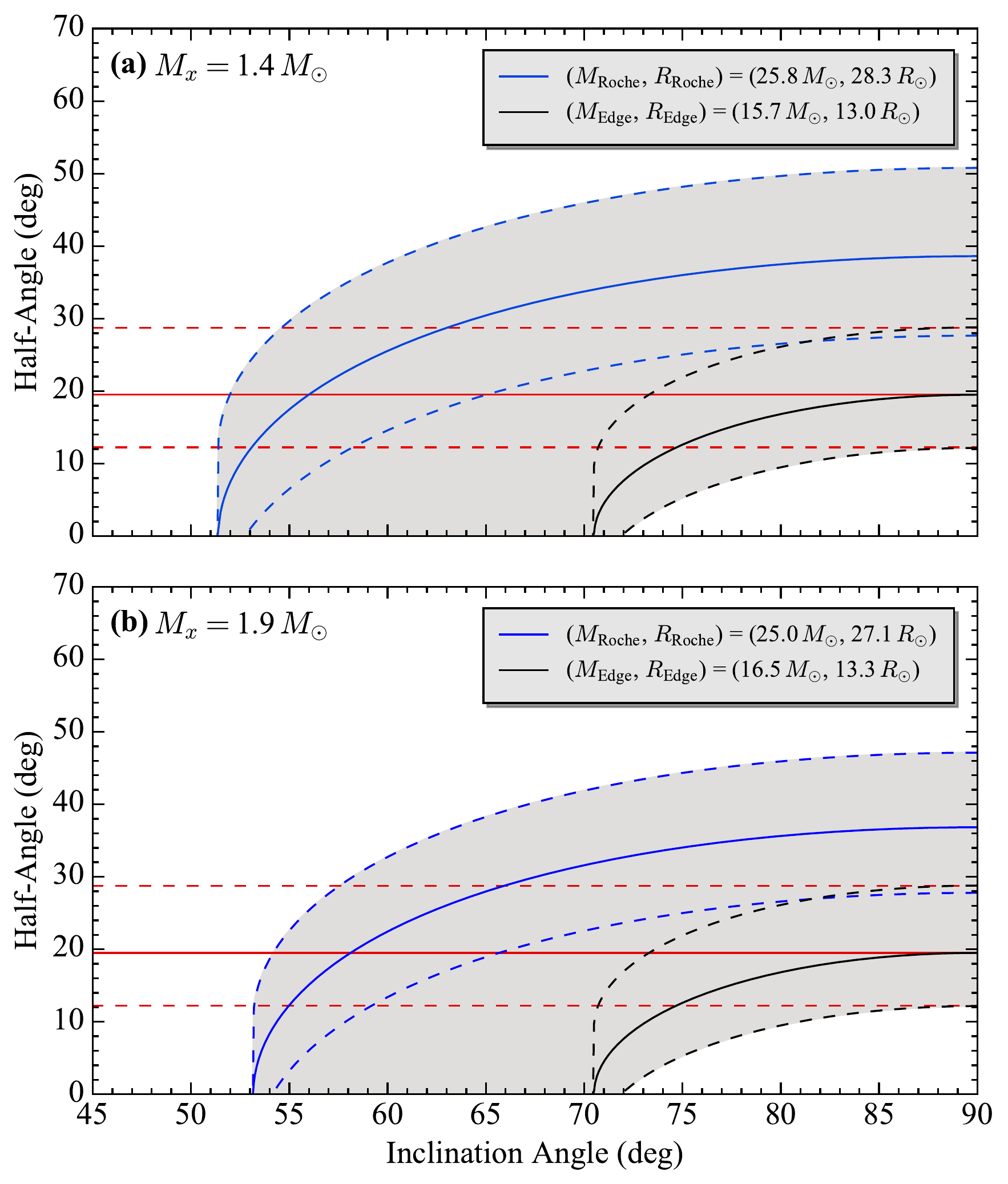}
			\label{Figure:Figure11a}
		}
		
		&
		
		\subfigure
		{
			\includegraphics[trim=0cm 0cm 0cm 0cm, clip=false, scale=0.46, angle=0]{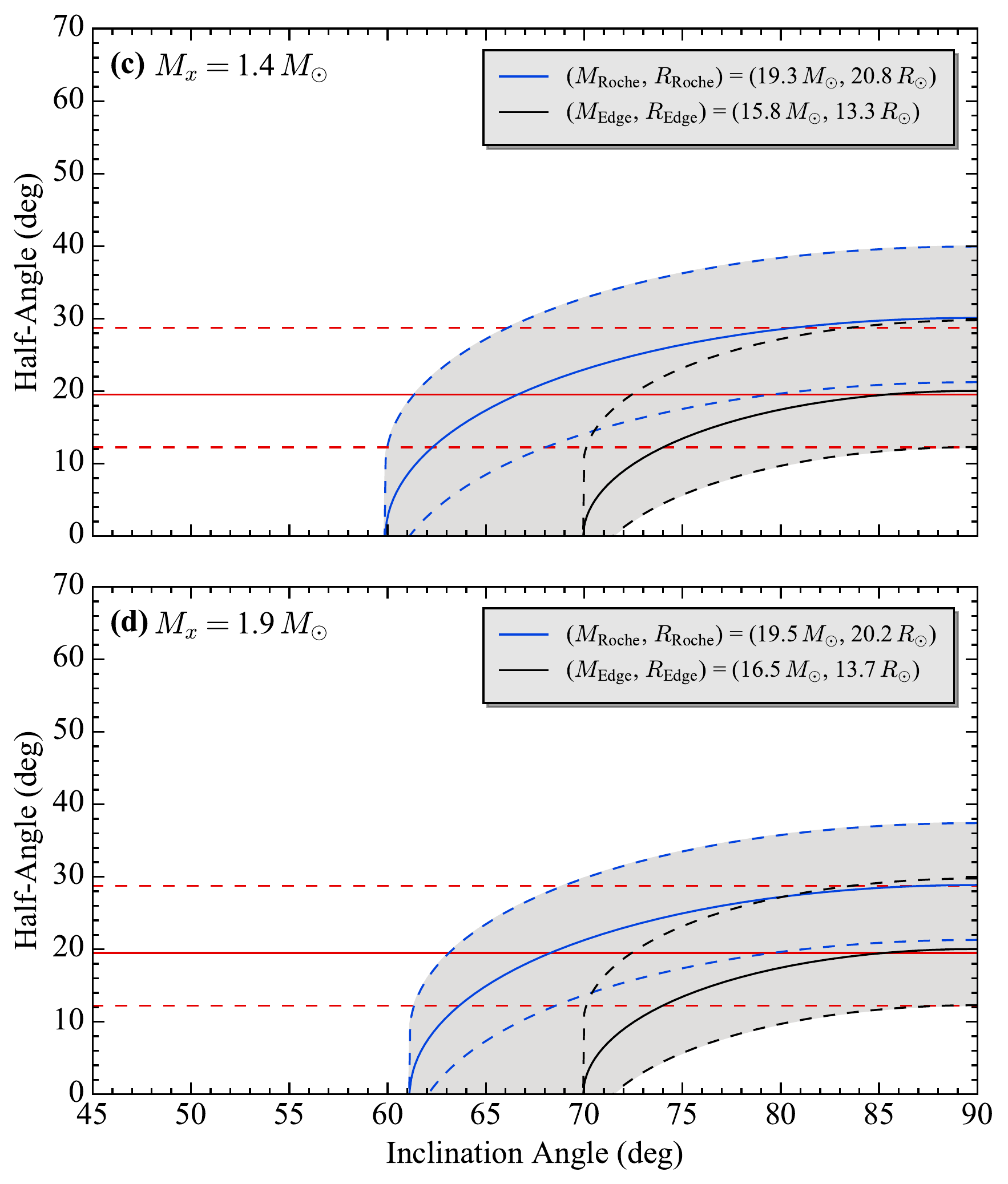}
			\label{Figure:Figure11b}
		}
		
	\end{tabular}
	
	\caption{Predicted eclipse \text{half-angle} of \text{IGR~J16493--4348} as a function of inclination angle, assuming neutron star masses of~1.4\,$M_{\odot}$ in~(a) and~(c) and~1.9\,$M_{\odot}$ in~(b) and~(d). These constraints are obtained using the orbital parameters from (left)~circular solution~1 and (right)~eccentric solution~1 in Table~\ref{Table:OrbitalParameters}, together with the asymmetric eclipse model parameters in Table~\ref{Table:AsymmetricEclipseModel} from fitting the \textit{Swift}~BAT~transient monitor~\text{(15--50\,keV)} orbital profile. The solid blue curves are derived using supergiant mass and radius values corresponding to where the donor fills its Roche lobe, and the solid black curves are obtained using supergiant mass and radius values derived for an \text{edge-on} orbit. The solid red lines indicate the measured eclipse \text{half-angle} in Table~\ref{Table:AsymmetricEclipseModel} from fitting the BAT~transient monitor orbital profile. The dashed curves correspond to 1$\sigma$~uncertainties on the eclipse \text{half-angles}. The grey shaded regions show the allowed parameter space.}
	\label{Figure:Figure11}
\end{figure*}

The constraints on the inclination angle are further visualized in Figure~\ref{Figure:Figure11}, together with our measurement of the eclipse \text{half-angle} in Table~\ref{Table:AsymmetricEclipseModel} from fitting the BAT~transient monitor orbital profile. We show the predicted eclipse \text{half-angle} of \text{IGR~J16493--4348} as a function of inclination angle using Equation~(\ref{Equation:CompanionRadius}) with supergiant mass and radius values corresponding to an \text{edge-on} orbit and where the donor star fills its Roche lobe. This behavior is shown for neutron star masses of~1.4\,$M_{\odot}$ and~1.9\,$M_{\odot}$ using the asymmetric eclipse model parameters from the folded BAT~transient monitor light curve and the orbital parameters from circular solution~1 and eccentric solution~1. The allowed parameter space is indicated by the grey shaded regions. For an eclipse \text{half-angle} of 19.5$^{\circ}$, shown by the solid red lines in Figure~\ref{Figure:Figure11}, we find that Roche lobe overflow would occur at inclination angles of $i$\,$\approx$\,57$^{\circ}$ and $i$\,$\approx$\,67$^{\circ}$ using the orbital parameters in circular solution~1 and eccentric solution~1, respectively.

\begin{figure*}[t]
	\centering
	\begin{tabular}{cc}
		
		\subfigure
		{
			\includegraphics[trim=0cm 0cm 0cm 0cm, clip=false, scale=0.46, angle=0]{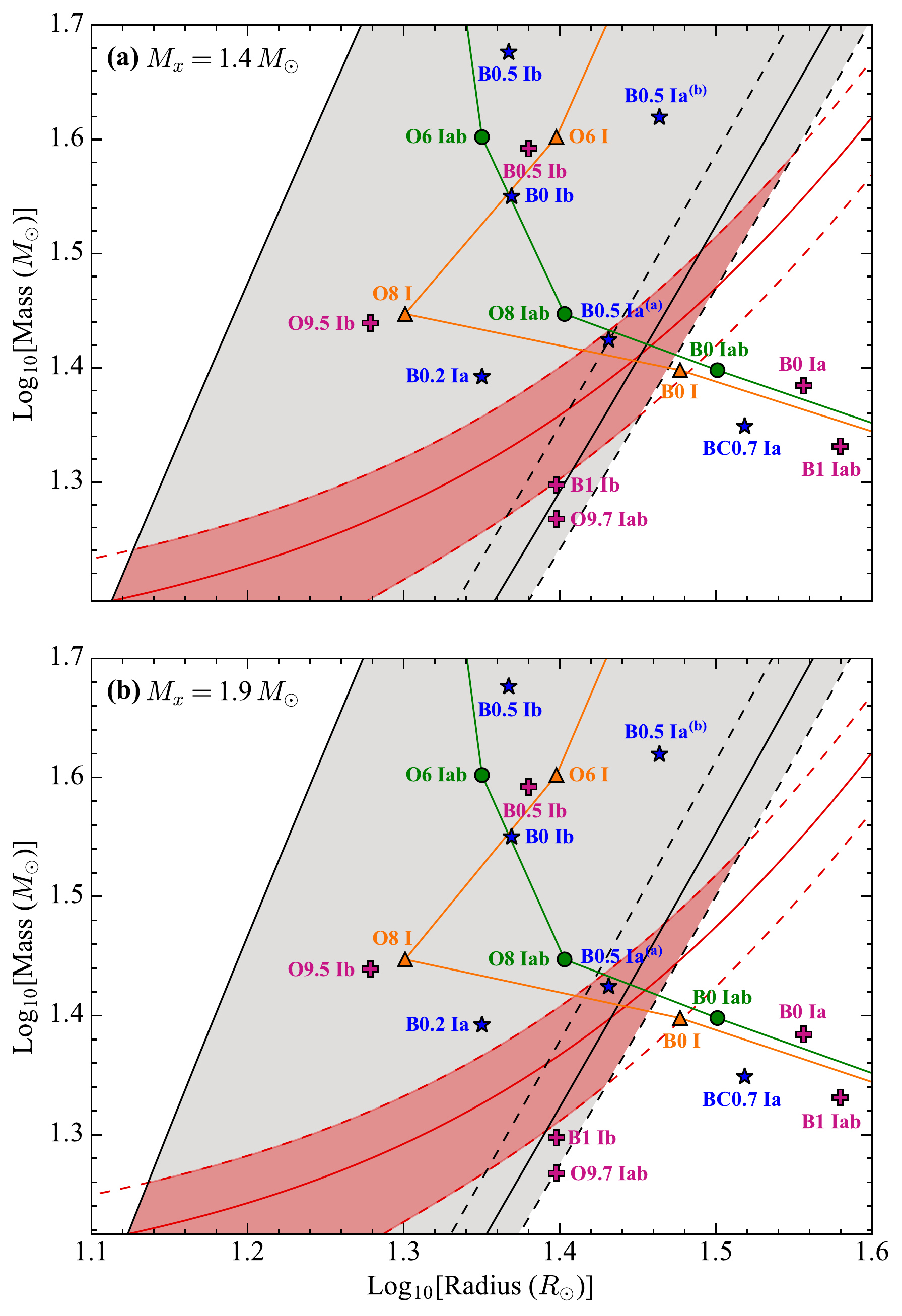}
			\label{Figure:Figure12a}
		}
		
		&
		
		\subfigure
		{
			\includegraphics[trim=0cm 0cm 0cm 0cm, clip=false, scale=0.46, angle=0]{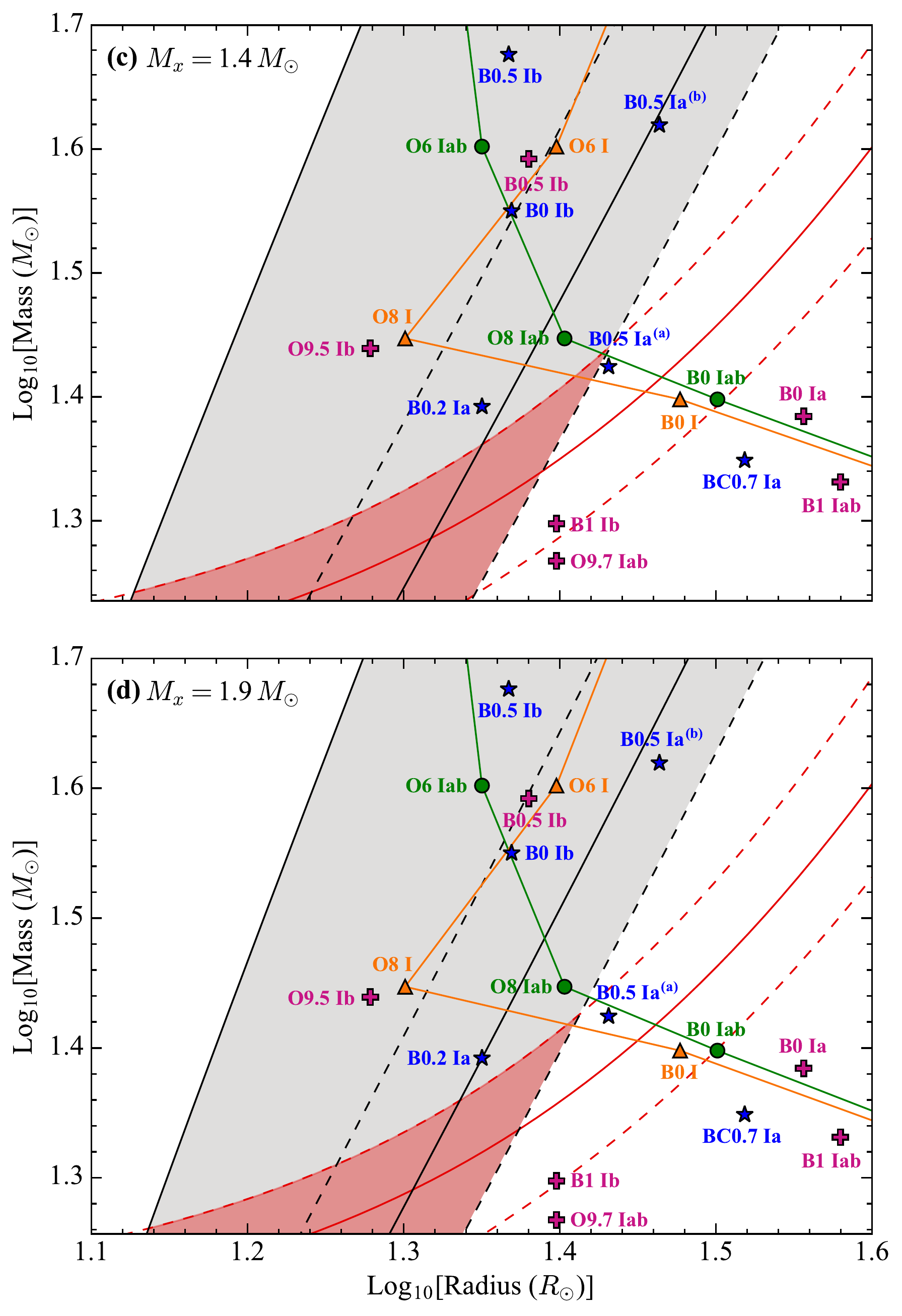}
			\label{Figure:Figure12b}
		}
		
	\end{tabular}
	
	\caption{\text{Log-log} plots of stellar mass as a function of stellar radius for \text{IGR~J16493--4348's} supergiant companion using the orbital parameters from (left)~circular solution~1 and (right)~eccentric solution~1 in Table~\ref{Table:OrbitalParameters}. We assume neutron star masses of~1.4\,$M_{\odot}$ in~(a) and~(c) and~1.9\,$M_{\odot}$ in~(b) and~(d). The left and right solid black curves show constraints corresponding to an \text{edge-on} orbit and where the supergiant fills its Roche lobe, respectively. The solid red curves show constraints obtained using the orbital parameters in Table~\ref{Table:OrbitalParameters} and the asymmetric eclipse model parameters in Table~\ref{Table:AsymmetricEclipseModel} from fitting the \textit{Swift}~BAT~transient monitor~\text{(15--50\,keV)} orbital profile. The dashed curves indicate 1$\sigma$~uncertainties on these constraints. The grey shaded regions correspond to the allowed parameter space for inclination angles between Roche lobe overflow and an \text{edge-on} orbit, and the red shaded areas indicate the joint allowed region also satisfying constraints from the asymmetric eclipse and timing models. The green circles, orange triangles, blue stars, and magenta crosses correspond to supergiant spectral types from~\citet{Carroll2006},~\citet{Cox2000},~\citet{Searle2008}, and~\citet{Lefever2007}, respectively. The~B0.5~Ia$^{\text{(a)}}$ and~B0.5~Ia$^{\text{(b)}}$ labels are used to distinguish between the two B0.5~Ia Galactic~B~supergiants with different masses and radii in Table~3 of~\citet{Searle2008}. We favor a spectral type of B0.5~Ia for the supergiant donor since this is the only spectral type that lies in the joint allowed regions derived using the orbital parameters from circular solution~1.}
	\label{Figure:Figure12}
\end{figure*}

Next, we constrain the spectral type of \text{IGR~J16493--4348's} supergiant companion using the stellar \text{mass-radius} diagrams in Figure~\ref{Figure:Figure12}. The relationship between the supergiant's mass and radius is shown for neutron star masses of~1.4\,$M_{\odot}$ and~1.9\,$M_{\odot}$. Constraints are derived using the asymmetric eclipse model parameters from the BAT~transient monitor orbital profile and the orbital parameters from circular solution~1 and eccentric solution~1. The grey shaded regions show the allowed parameter space for inclination angles between Roche lobe overflow and an \text{edge-on} orbit, and the red shaded areas correspond to the joint allowed region also satisfying constraints from the asymmetric eclipse and timing models. Supergiant spectral types from~\citet{Carroll2006},~\citet{Cox2000},~\citet{Searle2008}, and~\citet{Lefever2007} are labeled using green circles, orange triangles, blue stars, and magenta crosses, respectively. We spectrally classify the companion of \text{IGR~J16493--4348} as a B0.5~Ia supergiant since this is the only spectral type that lies in the joint allowed regions obtained using the orbital parameters from circular solution~1. This spectral type is consistent with the previous spectral classification by~\citet{Nespoli2010} from \text{$K_{\text{S}}$-band} spectroscopy of \text{IGR~J16493--4348's} infrared counterpart. There are no supergiant spectral types from~\citet{Carroll2006}, \citet{Cox2000}, \citet{Searle2008}, or~\citet{Lefever2007} inside the joint allowed regions derived using eccentric solution~1, which may be due to the few degrees of freedom in the fit.

In Table~\ref{Table:DonorStarParametersSpectralTypes}, we assume a neutron star mass of~1.4\,$M_{\odot}$ and present supergiant donor parameters for selected spectral types from~\citet{Carroll2006}, \citet{Searle2008}, and~\citet{Lefever2007}, along with estimates of the source distance and hydrogen column density. The inclination angles were calculated from Equation~(\ref{Equation:CompanionRadius}) using published values for the companion masses and radii and the measured eclipse \text{half-angle} in Table~\ref{Table:AsymmetricEclipseModel} from fitting the BAT~transient monitor orbital profile. The B0.5~Ia spectral type from~\citet{Searle2008}, which lies in the joint allowed region of the stellar \text{mass-radius} diagrams in Figures~\ref{Figure:Figure12}(a) and~\ref{Figure:Figure12}(b), is highlighted in bold.

\begin{figure*}[t]
	\centering
	\includegraphics[trim=0cm 0cm 0cm 0cm, clip=false, scale=0.46, angle=0]{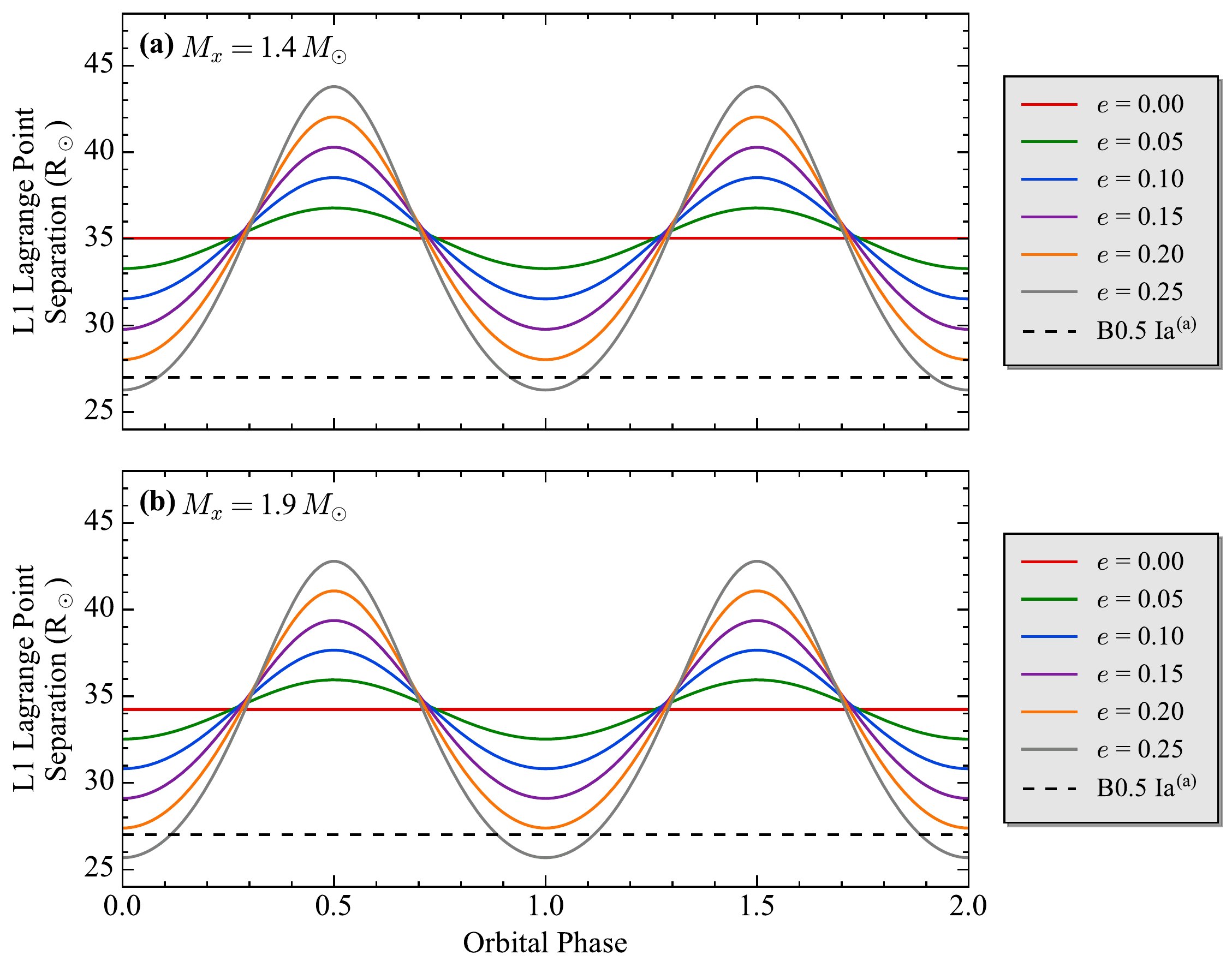}
	\caption{L1~Lagrange point separation from \text{IGR~J16493--4348's} supergiant companion as a function of orbital phase. The solid curves indicate the separation for different eccentricities between~0 and~0.25, and the horizontal dashed lines correspond to a supergiant radius of 27\,$R_{\odot}$ for the favored B0.5~Ia spectral type from~\citet{Searle2008}. For eccentric orbits with $e$\,$\gtrsim$\,0.20, Roche lobe overflow will be induced during orbital phases where the L1~Lagrange point separation is inside the supergiant.}
	\label{Figure:Figure13}
\end{figure*}

Mass transfer in close eccentric binaries is expected to occur at or near periastron, where the effective Roche lobes of the constituent stars are smallest~\citep{Sepinksy2010}. We show the variation in the L1~Lagrange point separation from the supergiant companion as a function of orbital phase in Figure~\ref{Figure:Figure13} for a range of eccentricities between~0 and~0.25. The horizontal dashed lines correspond to a companion radius of~27\,$R_{\odot}$ for the B0.5~Ia spectral type from~\citet{Searle2008}. We find that an eccentric orbit with $e$\,$\gtrsim$\,0.20 would induce Roche lobe overflow during orbital phases when the L1~Lagrange point is inside the supergiant.



\addtocounter{table}{1}

\begin{deluxetable*}{lcc}
	\tablenum{7}
	\tabletypesize{\scriptsize}
	\tablecolumns{3}
	\tablewidth{0pt}
	\tablecaption{\textsc{Supergiant Donor Parameters of \text{IGR~J16493--4348}}}
	\tablehead{
		\colhead{Parameter} &
		\colhead{Roche Lobe Overflow} &
		\colhead{Edge-On}
	}
	$i$\,(deg)$^{\mathrm{a}}$ & 56.0$^{\text{+6.4}}_{\text{--5.8}}$ & 90.0 \\
	$M_{x}$\,$(M_{\odot})$$^{\mathrm{b}}$ & 1.4 & 1.4 \\
	$M_{c}$\,$(M_{\odot})$$^{\mathrm{c}}$ & 25.8$^{\text{+6.3}}_{\text{--5.9}}$ & 15.7$^{\text{+2.4}}_{\text{--2.5}}$ \\
	$q$$^{\mathrm{d}}$ & 0.05\,$\pm$\,0.01 & 0.09\,$\pm$\,0.01 \\
	$R_{c}$\,$(R_{\odot})$$^{\mathrm{e}}$ & 28.3$^{\text{+5.7}}_{\text{--5.1}}$ & 13.0$^{\text{+5.9}}_{\text{--4.7}}$ \\
	$R_{L}$\,$(R_{\odot})$$^{\mathrm{f}}$ & 28.3$^{\text{+2.8}}_{\text{--2.6}}$ & 22.8\,$\pm$\,1.2 \\
	$\beta$$^{\mathrm{g}}$ & 1.00$^{\text{+0.22}}_{\text{--0.20}}$ & 0.57$^{\text{+0.26}}_{\text{--0.21}}$ \\
	\tableline
	$i$\,(deg)$^{\mathrm{a}}$ & 58.1$^{\text{+5.9}}_{\text{--5.1}}$ & 90.0 \\
	$M_{x}$\,$(M_{\odot})$$^{\mathrm{b}}$ & 1.9 & 1.9 \\
	$M_{c}$\,$(M_{\odot})$$^{\mathrm{c}}$ & 25.0$^{\text{+5.1}}_{\text{--4.8}}$ & 16.5\,$\pm$\,2.5 \\
	$q$$^{\mathrm{d}}$ & 0.08$^{\text{+0.02}}_{\text{--0.01}}$ & 0.12\,$\pm$\,0.02 \\
	$R_{c}$\,$(R_{\odot})$$^{\mathrm{e}}$ & 27.1$^{\text{+5.0}}_{\text{--4.4}}$ & 13.3$^{\text{+6.1}}_{\text{--4.8}}$ \\
	$R_{L}$\,$(R_{\odot})$$^{\mathrm{f}}$ & 27.1$^{\text{+2.2}}_{\text{--2.1}}$ & 22.6$^{\text{+1.1}}_{\text{--1.2}}$ \\
	$\beta$$^{\mathrm{g}}$ & 1.00$^{\text{+0.20}}_{\text{--0.18}}$ & 0.59$^{\text{+0.27}}_{\text{--0.22}}$
	\enddata
	\tablecomments{Parameter values were obtained using the orbital parameters from circular solution~1 in Table~\ref{Table:OrbitalParameters} and the asymmetric eclipse model parameters from the \textit{Swift}~BAT~transient monitor~\text{(15--50\,keV)} orbital profile in Table~\ref{Table:AsymmetricEclipseModel}. We quote 1$\sigma$~uncertainties on each parameter, if applicable. \\
		$^{\mathrm{a}}$ Inclination angles where the supergiant donor fills its Roche lobe and where the binary system is viewed \text{edge-on}~($i$\,$=$\,90$^{\circ}$). \\
		$^{\mathrm{b}}$ Assumed mass of the neutron star. \\
		$^{\mathrm{c}}$ Mass of the supergiant donor calculated using Equation~(\ref{Equation:XrayMassFunction}). \\
		$^{\mathrm{d}}$ Mass ratio,~$q$\,$=$\,$M_{x}/M_{c}$, where $M_{x}$~is the mass of the neutron star and $M_{c}$~is the mass of the supergiant companion. \\
		$^{\mathrm{e}}$ Radius of the supergiant donor obtained using Equation~(\ref{Equation:CompanionRadius}). \\
		$^{\mathrm{f}}$ Roche lobe radius calculated using Equation~(\ref{Equation:RocheLobeRadius}). \\
		$^{\mathrm{g}}$ Roche lobe filling factor,~$\beta$\,$=$\,$R_{c}/R_{L}$, where $R_{c}$~is the radius of the supergiant companion and $R_{L}$~is the Roche lobe radius.}
	\label{Table:DonorStarParameters}
\end{deluxetable*}



\addtocounter{table}{1}

\begin{deluxetable*}{cccccccccccc}
	\tablenum{8}
	\tabletypesize{\small}
	\tablecolumns{12}
	\tablewidth{1.07\textwidth}
	\tablecaption{\textsc{System Parameters of \text{IGR~J16493--4348} for Selected Spectral Types}}
	\tablehead{
		\colhead{Spectral Type} &
		\colhead{$M_{c}$$^{\mathrm{a}}$} &
		\colhead{$q$$^{\mathrm{b}}$} &
		\colhead{$R_{c}$$^{\mathrm{c}}$} &
		\colhead{$R_{L}$$^{\mathrm{d}}$} &
		\colhead{$\beta$$^{\mathrm{e}}$} &
		\colhead{$i$$^{\mathrm{f}}$} &
		\colhead{$M_{V}$$^{\mathrm{g}}$} &
		\colhead{$(J-K)_{0}$$^{\mathrm{h}}$} &
		\colhead{$E(J-K)$$^{\mathrm{i}}$} &
		\colhead{$d$$^{\mathrm{j}}$} &
		\colhead{$N_{\text{H}}$$^{\mathrm{k}}$} \\
		\colhead{} &
		\colhead{$(M_{\odot})$} &
		\colhead{} &
		\colhead{$(R_{\odot})$} &
		\colhead{$(R_{\odot})$} &
		\colhead{} &
		\colhead{(deg)} &
		\colhead{(mag)} &
		\colhead{(mag)} &
		\colhead{(mag)} &
		\colhead{(kpc)} &
		\colhead{(10$^{\text{22}}$ cm$^{\text{2}}$)}
	}
	\startdata
	$^{\mathrm{l}}$O8 Iab & 28.0 & 0.05 & 25.3 & 29.3 & 0.86 & 62.9$^{\text{+6.4}}_{\text{--5.0}}$ & --6.6 & --0.18\,$\pm$\,0.13 & 2.84\,$\pm$\,0.15 & 14.8\,$\pm$\,1.2 & 3.00\,$\pm$\,0.25 \\
	$^{\mathrm{m}}$B0.2 Ia & 24.7\,$\pm$\,7.1 & 0.06\,$\pm$\,0.02 & 22.4\,$\pm$\,3.2 & 27.8\,$\pm$\,2.7 & 0.81\,$\pm$\,0.14 & 66.7$^{\text{+10.7}}_{\text{--9.7}}$ & --6.07\,$\pm$\,0.30 & --0.13\,$\pm$\,0.13 & 2.79\,$\pm$\,0.15 & 12.9\,$\pm$\,2.1 & 2.94\,$\pm$\,0.24 \\
	$^{\mathrm{m}}$\textbf{B0.5~Ia$^{\text{(a)}}$} & 26.6\,$\pm$\,2.4 & 0.053\,$\pm$\,0.005 & 27.0\,$\pm$\,1.2 & 28.7\,$\pm$\,0.9 & 0.94\,$\pm$\,0.05 & 59.0$^{+6.1}_{-5.1}$ & --6.48\,$\pm$\,0.10 & --0.12\,$\pm$\,0.13 & 2.78\,$\pm$\,0.15 & 16.1\,$\pm$\,1.5 & 2.93\,$\pm$\,0.24 \\
	$^{\mathrm{m}}$B0.5~Ia$^{\text{(b)}}$ & 41.6\,$\pm$\,3.8 & 0.034\,$\pm$\,0.003 & 29.1\,$\pm$\,1.3 & 34.7\,$\pm$\,1.1 & 0.84\,$\pm$\,0.05 & 62.3$^{\text{+6.8}}_{\text{--5.6}}$ & --6.54\,$\pm$\,0.10 & --0.12\,$\pm$\,0.13 & 2.78\,$\pm$\,0.15 & 16.5\,$\pm$\,1.6 & 2.93\,$\pm$\,0.24 \\
	$^{\mathrm{m}}$B0.5 Ib & 47.5\,$\pm$\,8.8 & 0.029\,$\pm$\,0.005 & 23.3\,$\pm$\,2.2 & 36.6\,$\pm$\,2.3 & 0.64\,$\pm$\,0.07 & 74.1$^{\text{+12.4}}_{\text{--10.2}}$ & --6.36\,$\pm$\,0.20 & --0.12\,$\pm$\,0.13 & 2.78\,$\pm$\,0.15 & 15.3\,$\pm$\,1.9 & 2.93\,$\pm$\,0.24 \\
	$^{\mathrm{n}}$B1 Ib & 19.8 & 0.07 & 25.0 & 25.3 & 0.99 & 58.2$^{\text{+5.3}}_{\text{--4.2}}$ & --5.8 & --0.13\,$\pm$\,0.13 & 2.79\,$\pm$\,0.15 & 11.9\,$\pm$\,1.0 & 2.94\,$\pm$\,0.24
	\enddata
	\tablecomments{System parameters for selected spectral types from~\citet{Carroll2006}, \citet{Searle2008}, and~\citet{Lefever2007}. We assumed a canonical neutron star mass of~1.4\,$M_{\odot}$ and used the orbital parameters from circular solution~1 in Table~\ref{Table:OrbitalParameters} and the eclipse \text{half-angle} in Table~\ref{Table:AsymmetricEclipseModel} from fitting the \textit{Swift}~BAT~transient monitor~\text{(15--50\,keV)} orbital profile with an asymmetric eclipse model. We report 1$\sigma$~uncertainties on these parameters, if applicable. The favored supergiant spectral type from the stellar \text{mass-radius} diagrams in Figure~\ref{Figure:Figure12} is highlighted in bold. \\
		$^{\mathrm{a}}$ Mass of the supergiant companion from~\citet{Carroll2006}, \citet{Searle2008}, and~\citet{Lefever2007}. \\
		$^{\mathrm{b}}$ Mass ratio, $q$\,$=$\,$M_{x}/M_{c}$, where $M_{x}$~is the neutron star mass and $M_{c}$~is the mass of the supergiant donor. \\
		$^{\mathrm{c}}$ Radius of the supergiant companion from~\citet{Carroll2006}, \citet{Searle2008}, and~\citet{Lefever2007}. \\
		$^{\mathrm{d}}$ Roche lobe radius calculated using Equation~(\ref{Equation:RocheLobeRadius}). \\
		$^{\mathrm{e}}$ Roche lobe filling factor, $\beta$\,$=$\,$R_{c}/R_{L}$, where $R_{c}$~is radius of the supergiant donor and $R_{L}$~is the Roche lobe radius. \\
		$^{\mathrm{f}}$ Inclination angle calculated using Equation~(\ref{Equation:CompanionRadius}). \\
		$^{\mathrm{g}}$ Absolute magnitude from~\citet{Carroll2006}, \citet{Searle2008}, and~\citet{Lefever2007}. \\
		$^{\mathrm{h}}$ \mbox{$(J-K)_{0}$}~intrinsic color index calculated using the~\mbox{$(J-V)_{0}$} and \mbox{$(K-V)_{0}$}~color indices from~\citet{Wegner1994}. The uncertainties on the intrinsic color indices were calculated using Equation~(6) in~\citet{Wegner1994}. \\
		$^{\mathrm{i}}$ Excess color calculated by subtracting the intrinsic \mbox{$(J-K)_{0}$}~color from the \mbox{$(J-K)_{\text{2MASS}}$}~color. \\
		$^{\mathrm{j}}$ Distance to the source calculated using the distance modulus with an apparent 2MASS~\text{$K$-band}~magnitude from~\citet{Cutri2003} and \text{$K$-band}~extinction determined from~\citet{Rieke1985}. \\
		$^{\mathrm{k}}$ Hydrogen column density calculated from the correlation with visual extinction in~\citet{Predehl1995}. \\
		$^{\mathrm{l}}$ Supergiant star from~\citet{Carroll2006}. \\
		$^{\mathrm{m}}$ Supergiant star from~\citet{Searle2008}. \\
		$^{\mathrm{n}}$ Supergiant star from~\citet{Lefever2007}. \\}
	\label{Table:DonorStarParametersSpectralTypes}
\end{deluxetable*}



\section{Discussion}
\label{Section:Discussion}

\subsection{Donor Star Spectral Type}
\label{Section:Spectral_Type}

Our measurements of the 6.78\,day~orbital period and 1093\,s~pulse period firmly place \text{IGR~J16493--4348} in the \text{wind-fed}~sgHMXB region of the \text{$P_{\text{orb}}$--$P_{\text{pulse}}$}~Corbet diagram~\citep{Corbet1984, Corbet1986}. This is further supported by comparing our \text{X-ray} mass function of $f_{x}(M)$\,$=$\,13.2$^{\text{+2.4}}_{\text{--2.5}}$\,$M_{\odot}$ to the \text{X-ray} mass functions of other sgHMXBs~\citep{Townsend2011}. \citet{Nespoli2010} estimated the spectral type of the donor star to be a \text{B0.5-1~Ia-Ib}~supergiant by comparing the relative strength of He~I~lines from \text{$K_{\text{S}}$-band} spectroscopy of \text{IGR~J16493--4348's} infrared companion to those reported in~\citet{Hanson1996}. We find a spectral type of~B0.5~Ia for the supergiant companion using constraints derived in the stellar \text{mass-radius} diagrams shown in Figure~\ref{Figure:Figure12}. We assumed neutron star masses of~1.4\,$M_{\odot}$ and~1.9\,$M_{\odot}$ since neither optical nor infrared radial velocity \text{semi-amplitude} measurements were available. In both cases, we obtain a spectral type that is consistent with the result in~\citet{Nespoli2010}, but we note that a compact object of~1.9\,$M_{\odot}$ would make it one of the most massive neutron stars in an \text{X-ray} binary~\citep{vanKerkwijk1995, Kaper2006}. Two Galactic~B~supergiants with spectral types of~B0.5~Ia from~\citet{Searle2008} are shown in Figure~\ref{Figure:Figure12}, but constraints on the allowed mass and radius from our timing models exclude the more massive, larger donor. From our spectral classification, we estimate the surface effective temperature and luminosity to be approximately~26,000\,K and~3.0\,$\times$\,10$^{5}\,L_{\odot}$, respectively.

We estimate the distance to the source using the supergiant's B0.5~Ia~spectral type and the reported parameters in Table~3 of~\citet{Searle2008}. The infrared counterpart has an apparent \text{$K$-band} magnitude of~$m_{K}$\,$=$\,11.94\,$\pm$\,0.04 from 2MASS~photometry~\citep{Cutri2003}. An absolute \text{$K$-band} magnitude of~\mbox{$M_{K}$\,$=$\,--5.93\,$\pm$\,0.14} was derived from the absolute \text{$V$-band} magnitude of \text{$M_{V}$\,$=$\,--6.48\,$\pm$\,0.10} in~\citet{Searle2008} and the intrinsic \mbox{$(K-V)_{0}$}~color index of~0.55\,$\pm$\,0.10 in~\citet{Wegner1994}. Next, we found the instrinsic~\mbox{$(J-K)_{0}$} color index to be~\text{--0.12\,$\pm$\,0.13} using the~\mbox{$(J-V)_{0}$} and \mbox{$(K-V)_{0}$}~color indices from~\citet{Wegner1994}. An \mbox{$E(J-K)$}~color excess of~2.78\,$\pm$\,0.15 was obtained by subtracting the intrinsic color \mbox{$(J-K)_{0}$} from the \mbox{$(J-K)_{\text{2MASS}}$} color. Assuming an average extincition of~$R_{V}$\,$=$\,3.09\,$\pm$\,0.03, we found a \text{$V$-band} extinction magnitude~of~$A_{V}$\,$=$\,16.4\,$\pm$\,1.3 from the relation \mbox{$A_{V}/E(J-K)$\,$=$\,5.90\,$\pm$\,0.36}~\citep{Rieke1985}. This yielded an extinction magnitude of~$A_{K}$\,$=$\,1.8\,$\pm$\,0.1 at \text{$K$-band} using $A_{K}/A_{V}$\,$=$\,0.112 from Table~3 in~\citet{Rieke1985}. From the distance modulus, $M_{K}$\,$=$\,$m_{K}$\,$+$\,$5$\,$-$\,$5$\,$\mathrm{log}\,d$\,$-$\,$A_{K}$, we find that \text{IGR~J16493--4348} lies at a distance of 16.1\,$\pm$\,1.5\,kpc, which is consistent with the \text{6--26\,kpc} distance estimate in~\citet{Nespoli2010}. Our distance measurement is also in agreement with the {7.5--22\,kpc} distance reported by~\citet{Hill2008} from infrared spectral energy distribution measurements of the supergiant companion.

Next, we calculate a hydrogen column density of $N_{\text{H}}$\,$=$\,(2.93\,$\pm$\,0.24)\,$\times$\,10$^{\text{22}}$\,cm$^{\text{--2}}$ from the correlation between visual extinction and hydrogen column density in~\citet{Predehl1995}, which is consistent with the estimate given in~\citet{Nespoli2010}. If we instead use the more recently measured correlation between optical extinction and~$N_{\text{H}}$ in~\citet{Guver2009}, we obtain a hydrogen column density of $N_{\text{H}}$\,$=$\,(3.62\,$\pm$\,0.33)\,$\times$\,10$^{\text{22}}$\,cm$^{\text{--2}}$. Using the procedure in~\citet{Willingale2013}, we find a total hydrogen column density of~$N_{\text{H}}$\,$=$\,1.56\,$\times$\,10$^{\text{22}}$\,cm$^{\text{--2}}$, which is comparable to the $N_{\text{H}}$ values of~1.42\,$\times$\,10$^{\text{22}}$\,cm$^{\text{--2}}$ and~1.82\,$\times$\,10$^{\text{22}}$\,cm$^{\text{--2}}$ obtained from the Leiden/Argentine/Bonn survey~\citep{Kalberla2005} and~\citet{Dickey1990}, respectively, using measurements of~H~I~in the Galaxy. We note that all of these values are smaller than the observed hydrogen column densities measured by~\citet{Hill2008}, \citet{Morris2009}, \citet{DAi2011}, and~\citet{Coley2018}. Spectral analyses in the soft and hard \text{X-ray} bands have found hydrogen column densities ranging between roughly \text{5--10}\,$\times$\,10$^{\text{22}}$\,cm$^{\text{--2}}$ on average. This suggests that there may be an additional component of the hydrogen absorbing column that is intrinsic to the system.


\subsection{Eclipse Asymmetry}
\label{Section:Eclipse_Asymmetry}

Asymmetry in the \text{X-ray} eclipse profile is often a signature of a photoionization wake~\citep{Fransson1980, Feldmeier1996}, accretion bow shock and/or accretion wake trailing the neutron star~\citep{Blondin1990, Blondin1991}, or other complex structure in the stellar wind. We discuss these phenomena and argue that a strong photoionization or accretion wake is not supported by the \text{X-ray}~emission observed from \text{IGR~J16493--4348}.

Mass transfer onto the neutron star occurs through the radiatively powered stellar wind of the supergiant companion. \text{X-ray}~photoionization can result in collisions between the compressed, ionized gas and the accelerating wind, which causes shocks and dense regions of compressed gas from the wind to trail the \text{X-ray}~source in its orbit around the supergiant~\citep{Jackson1975, Fransson1980}. In systems with high \text{X-ray}~luminosities, the wind is highly ionized in the vicinity of the \text{X-ray}~source, and the radiative driving force powering the stellar wind is significantly reduced near the surface of the supergiant~\citep{Feldmeier1996}. As seen in Vela~X-1~\citep{Feldmeier1996}, the dense gas trailing the photoionization wake can lead to \text{X-ray}~photoelectric absorption at orbital phases prior to the eclipse and \text{X-ray}~scattering into the observer's line of sight after eclipse ingress. This can produce ingress durations that are longer than those observed at egress.

Dense regions of compressed gas in the accretion bow shock and/or accretion wake of the compact object can also induce \text{phase-dependent} photoelectric absorption~\citep{Jackson1975, Blondin1990, Blondin1991}. This leads to an enhancement in the hydrogen column density and absorption of the \text{X-ray}~emission prior to the eclipse~\citep{Manousakis2015}. No apparent increase in the hydrogen column density is observable during egress when the accretion wake is located beyond the compact object. We do not find evidence of a strong photoionization or accretion wake since there are no statistically significant differences between the ingress and egress durations or the count rates near ingress and egress in the PCA~scan or BAT~orbital profiles.

The eclipse profile structure is often dependent on \text{X-ray}~photon energy. For example, \citet{Jain2009} found that the \text{X-ray}~eclipses from the~SFXT \text{IGR~J16479--4514} were more evident and exhibited sharper transitions at higher energies using the \textit{Swift}~BAT compared to observations at lower energies with the \textit{RXTE}~All-Sky Monitor~(ASM). This type of behavior has also been observed from various other eclipsing systems in the hard \text{X-ray} band~(e.g.,~\citealt{Falanga2015}). These effects are often linked to the absorbing column density, which causes increased \text{X-ray}~absorption and scattering at softer \text{X-ray}~energies. Although the eclipse duration of \text{IGR~J16493--4348} was consistent between the PCA~scan and BAT~orbital profiles, there are observable differences in the eclipse profile structure across the \text{X-ray}~energy band~(see~Figure~\ref{Figure:Figure7}). While these differences may be indicative of energy dependent structure in the eclipses, systematic effects from binning could also affect the observed eclipse shape.


\subsection{Orbital Eccentricity}
\label{Section:Orbital_Eccentricity}

Previous estimates by~\citet{Cusumano2010} suggest that the orbital eccentricity cannot exceed~0.15 based on \text{IGR~J16493--4348's} classification as a \text{wind-fed}~sgHMXB. An eccentricity of $e$\,$=$\,0.17\,$\pm$\,0.09 was measured using the timing model in eccentric solution~1. While this eccentricity is consistent with the upper limit presented in~\citet{Cusumano2010}, we suspect that the orbit is nearly circular since the~ToAs are well modeled by circular solution~1 and a B0.5~Ia spectral type fell within the joint allowed parameter space in the corresponding stellar \text{mass-radius} diagrams in Figure~\ref{Figure:Figure12}. This spectral type is also consistent with the spectral classification given by~\citet{Nespoli2010}. Additionally, no spectral types were found inside the joint allowed regions obtained using eccentric solution~1. If the orbit were highly eccentric~($e$\,$\gtrsim$\,0.20), then the L1~Lagrange point separation from the supergiant would be located inside the donor during a fraction of the orbit, which would lead to Roche lobe overflow and inhibit mass transfer via the stellar wind. Since the eccentric timing model fits have only a few degrees of freedom, higher cadence pulsar timing observations over multiple orbital cycles are needed to measure the system's eccentricity and longitude of periastron more accurately.

We compare the \text{mid-eclipse} time predicted by eccentric solution~1 in Table~\ref{Table:OrbitalParameters} to the measured \text{mid-eclipse} times from the PCA~scan and BAT~orbital profiles in Tables~\ref{Table:AsymmetricEclipseModel} and~\ref{Table:SymmetricEclipseModel}. To first order in $e$, the time of \text{mid-eclipse} in an eccentric orbit is given by~\citep{vanDerKlis1984, Falanga2015}:
\begin{equation}
T_{\text{mid}}=T_{\pi/2}-\frac{eP_{\text{orb}}}{\pi}\cos\omega
\label{Equation:EccentricMidEclipse}
\end{equation}
Here, $T_{\pi/2}$ is calculated from the periastron passage time, $T_{\text{peri}}$, using~\citep{vanDerKlis1984}:
\begin{equation}
T_{\pi/2}=T_{\text{peri}}+\frac{P_{\text{orb}}}{2\pi}\left(\frac{\pi}{2}-\omega\right)
\label{Equation:EccentricT90}
\end{equation}
If the orbit is circular, the values of $T_{\pi/2}$ and $T_{\text{mid}}$ will coincide.

Substituting the orbital parameters from eccentric solution~1 into Equations~(\ref{Equation:EccentricMidEclipse}) and~(\ref{Equation:EccentricT90}), we find that $T_{\pi/2}$\,$=$\,MJD\,55850.9\,$\pm$\,0.8 and $T_{\text{mid}}$\,$=$\,MJD\,55851.0\,$\pm$\,0.8 are consistent with each other. The large uncertainties in these calculated values are attributed to the broad posterior distributions measured for the periastron passage time and longitude of periastron in eccentric solution~1. This calculated \text{mid-eclipse} time agrees with all of the measured \text{mid-eclipse} times in Tables~\ref{Table:AsymmetricEclipseModel} and~\ref{Table:SymmetricEclipseModel} to within 1$\sigma$. In addition, the value of $T_{\pi/2}$ derived from eccentric solution~1 is consistent with the values of $T_{\pi/2}$ measured using the circular timing models in Table~\ref{Table:OrbitalParameters}. This further supports the notion that the orbit is likely not highly eccentric.


\subsection{Superorbital Mechanisms}
\label{Section:Superorbital_Mechanisms}

\text{IGR~J16493--4348} is one of only five \text{wind-fed}~sgHMXB systems in which superorbital modulation has been definitively observed~(e.g.,~2S~0114+650, 4U~1909+07, \text{IGR~J16418--4532}, and~\text{IGR~J16479--4514};~\citealt{Corbet2013}). In addition, \citet{Corbet2018}~recently reported apparent superorbital modulation from the eclipsing~sgHMXB, 4U~1538--52. Superorbital variability from other \text{X-ray}~pulsar binaries, such as~\text{Her~X-1}, \text{SMC~X-1}, and~\text{LMC~X-4}, has been linked to mass flow onto the accretion disk of the neutron star via Roche lobe overflow~\citep{Clarkson2003a, Clarkson2003b}. Accretion flow onto the surface of a freely precessing neutron star with a complex \text{non-dipole} magnetic field has also been suggested to explain the 35\,day~superorbital period of~\text{Her~X-1}~\citep{Postnov2013}. Alternatively, the periodic superorbital behavior in these systems could be caused by a twisted, warped precessing accretion disk~\citep{Petterson1975, Wojdowski1998, Ogilvie2001, Hung2010}.

We detected coherent superorbital modulation at a period of~20.07\,days from \text{semi-weighted}~DFTs of the~BAT and~PCA~scan light curves. While superorbital periods of similar length have been detected in other wind accreting~sgHMXBs, such as~2S~0114+650, the mechanism responsible for the variability still has not been clearly identified~\citep{Farrell2006, Masetti2006, Farrell2008, Hu2017}. It may be possible that tidal oscillations from \text{IGR~J16493--4348's} B0.5~Ia~supergiant companion are driving the variability if the orbit is indeed circular~\citep{Koenigsberger2003, Moreno2005, Koenigsberger2006}. Using a tidal interaction model, \citet{Koenigsberger2006} found that these oscillations would produce modulation on superorbital timescales in binary systems with circular orbits, while orbital period length variability would be observed if these oscillations occurred in eccentric orbits. In both cases, suborbital variability was also predicted on shorter timescales. This may suggest that the mechanism responsible for the superorbital modulation is the structured stellar wind of the supergiant companion, possibly along with \text{X-ray}~emission generated by strong perturbations on the surface layers of the donor star.

Alternatively, the superorbital variability may be related to the presence of corotating interaction regions~(CIRs) in the stellar wind of the supergiant~\citep{Bozzo2017}. These structures are thought to form from irregularities on the surface of the donor star and are located at radial distances of tens of stellar radii~\citep{Mullan1984, Cranmer1996}. We find that \text{IGR~J16493-4348's} superorbital period is persistently detected in \text{X-ray} observations spanning several years, but its strength is variable in time~\citep{Coley2018}. This implies that these~CIRs would have to be stable over long timescales if this is the dominant mechanism driving the variability, which has not yet been established. A detailed discussion of other possible mechanisms responsible for the superorbital modulation is presented in~\citet{Coley2018}.



\section{Conclusion}
\label{Section:Conclusion}

\text{IGR~J16493--4348} is an eclipsing, \text{wind-fed}~sgHMXB with an early \text{B-type} supergiant companion. We refine the superorbital period to~20.067\,$\pm$\,0.009\,days from a \text{semi-weighted}~DFT of the BAT~transient monitor light curve. An improved orbital period measurement of~6.7828\,$\pm$\,0.0004\,days is obtained from an \text{O--C}~analysis using the PCA~scan and BAT~transient monitor data. Asymmetric and symmetric eclipse models were fit to the PCA~scan and BAT~orbital profiles, and no evidence of a strong photoionization or accretion wake was found.

Pulsations were detected in the unweighted power spectrum of the pointed~PCA light curve after the removal of low frequency noise. We refine the pulse period to~1093.1036\,$\pm$\,0.0004\,s from a pulsar timing analysis using the pointed~PCA data. The system's Keplerian binary orbital parameters were measured by fitting circular and eccentric timing models to the~ToAs. We find that the orbit is likely nearly circular, and no significant change in the rotational period of the pulsar was observed. A mass function of $f_{x}(M)$\,$=$\,13.2$^{\text{+2.4}}_{\text{--2.5}}$\,$M_{\odot}$ was derived from the binary orbital parameters, which allows us to definitively classify \text{IGR~J16493--4348} as an~sgHMXB. This is further supported by its updated placement in the \text{wind-fed} sgHMXB region of the \text{$P_{\text{orb}}$--$P_{\text{pulse}}$} Corbet diagram. We derive new constraints on the mass and radius of the donor star, which indicate a B0.5~Ia spectral type for the supergiant companion. Additional parameters describing the system geometry are also provided, which give insight into possible inclination angles and Roche lobe sizes.

Although we argue that the binary follows a nearly circular orbit, additional~ToAs are needed to provide improved constraints on the system's eccentricity and longitude of periastron. Optical or infrared radial velocity measurements would directly determine the pulsar's neutron star mass, which would allow the system to be classified as a \text{double-lined} eclipsing binary. The driving mechanism behind the superorbital modulation remains unexplained, but it is currently thought to be linked with the stellar wind of the supergiant companion.


\section*{Acknowledgments}

We thank the referee for useful suggestions that helped us improve this paper. A.~B.~Pearlman acknowledges support by the Department of Defense~(DoD) through the National Defense Science and Engineering Graduate Fellowship~(NDSEG) Program and by the National Science Foundation Graduate Research Fellowship under Grant~No.~DGE-1144469. This work was partially supported by NASA~grant~NNX15AI74G.


\section*{Appendix}
\label{Section:Appendix}

In Table~\ref{Table:DonorStarParametersAppendix}, we provide a complete list of calculated values for the companion mass, mass ratio, companion radius, Roche lobe radius, and Roche lobe filling factor using the orbital parameters from each timing solution in Table~\ref{Table:OrbitalParameters} and the asymmetric and symmetric eclipse model parameters in Tables~\ref{Table:AsymmetricEclipseModel} and~\ref{Table:SymmetricEclipseModel} from fitting the folded BAT~transient monitor and PCA~scan orbital profiles. These values were determined at inclination angles corresponding to Roche lobe overflow and an \text{edge-on} orbit. We assumed neutron star masses of 1.4\,$M_{\odot}$ and 1.9\,$M_{\odot}$ in these calculations.

\clearpage



\addtocounter{table}{1}

\begin{deluxetable*}{lcccccccccc}
	\tablenum{9}
	\tabletypesize{\scriptsize}
	\tablecolumns{11}
	\tablewidth{0pt}
	\tablecaption{\textsc{Supergiant Donor Parameters of \text{IGR~J16493--4348}}}
	\tablehead{
		\colhead{Parameter} &
		\multicolumn{2}{c}{\textbf{Circular Solution~1}} &
		\multicolumn{2}{c}{Circular Solution~2} &
		\multicolumn{2}{c}{Circular Solution~3} &
		\multicolumn{2}{c}{Eccentric Solution~1} &
		\multicolumn{2}{c}{Eccentric Solution~2}
	}
	\startdata
	\tableline
	\tableline
	\tableline
	\multicolumn{11}{c}{\textit{Swift}~BAT Transient Monitor~\text{(15--50\,keV),} Asymmetric Eclipse Model} \\
	\tableline
	\tableline
	\tableline
	$i$\,(deg)$^{\mathrm{a}}$ & 56.0$^{\text{+6.4}}_{\text{--5.8}}$ & 90.0 & 56.0$^{\text{+6.1}}_{\text{--5.6}}$ & 90.0 & 56.4$^{\text{+5.9}}_{\text{--5.3}}$ & 90.0 & 67.8$^{\text{+9.2}}_{\text{--8.3}}$ & 90.0 & 68.2$^{\text{+10.7}}_{\text{--9.7}}$ & 90.0 \\
	$M_{x}$\,$(M_{\odot})$$^{\mathrm{b}}$ & 1.4 & 1.4 & 1.4 & 1.4 & 1.4 & 1.4 & 1.4 & 1.4 & 1.4 & 1.4 \\
	$M_{c}$\,$(M_{\odot})$$^{\mathrm{c}}$ & 25.8$^{\text{+6.3}}_{\text{--5.9}}$ & 15.7$^{\text{+2.4}}_{\text{--2.5}}$ & 25.7$^{\text{+6.2}}_{\text{--5.9}}$ & 15.6$^{\text{+2.8}}_{\text{--2.9}}$ & 24.3$^{\text{+5.5}}_{\text{--5.2}}$ & 15.0$^{\text{+2.6}}_{\text{--2.7}}$ & 19.3$^{\text{+4.2}}_{\text{--3.9}}$ & 15.8$^{\text{+2.5}}_{\text{--2.6}}$ & 18.8$^{\text{+4.8}}_{\text{--4.5}}$ & 15.5\,$\pm$\,2.7 \\
	$q$$^{\mathrm{d}}$ & 0.05\,$\pm$\,0.01 & 0.09\,$\pm$\,0.01 & 0.05\,$\pm$\,0.01 & 0.09\,$\pm$\,0.02 & 0.06\,$\pm$\,0.01 & 0.09\,$\pm$\,0.02 & 0.07$^{\text{+0.02}}_{\text{--0.01}}$ & 0.09\,$\pm$\,0.01 & 0.07\,$\pm$\,0.02 & 0.09\,$\pm$\,0.02 \\
	$R_{c}$\,$(R_{\odot})$$^{\mathrm{e}}$ & 28.3$^{\text{+5.7}}_{\text{--5.1}}$ & 13.0$^{\text{+5.9}}_{\text{--4.7}}$ & 28.3$^{\text{+5.4}}_{\text{--4.9}}$ & 13.0$^{\text{+5.9}}_{\text{--4.7}}$ & 27.6$^{\text{+5.1}}_{\text{--4.6}}$ & 12.8$^{\text{+5.8}}_{\text{--4.6}}$ & 20.8$^{\text{+6.4}}_{\text{--5.7}}$ & 13.3$^{\text{+6.2}}_{\text{--5.0}}$ & 20.4$^{\text{+7.4}}_{\text{--6.5}}$ & 13.3$^{\text{+6.2}}_{\text{--4.9}}$ \\
	$R_{L}$\,$(R_{\odot})$$^{\mathrm{f}}$ & 28.3$^{\text{+2.8}}_{\text{--2.6}}$ & 22.8\,$\pm$\,1.2 & 28.3$^{\text{+2.7}}_{\text{--2.6}}$ & 22.8$^{\text{+1.3}}_{\text{--1.4}}$ & 27.6$^{\text{+2.5}}_{\text{--2.3}}$ & 22.3\,$\pm$\,1.3 & 20.8\,$\pm$\,2.5 & 19.0$^{\text{+2.2}}_{\text{--2.4}}$ & 20.4$^{\text{+2.7}}_{\text{--2.8}}$ & 18.8$^{\text{+2.1}}_{\text{--2.3}}$ \\
	$\beta$$^{\mathrm{g}}$ & 1.00$^{\text{+0.22}}_{\text{--0.20}}$ & 0.57$^{\text{+0.26}}_{\text{--0.21}}$ & 1.00$^{\text{+0.22}}_{\text{--0.20}}$ & 0.57$^{\text{+0.26}}_{\text{--0.21}}$ & 1.00$^{\text{+0.21}}_{\text{--0.19}}$ & 0.57$^{\text{+0.26}}_{\text{--0.21}}$ & 1.00$^{\text{+0.33}}_{\text{--0.30}}$ & 0.70$^{\text{+0.34}}_{\text{--0.27}}$ & 1.00$^{\text{+0.39}}_{\text{--0.35}}$ & 0.71$^{\text{+0.34}}_{\text{--0.28}}$ \\
	\tableline
	$i$\,(deg)$^{\mathrm{a}}$ & 58.1$^{\text{+5.9}}_{\text{--5.1}}$ & 90.0 & 58.1$^{\text{+7.8}}_{\text{--6.9}}$ & 90.0 & 58.5$^{\text{+7.5}}_{\text{--6.6}}$ & 90.0 & 69.4$^{\text{+8.6}}_{\text{--7.6}}$ & 90.0 & 69.9$^{\text{+10.1}}_{\text{--9.0}}$ & 90.0 \\
	$M_{x}$\,$(M_{\odot})$$^{\mathrm{b}}$ & 1.9 & 1.9 & 1.9 & 1.9 & 1.9 & 1.9 & 1.9 & 1.9 & 1.9 & 1.9 \\
	$M_{c}$\,$(M_{\odot})$$^{\mathrm{c}}$ & 25.0$^{\text{+5.1}}_{\text{--4.8}}$ & 16.5\,$\pm$\,2.5 & 24.9$^{\text{+7.0}}_{\text{--6.6}}$ & 16.4$^{\text{+2.8}}_{\text{--2.9}}$ & 23.6$^{\text{+6.2}}_{\text{--5.8}}$ & 15.7\,$\pm$\,2.7 & 19.5$^{\text{+3.6}}_{\text{--3.3}}$ & 16.5\,$\pm$\,2.6 & 19.0$^{\text{+4.1}}_{\text{--3.9}}$ & 16.3\,$\pm$\,2.7 \\
	$q$$^{\mathrm{d}}$ & 0.08$^{\text{+0.02}}_{\text{--0.01}}$ & 0.12\,$\pm$\,0.02 & 0.08\,$\pm$\,0.02 & 0.12\,$\pm$\,0.02 & 0.08\,$\pm$\,0.02 & 0.12\,$\pm$\,0.02 & 0.10\,$\pm$\,0.02 & 0.11\,$\pm$\,0.02 & 0.10\,$\pm$\,0.02 & 0.12\,$\pm$\,0.02 \\
	$R_{c}$\,$(R_{\odot})$$^{\mathrm{e}}$ & 27.1$^{\text{+5.0}}_{\text{--4.4}}$ & 13.3$^{\text{+6.1}}_{\text{--4.8}}$ & 27.0$^{\text{+6.6}}_{\text{--5.9}}$ & 13.3$^{\text{+6.1}}_{\text{--4.8}}$ & 26.4$^{\text{+6.3}}_{\text{--5.5}}$ & 13.1$^{\text{+6.0}}_{\text{--4.7}}$ & 20.2$^{\text{+5.9}}_{\text{--5.1}}$ & 13.7$^{\text{+6.4}}_{\text{--5.1}}$ & 19.9$^{\text{+6.8}}_{\text{--5.9}}$ & 13.6$^{\text{+6.3}}_{\text{--5.0}}$ \\
	$R_{L}$\,$(R_{\odot})$$^{\mathrm{f}}$ & 27.1$^{\text{+2.2}}_{\text{--2.1}}$ & 22.6$^{\text{+1.1}}_{\text{--1.2}}$ & 27.0$^{\text{+3.0}}_{\text{--2.8}}$ & 22.5\,$\pm$\,1.3 & 26.4$^{\text{+2.8}}_{\text{--2.6}}$ & 22.1\,$\pm$\,1.2 & 20.2\,$\pm$\,2.1 & 18.8$^{\text{+2.2}}_{\text{--2.3}}$ & 19.9$^{\text{+2.3}}_{\text{--2.4}}$ & 18.6$^{\text{+2.0}}_{\text{--2.3}}$ \\
	$\beta$$^{\mathrm{g}}$ & 1.00$^{\text{+0.20}}_{\text{--0.18}}$ & 0.59$^{\text{+0.27}}_{\text{--0.22}}$ & 1.00$^{\text{+0.27}}_{\text{--0.24}}$ & 0.59$^{\text{+0.27}}_{\text{--0.22}}$ & 1.00$^{\text{+0.26}}_{\text{--0.23}}$ & 0.59$^{\text{+0.27}}_{\text{--0.22}}$ & 1.00$^{\text{+0.31}}_{\text{--0.27}}$ & 0.73$^{\text{+0.35}}_{\text{--0.28}}$ & 1.00$^{\text{+0.36}}_{\text{--0.32}}$ & 0.73$^{\text{+0.35}}_{\text{--0.29}}$ \\
	\\
	\tableline
	\tableline
	\tableline
	\multicolumn{11}{c}{\textit{RXTE}~PCA Galactic Bulge Scans~\text{(2--10\,keV),} Asymmetric Eclipse Model} \\
	\tableline
	\tableline
	\tableline
	$i$\,(deg)$^{\mathrm{a}}$ & 58.5$^{\text{+4.9}}_{\text{--5.1}}$ & 90.0 & 58.6$^{\text{+4.7}}_{\text{--4.8}}$ & 90.0 & 59.0\,$\pm$\,5.5 & 90.0 & 71.8$^{\text{+9.7}}_{\text{--10.1}}$ & 90.0 & 72.3$^{\text{+6.9}}_{\text{--7.4}}$ & 90.0 \\
	$M_{x}$\,$(M_{\odot})$$^{\mathrm{b}}$ & 1.4 & 1.4 & 1.4 & 1.4 & 1.4 & 1.4 & 1.4 & 1.4 & 1.4 & 1.4 \\
	$M_{c}$\,$(M_{\odot})$$^{\mathrm{c}}$ & 23.9$^{\text{+4.7}}_{\text{--4.9}}$ & 15.7$^{\text{+2.4}}_{\text{--2.5}}$ & 23.8$^{\text{+4.5}}_{\text{--4.6}}$ & 15.6$^{\text{+2.8}}_{\text{--2.9}}$ & 22.5$^{\text{+4.9}}_{\text{--5.0}}$ & 15.0$^{\text{+2.6}}_{\text{--2.7}}$ & 18.0$^{\text{+3.9}}_{\text{--4.0}}$ & 15.8$^{\text{+2.5}}_{\text{--2.6}}$ & 17.6$^{\text{+2.8}}_{\text{--2.9}}$ & 15.5\,$\pm$\,2.7 \\
	$q$$^{\mathrm{d}}$ & 0.06\,$\pm$\,0.01 & 0.09\,$\pm$\,0.01 & 0.06\,$\pm$\,0.01 & 0.09\,$\pm$\,0.02 & 0.06\,$\pm$\,0.01 & 0.09\,$\pm$\,0.02 & 0.08\,$\pm$\,0.02 & 0.09\,$\pm$\,0.01 & 0.08\,$\pm$\,0.01 & 0.09\,$\pm$\,0.02 \\
	$R_{c}$\,$(R_{\odot})$$^{\mathrm{e}}$ & 27.4$^{\text{+3.8}}_{\text{--4.0}}$ & 15.1$^{\text{+1.8}}_{\text{--1.9}}$ & 27.3$^{\text{+3.6}}_{\text{--3.7}}$ & 15.1$^{\text{+1.9}}_{\text{--2.0}}$ & 26.7$^{\text{+4.1}}_{\text{--4.2}}$ & 14.9$^{\text{+1.8}}_{\text{--1.9}}$ & 20.2$^{\text{+5.1}}_{\text{--5.3}}$ & 15.6$^{\text{+2.3}}_{\text{--2.4}}$ & 19.8$^{\text{+3.6}}_{\text{--3.8}}$ & 15.5\,$\pm$\,2.3 \\
	$R_{L}$\,$(R_{\odot})$$^{\mathrm{f}}$ & 27.4$^{\text{+2.1}}_{\text{--2.2}}$ & 22.8\,$\pm$\,1.2 & 27.3$^{\text{+2.0}}_{\text{--2.1}}$ & 22.8$^{\text{+1.3}}_{\text{--1.4}}$ & 26.7$^{\text{+2.2}}_{\text{--2.3}}$ & 22.3\,$\pm$\,1.3 & 20.2$^{\text{+2.7}}_{\text{--2.8}}$ & 19.0$^{\text{+2.2}}_{\text{--2.4}}$ & 19.8$^{\text{+1.8}}_{\text{--2.0}}$ & 18.8$^{\text{+2.1}}_{\text{--2.3}}$ \\
	$\beta$$^{\mathrm{g}}$ & 1.00$^{\text{+0.16}}_{\text{--0.17}}$ & 0.66\,$\pm$\,0.09 & 1.00$^{\text{+0.15}}_{\text{--0.16}}$ & 0.66\,$\pm$\,0.09 & 1.00$^{\text{+0.17}}_{\text{--0.18}}$ & 0.67\,$\pm$\,0.09 & 1.00$^{\text{+0.29}}_{\text{--0.30}}$ & 0.82$^{\text{+0.15}}_{\text{--0.16}}$ & 1.00$^{\text{+0.20}}_{\text{--0.22}}$ & 0.82$^{\text{+0.15}}_{\text{--0.16}}$ \\
	\tableline
	$i$\,(deg)$^{\mathrm{a}}$ & 60.9$^{\text{+4.2}}_{\text{--4.4}}$ & 90.0 & 60.9$^{\text{+5.7}}_{\text{--5.9}}$ & 90.0 & 61.3$^{\text{+4.9}}_{\text{--5.0}}$ & 90.0 & 73.8$^{\text{+7.5}}_{\text{--7.8}}$ & 90.0 & 74.4$^{\text{+7.4}}_{\text{--7.9}}$ & 90.0 \\
	$M_{x}$\,$(M_{\odot})$$^{\mathrm{b}}$ & 1.9 & 1.9 & 1.9 & 1.9 & 1.9 & 1.9 & 1.9 & 1.9 & 1.9 & 1.9 \\
	$M_{c}$\,$(M_{\odot})$$^{\mathrm{c}}$ & 23.2$^{\text{+3.7}}_{\text{--3.9}}$ & 16.5\,$\pm$\,2.5 & 23.1$^{\text{+5.1}}_{\text{--5.3}}$ & 16.4$^{\text{+2.8}}_{\text{--2.9}}$ & 21.9$^{\text{+4.1}}_{\text{--4.2}}$ & 15.7\,$\pm$\,2.7 & 18.3\,$\pm$\,2.8 & 16.5\,$\pm$\,2.6 & 17.9\,$\pm$\,2.8 & 16.3\,$\pm$\,2.7 \\
	$q$$^{\mathrm{d}}$ & 0.08\,$\pm$\,0.01 & 0.12\,$\pm$\,0.02 & 0.08\,$\pm$\,0.02 & 0.12\,$\pm$\,0.02 & 0.09\,$\pm$\,0.02 & 0.12\,$\pm$\,0.02 & 0.10\,$\pm$\,0.02 & 0.11\,$\pm$\,0.02 & 0.11\,$\pm$\,0.02 & 0.12\,$\pm$\,0.02 \\
	$R_{c}$\,$(R_{\odot})$$^{\mathrm{e}}$ & 26.2$^{\text{+3.1}}_{\text{--3.2}}$ & 15.5$^{\text{+1.8}}_{\text{--1.9}}$ & 26.2$^{\text{+4.2}}_{\text{--4.3}}$ & 15.5$^{\text{+1.9}}_{\text{--2.0}}$ & 25.6\,$\pm$\,3.5 & 15.3$^{\text{+1.8}}_{\text{--1.9}}$ & 19.7$^{\text{+3.7}}_{\text{--3.8}}$ & 15.9$^{\text{+2.3}}_{\text{--2.4}}$ & 19.3$^{\text{+3.6}}_{\text{--3.8}}$ & 15.9$^{\text{+2.3}}_{\text{--2.4}}$ \\
	$R_{L}$\,$(R_{\odot})$$^{\mathrm{f}}$ & 26.2$^{\text{+1.6}}_{\text{--1.7}}$ & 22.6$^{\text{+1.1}}_{\text{--1.2}}$ & 26.2$^{\text{+2.2}}_{\text{--2.3}}$ & 22.5\,$\pm$\,1.3 & 25.6\,$\pm$\,1.8 & 22.1\,$\pm$\,1.2 & 19.7$^{\text{+1.9}}_{\text{--2.0}}$ & 18.8$^{\text{+2.2}}_{\text{--2.3}}$ & 19.3$^{\text{+1.7}}_{\text{--2.0}}$ & 18.6$^{\text{+2.0}}_{\text{--2.3}}$ \\
	$\beta$$^{\mathrm{g}}$ & 1.00$^{\text{+0.13}}_{\text{--0.14}}$ & 0.69\,$\pm$\,0.09 & 1.00$^{\text{+0.18}}_{\text{--0.19}}$ & 0.69$^{\text{+0.09}}_{\text{--0.10}}$ & 1.00$^{\text{+0.15}}_{\text{--0.16}}$ & 0.69$^{\text{+0.09}}_{\text{--0.10}}$ & 1.00$^{\text{+0.21}}_{\text{--0.22}}$ & 0.85\,$\pm$\,0.16 & 1.00$^{\text{+0.21}}_{\text{--0.22}}$ & 0.85$^{\text{+0.16}}_{\text{--0.17}}$ \\
	\\
	\tableline
	\tableline
	\tableline
	\multicolumn{11}{c}{\textit{Swift}~BAT Transient Monitor~\text{(15--50\,keV),} Symmetric Eclipse Model} \\
	\tableline
	\tableline
	\tableline
	$i$\,(deg)$^{\mathrm{a}}$ & 56.3$^{\text{+5.6}}_{\text{--5.4}}$ & 90.0 & 56.3$^{\text{+5.7}}_{\text{--5.5}}$ & 90.0 & 56.6$^{\text{+6.4}}_{\text{--6.1}}$ & 90.0 & 68.2$^{\text{+7.7}}_{\text{--7.4}}$ & 90.0 & 68.6$^{\text{+9.0}}_{\text{--8.8}}$ & 90.0 \\
	$M_{x}$\,$(M_{\odot})$$^{\mathrm{b}}$ & 1.4 & 1.4 & 1.4 & 1.4 & 1.4 & 1.4 & 1.4 & 1.4 & 1.4 & 1.4 \\
	$M_{c}$\,$(M_{\odot})$$^{\mathrm{c}}$ & 25.6\,$\pm$\,5.8 & 15.7$^{\text{+2.4}}_{\text{--2.5}}$ & 25.5\,$\pm$\,5.9 & 15.6$^{\text{+2.8}}_{\text{--2.9}}$ & 24.1$^{\text{+6.2}}_{\text{--6.1}}$ & 15.0$^{\text{+2.6}}_{\text{--2.7}}$ & 19.2$^{\text{+3.7}}_{\text{--3.6}}$ & 15.8$^{\text{+2.5}}_{\text{--2.6}}$ & 18.7$^{\text{+4.3}}_{\text{--4.2}}$ & 15.5\,$\pm$\,2.7 \\
	$q$$^{\mathrm{d}}$ & 0.05\,$\pm$\,0.01 & 0.09\,$\pm$\,0.01 & 0.05\,$\pm$\,0.01 & 0.09\,$\pm$\,0.02 & 0.06$^{\text{+0.02}}_{\text{--0.01}}$ & 0.09\,$\pm$\,0.02 & 0.07\,$\pm$\,0.01 & 0.09\,$\pm$\,0.01 & 0.08\,$\pm$\,0.02 & 0.09\,$\pm$\,0.02 \\
	$R_{c}$\,$(R_{\odot})$$^{\mathrm{e}}$ & 28.2$^{\text{+4.9}}_{\text{--4.7}}$ & 13.2$^{\text{+4.0}}_{\text{--3.4}}$ & 28.2$^{\text{+4.9}}_{\text{--4.7}}$ & 13.2$^{\text{+4.0}}_{\text{--3.4}}$ & 27.5$^{\text{+5.4}}_{\text{--5.1}}$ & 13.0$^{\text{+3.9}}_{\text{--3.4}}$ & 20.7$^{\text{+5.2}}_{\text{--4.9}}$ & 13.6$^{\text{+4.2}}_{\text{--3.7}}$ & 20.4$^{\text{+5.9}}_{\text{--5.7}}$ & 13.6$^{\text{+4.2}}_{\text{--3.7}}$ \\
	$R_{L}$\,$(R_{\odot})$$^{\mathrm{f}}$ & 28.2\,$\pm$\,2.5 & 22.8\,$\pm$\,1.2 & 28.2\,$\pm$\,2.6 & 22.8$^{\text{+1.3}}_{\text{--1.4}}$ & 27.5$^{\text{+2.8}}_{\text{--2.7}}$ & 22.3\,$\pm$\,1.3 & 20.7$^{\text{+2.3}}_{\text{--2.4}}$ & 19.0$^{\text{+2.2}}_{\text{--2.4}}$ & 20.4$^{\text{+2.5}}_{\text{--2.7}}$ & 18.8$^{\text{+2.1}}_{\text{--2.3}}$ \\
	$\beta$$^{\mathrm{g}}$ & 1.00\,$\pm$\,0.19 & 0.58$^{\text{+0.18}}_{\text{--0.15}}$ & 1.00$^{\text{+0.20}}_{\text{--0.19}}$ & 0.58$^{\text{+0.18}}_{\text{--0.15}}$ & 1.00$^{\text{+0.22}}_{\text{--0.21}}$ & 0.58$^{\text{+0.18}}_{\text{--0.15}}$ & 1.00$^{\text{+0.27}}_{\text{--0.26}}$ & 0.72$^{\text{+0.24}}_{\text{--0.21}}$ & 1.00$^{\text{+0.32}}_{\text{--0.31}}$ & 0.72$^{\text{+0.24}}_{\text{--0.21}}$ \\
	\tableline
	$i$\,(deg)$^{\mathrm{a}}$ & 58.4$^{\text{+5.7}}_{\text{--5.5}}$ & 90.0 & 58.4$^{\text{+5.3}}_{\text{--5.0}}$ & 90.0 & 58.8$^{\text{+6.4}}_{\text{--6.0}}$ & 90.0 & 69.9$^{\text{+7.9}}_{\text{--7.6}}$ & 90.0 & 70.3$^{\text{+9.3}}_{\text{--9.0}}$ & 90.0 \\
	$M_{x}$\,$(M_{\odot})$$^{\mathrm{b}}$ & 1.9 & 1.9 & 1.9 & 1.9 & 1.9 & 1.9 & 1.9 & 1.9 & 1.9 & 1.9 \\
	$M_{c}$\,$(M_{\odot})$$^{\mathrm{c}}$ & 24.8\,$\pm$\,5.3 & 16.5\,$\pm$\,2.5 & 24.7\,$\pm$\,5.0 & 16.4$^{\text{+2.8}}_{\text{--2.9}}$ & 23.4$^{\text{+5.7}}_{\text{--5.5}}$ & 15.7\,$\pm$\,2.7 & 19.4$^{\text{+3.5}}_{\text{--3.4}}$ & 16.5\,$\pm$\,2.6 & 18.9$^{\text{+4.1}}_{\text{--4.0}}$ & 16.3\,$\pm$\,2.7 \\
	$q$$^{\mathrm{d}}$ & 0.08\,$\pm$\,0.02 & 0.12\,$\pm$\,0.02 & 0.08\,$\pm$\,0.02 & 0.12\,$\pm$\,0.02 & 0.08\,$\pm$\,0.02 & 0.12\,$\pm$\,0.02 & 0.10\,$\pm$\,0.02 & 0.11\,$\pm$\,0.02 & 0.10\,$\pm$\,0.02 & 0.12\,$\pm$\,0.02 \\
	$R_{c}$\,$(R_{\odot})$$^{\mathrm{e}}$ & 27.0$^{\text{+4.8}}_{\text{--4.5}}$ & 13.6$^{\text{+4.1}}_{\text{--3.5}}$ & 26.9$^{\text{+4.4}}_{\text{--4.2}}$ & 13.5$^{\text{+4.1}}_{\text{--3.5}}$ & 26.3$^{\text{+5.2}}_{\text{--4.9}}$ & 13.4$^{\text{+4.0}}_{\text{--3.4}}$ & 20.2$^{\text{+5.1}}_{\text{--4.8}}$ & 13.9$^{\text{+4.3}}_{\text{--3.7}}$ & 19.8$^{\text{+6.0}}_{\text{--5.7}}$ & 13.9$^{\text{+4.3}}_{\text{--3.7}}$ \\
	$R_{L}$\,$(R_{\odot})$$^{\mathrm{f}}$ & 27.0$^{\text{+2.3}}_{\text{--2.2}}$ & 22.6$^{\text{+1.1}}_{\text{--1.2}}$ & 26.9\,$\pm$\,2.1 & 22.5\,$\pm$\,1.3 & 26.3$^{\text{+2.5}}_{\text{--2.4}}$ & 22.1\,$\pm$\,1.2 & 20.2\,$\pm$\,2.2 & 18.8$^{\text{+2.2}}_{\text{--2.3}}$ & 19.8$^{\text{+2.4}}_{\text{--2.6}}$ & 18.6$^{\text{+2.0}}_{\text{--2.3}}$ \\
	$\beta$$^{\mathrm{g}}$ & 1.00$^{\text{+0.20}}_{\text{--0.19}}$ & 0.60$^{\text{+0.18}}_{\text{--0.16}}$ & 1.00$^{\text{+0.18}}_{\text{--0.17}}$ & 0.60$^{\text{+0.18}}_{\text{--0.16}}$ & 1.00$^{\text{+0.22}}_{\text{--0.21}}$ & 0.60$^{\text{+0.18}}_{\text{--0.16}}$ & 1.00$^{\text{+0.28}}_{\text{--0.26}}$ & 0.74$^{\text{+0.25}}_{\text{--0.22}}$ & 1.00$^{\text{+0.32}}_{\text{--0.31}}$ & 0.75$^{\text{+0.25}}_{\text{--0.22}}$ \\
	\\
	\tableline
	\tableline
	\tableline
	\multicolumn{11}{c}{\textit{RXTE}~PCA Galactic Bulge Scans~\text{(2--10\,keV),} Symmetric Eclipse Model} \\
	\tableline
	\tableline
	\tableline
	$i$\,(deg)$^{\mathrm{a}}$ & 60.0$^{\text{+4.4}}_{\text{--4.5}}$ & 90.0 & 60.1\,$\pm$\,5.7 & 90.0 & 60.5$^{\text{+5.1}}_{\text{--5.0}}$ & 90.0 & 74.4$^{\text{+11.2}}_{\text{--11.4}}$ & 90.0 & 75.0$^{\text{+8.3}}_{\text{--8.6}}$ & 90.0 \\
	$M_{x}$\,$(M_{\odot})$$^{\mathrm{b}}$ & 1.4 & 1.4 & 1.4 & 1.4 & 1.4 & 1.4 & 1.4 & 1.4 & 1.4 & 1.4 \\
	$M_{c}$\,$(M_{\odot})$$^{\mathrm{c}}$ & 22.9$^{\text{+3.8}}_{\text{--3.9}}$ & 15.7$^{\text{+2.4}}_{\text{--2.5}}$ & 22.8$^{\text{+4.9}}_{\text{--5.0}}$ & 15.6$^{\text{+2.8}}_{\text{--2.9}}$ & 21.6\,$\pm$\,4.1 & 15.0$^{\text{+2.6}}_{\text{--2.7}}$ & 17.4\,$\pm$\,3.7 & 15.8$^{\text{+2.5}}_{\text{--2.6}}$ & 17.0$^{\text{+2.7}}_{\text{--2.8}}$ & 15.5\,$\pm$\,2.7 \\
	$q$$^{\mathrm{d}}$ & 0.06\,$\pm$\,0.01 & 0.09\,$\pm$\,0.01 & 0.06\,$\pm$\,0.01 & 0.09\,$\pm$\,0.02 & 0.06\,$\pm$\,0.01 & 0.09\,$\pm$\,0.02 & 0.08\,$\pm$\,0.02 & 0.09\,$\pm$\,0.01 & 0.08\,$\pm$\,0.01 & 0.09\,$\pm$\,0.02 \\
	$R_{c}$\,$(R_{\odot})$$^{\mathrm{e}}$ & 26.9\,$\pm$\,3.2 & 16.1$^{\text{+1.9}}_{\text{--1.7}}$ & 26.8\,$\pm$\,4.1 & 16.1$^{\text{+1.9}}_{\text{--1.8}}$ & 26.2$^{\text{+3.6}}_{\text{--3.5}}$ & 15.9$^{\text{+1.9}}_{\text{--1.7}}$ & 19.9$^{\text{+5.0}}_{\text{--5.1}}$ & 16.6$^{\text{+2.4}}_{\text{--2.3}}$ & 19.5$^{\text{+3.6}}_{\text{--3.7}}$ & 16.5$^{\text{+2.4}}_{\text{--2.2}}$ \\
	$R_{L}$\,$(R_{\odot})$$^{\mathrm{f}}$ & 26.9$^{\text{+1.7}}_{\text{--1.8}}$ & 22.8\,$\pm$\,1.2 & 26.8$^{\text{+2.2}}_{\text{--2.3}}$ & 22.8$^{\text{+1.3}}_{\text{--1.4}}$ & 26.2\,$\pm$\,1.9 & 22.3\,$\pm$\,1.3 & 19.9$^{\text{+2.5}}_{\text{--2.7}}$ & 19.0$^{\text{+2.2}}_{\text{--2.4}}$ & 19.5$^{\text{+1.8}}_{\text{--1.9}}$ & 18.8$^{\text{+2.1}}_{\text{--2.3}}$ \\
	$\beta$$^{\mathrm{g}}$ & 1.00$^{\text{+0.13}}_{\text{--0.14}}$ & 0.71$^{\text{+0.09}}_{\text{--0.08}}$ & 1.00\,$\pm$\,0.17 & 0.71\,$\pm$\,0.09 & 1.00\,$\pm$\,0.15 & 0.71\,$\pm$\,0.09 & 1.00$^{\text{+0.28}}_{\text{--0.29}}$ & 0.87\,$\pm$\,0.16 & 1.00\,$\pm$\,0.21 & 0.88\,$\pm$\,0.16 \\
	\tableline
	$i$\,(deg)$^{\mathrm{a}}$ & 62.5\,$\pm$\,4.2 & 90.0 & 62.5\,$\pm$\,5.1 & 90.0 & 62.9\,$\pm$\,4.6 & 90.0 & 76.8$^{\text{+11.7}}_{\text{--11.8}}$ & 90.0 & 77.5$^{\text{+9.1}}_{\text{--9.3}}$ & 90.0 \\
	$M_{x}$\,$(M_{\odot})$$^{\mathrm{b}}$ & 1.9 & 1.9 & 1.9 & 1.9 & 1.9 & 1.9 & 1.9 & 1.9 & 1.9 & 1.9 \\
	$M_{c}$\,$(M_{\odot})$$^{\mathrm{c}}$ & 22.3$^{\text{+3.3}}_{\text{--3.4}}$ & 16.5\,$\pm$\,2.5 & 22.2$^{\text{+4.1}}_{\text{--4.2}}$ & 16.4$^{\text{+2.8}}_{\text{--2.9}}$ & 21.1\,$\pm$\,3.5 & 15.7\,$\pm$\,2.7 & 17.7$^{\text{+3.3}}_{\text{--3.4}}$ & 16.5\,$\pm$\,2.6 & 17.3\,$\pm$\,2.6 & 16.3\,$\pm$\,2.7 \\
	$q$$^{\mathrm{d}}$ & 0.09\,$\pm$\,0.01 & 0.12\,$\pm$\,0.02 & 0.09\,$\pm$\,0.02 & 0.12\,$\pm$\,0.02 & 0.09\,$\pm$\,0.01 & 0.12\,$\pm$\,0.02 & 0.11\,$\pm$\,0.02 & 0.11\,$\pm$\,0.02 & 0.11\,$\pm$\,0.02 & 0.12\,$\pm$\,0.02 \\
	$R_{c}$\,$(R_{\odot})$$^{\mathrm{e}}$ & 25.8\,$\pm$\,2.9 & 16.5$^{\text{+1.9}}_{\text{--1.7}}$ & 25.7\,$\pm$\,3.5 & 16.5$^{\text{+1.9}}_{\text{--1.8}}$ & 25.1$^{\text{+3.1}}_{\text{--3.0}}$ & 16.3$^{\text{+1.9}}_{\text{--1.8}}$ & 19.4\,$\pm$\,4.6 & 17.0$^{\text{+2.4}}_{\text{--2.3}}$ & 19.1\,$\pm$\,3.5 & 16.9$^{\text{+2.4}}_{\text{--2.3}}$ \\
	$R_{L}$\,$(R_{\odot})$$^{\mathrm{f}}$ & 25.8\,$\pm$\,1.5 & 22.6$^{\text{+1.1}}_{\text{--1.2}}$ & 25.7\,$\pm$\,1.8 & 22.5\,$\pm$\,1.3 & 25.1\,$\pm$\,1.6 & 22.1\,$\pm$\,1.2 & 19.4$^{\text{+2.3}}_{\text{--2.4}}$ & 18.8$^{\text{+2.2}}_{\text{--2.3}}$ & 19.1$^{\text{+1.6}}_{\text{--1.8}}$ & 18.6$^{\text{+2.0}}_{\text{--2.3}}$ \\
	$\beta$$^{\mathrm{g}}$ & 1.00\,$\pm$\,0.13 & 0.73\,$\pm$\,0.09 & 1.00\,$\pm$\,0.15 & 0.73$^{\text{+0.10}}_{\text{--0.09}}$ & 1.00$^{\text{+0.14}}_{\text{--0.13}}$ & 0.74$^{\text{+0.10}}_{\text{--0.09}}$ & 1.00\,$\pm$\,0.27 & 0.90$^{\text{+0.17}}_{\text{--0.16}}$ & 1.00$^{\text{+0.20}}_{\text{--0.21}}$ & 0.91$^{\text{+0.16}}_{\text{--0.17}}$
	\enddata
	\tablecomments{We quote 1$\sigma$~uncertainties on each parameter, if applicable. \\
		$^{\mathrm{a}}$ Inclination angles where the supergiant donor fills its Roche lobe and where the binary system is viewed \text{edge-on}~($i$\,$=$\,90$^{\circ}$). \\
		$^{\mathrm{b}}$ Assumed mass of the neutron star. \\
		$^{\mathrm{c}}$ Mass of the supergiant donor calculated using Equation~(\ref{Equation:XrayMassFunction}). \\
		$^{\mathrm{d}}$ Mass ratio,~$q$\,$=$\,$M_{x}/M_{c}$, where $M_{x}$~is the mass of the neutron star and $M_{c}$~is the mass of the supergiant companion. \\
		$^{\mathrm{e}}$ Radius of the supergiant donor obtained using Equation~(\ref{Equation:CompanionRadius}). \\
		$^{\mathrm{f}}$ Roche lobe radius calculated using Equation~(\ref{Equation:RocheLobeRadius}). \\
		$^{\mathrm{g}}$ Roche lobe filling factor,~$\beta$\,$=$\,$R_{c}/R_{L}$, where $R_{c}$~is the radius of the supergiant companion and $R_{L}$~is the Roche lobe radius.}
	\label{Table:DonorStarParametersAppendix}
\end{deluxetable*}



\bibliographystyle{yahapj}
\bibliography{references}

\begin{thebibliography}{}
\providecommand\natexlab[1]{#1}
\providecommand\JournalTitle[1]{#1}

\bibitem[{{Barthelmy} {et~al.}(2005){Barthelmy}, {Barbier}, {Cummings},
  {Fenimore}, {Gehrels}, {Hullinger}, {Krimm}, {Markwardt}, {Palmer},
  {Parsons}, {Sato}, {Suzuki}, {Takahashi}, {Tashiro}, \&
  {Tueller}}]{Barthelmy2005}
{Barthelmy}, S.~D., {Barbier}, L.~M., {Cummings}, J.~R., {et~al.} 2005,
  \href{http://dx.doi.org/10.1007/s11214-005-5096-3}{\JournalTitle{\ssr}, 120,
  143}

\bibitem[{{Baumgartner} {et~al.}(2013){Baumgartner}, {Tueller}, {Markwardt},
  {Skinner}, {Barthelmy}, {Mushotzky}, {Evans}, \& {Gehrels}}]{Baumgartner2013}
{Baumgartner}, W.~H., {Tueller}, J., {Markwardt}, C.~B., {et~al.} 2013,
  \href{http://dx.doi.org/10.1088/0067-0049/207/2/19}{\JournalTitle{\apjs},
  207, 19}

\bibitem[{{Bird} {et~al.}(2004){Bird}, {Barlow}, {Bassani}, {Bazzano},
  {Bodaghee}, {Capitanio}, {Cocchi}, {Del Santo}, {Dean}, {Hill}, {Lebrun},
  {Malaguti}, {Malizia}, {Much}, {Shaw}, {Stephen}, {Terrier}, {Ubertini}, \&
  {Walter}}]{Bird2004}
{Bird}, A.~J., {Barlow}, E.~J., {Bassani}, L., {et~al.} 2004,
  \href{http://dx.doi.org/10.1086/421772}{\JournalTitle{\apjl}, 607, L33}

\bibitem[{{Bird} {et~al.}(2006){Bird}, {Barlow}, {Bassani}, {Bazzano},
  {B{\'e}langer}, {Bodaghee}, {Capitanio}, {Dean}, {Fiocchi}, {Hill}, {Lebrun},
  {Malizia}, {Mas-Hesse}, {Molina}, {Moran}, {Renaud}, {Sguera}, {Shaw},
  {Stephen}, {Terrier}, {Ubertini}, {Walter}, {Willis}, \&
  {Winkler}}]{Bird2006}
{Bird}, A.~J., {Barlow}, E.~J., {Bassani}, L., {et~al.} 2006,
  \href{http://dx.doi.org/10.1086/498090}{\JournalTitle{\apj}, 636, 765}

\bibitem[{{Bird} {et~al.}(2007){Bird}, {Malizia}, {Bazzano}, {Barlow},
  {Bassani}, {Hill}, {B{\'e}langer}, {Capitanio}, {Clark}, {Dean}, {Fiocchi},
  {G{\"o}tz}, {Lebrun}, {Molina}, {Produit}, {Renaud}, {Sguera}, {Stephen},
  {Terrier}, {Ubertini}, {Walter}, {Winkler}, \& {Zurita}}]{Bird2007}
{Bird}, A.~J., {Malizia}, A., {Bazzano}, A., {et~al.} 2007,
  \href{http://dx.doi.org/10.1086/513148}{\JournalTitle{\apjs}, 170, 175}

\bibitem[{{Bird} {et~al.}(2010){Bird}, {Bazzano}, {Bassani}, {Capitanio},
  {Fiocchi}, {Hill}, {Malizia}, {McBride}, {Scaringi}, {Sguera}, {Stephen},
  {Ubertini}, {Dean}, {Lebrun}, {Terrier}, {Renaud}, {Mattana}, {G{\"o}tz},
  {Rodriguez}, {Belanger}, {Walter}, \& {Winkler}}]{Bird2010}
{Bird}, A.~J., {Bazzano}, A., {Bassani}, L., {et~al.} 2010,
  \href{http://dx.doi.org/10.1088/0067-0049/186/1/1}{\JournalTitle{\apjs}, 186,
  1}

\bibitem[{{Bird} {et~al.}(2016){Bird}, {Bazzano}, {Malizia}, {Fiocchi},
  {Sguera}, {Bassani}, {Hill}, {Ubertini}, \& {Winkler}}]{Bird2016}
{Bird}, A.~J., {Bazzano}, A., {Malizia}, A., {et~al.} 2016,
  \href{http://dx.doi.org/10.3847/0067-0049/223/1/15}{\JournalTitle{\apjs},
  223, 15}

\bibitem[{{Blackburn}(1995)}]{Blackburn1995}
{Blackburn}, J.~K. 1995, in Astronomical Society of the Pacific Conference
  Series, Vol.~77, Astronomical Data Analysis Software and Systems IV, ed.
  R.~A. {Shaw}, H.~E. {Payne}, \& J.~J.~E. {Hayes}, 367

\bibitem[{{Blandford} \& {Teukolsky}(1976)}]{Blandford1976}
{Blandford}, R., \& {Teukolsky}, S.~A. 1976,
  \href{http://dx.doi.org/10.1086/154315}{\JournalTitle{\apj}, 205, 580}

\bibitem[{{Blondin} {et~al.}(1990){Blondin}, {Kallman}, {Fryxell}, \&
  {Taam}}]{Blondin1990}
{Blondin}, J.~M., {Kallman}, T.~R., {Fryxell}, B.~A., \& {Taam}, R.~E. 1990,
  \href{http://dx.doi.org/10.1086/168865}{\JournalTitle{\apj}, 356, 591}

\bibitem[{{Blondin} {et~al.}(1991){Blondin}, {Stevens}, \&
  {Kallman}}]{Blondin1991}
{Blondin}, J.~M., {Stevens}, I.~R., \& {Kallman}, T.~R. 1991,
  \href{http://dx.doi.org/10.1086/169934}{\JournalTitle{\apj}, 371, 684}

\bibitem[{{Bozzo} {et~al.}(2017){Bozzo}, {Oskinova}, {Lobel}, \&
  {Hamann}}]{Bozzo2017}
{Bozzo}, E., {Oskinova}, L., {Lobel}, A., \& {Hamann}, W.-R. 2017,
  \href{http://dx.doi.org/10.1051/0004-6361/201731930}{\JournalTitle{\aap},
  606, L10}

\bibitem[{{Carroll} \& {Ostlie}(2006)}]{Carroll2006}
{Carroll}, B.~W., \& {Ostlie}, D.~A. 2006, {An Introduction to Modern
  Astrophysics}, 2nd edn. ({Pearson, Addison-Wesley}), A

\bibitem[{{Chakrabarty}(1996)}]{Chakrabarty1996}
{Chakrabarty}, D. 1996, PhD thesis, California Insititute of Technology

\bibitem[{{Clarkson} {et~al.}(2003{\natexlab{a}}){Clarkson}, {Charles}, {Coe},
  \& {Laycock}}]{Clarkson2003a}
{Clarkson}, W.~I., {Charles}, P.~A., {Coe}, M.~J., \& {Laycock}, S.
  2003{\natexlab{a}},
  \href{http://dx.doi.org/10.1046/j.1365-8711.2003.06761.x}{\JournalTitle{\mnras},
  343, 1213}

\bibitem[{{Clarkson} {et~al.}(2003{\natexlab{b}}){Clarkson}, {Charles}, {Coe},
  {Laycock}, {Tout}, \& {Wilson}}]{Clarkson2003b}
{Clarkson}, W.~I., {Charles}, P.~A., {Coe}, M.~J., {et~al.} 2003{\natexlab{b}},
  \href{http://dx.doi.org/10.1046/j.1365-8711.2003.06176.x}{\JournalTitle{\mnras},
  339, 447}

\bibitem[{{Cochran}(1937)}]{Cochran1937}
{Cochran}, W.~G. 1937,
  \href{http://dx.doi.org/10.2307/2984123}{\JournalTitle{Supplement to the
  Journal of the Royal Statistical Society}, 4, 102}

\bibitem[{{Cochran}(1954)}]{Cochran1954}
{Cochran}, W.~G. 1954,
  \href{http://dx.doi.org/10.2307/3001666}{\JournalTitle{Biometrics}, 10, 101}

\bibitem[{{Coley} {et~al.}(2018){Coley}, {Corbet}, {F{\"u}rst}, {Huxtable},
  {Krimm}, {Pearlman}, \& {Pottschmidt}}]{Coley2018}
{Coley}, J.~B., {Corbet}, R.~H.~D., {F{\"u}rst}, F., {et~al.} 2018,
  \JournalTitle{\apj, Submitted}

\bibitem[{{Coley} {et~al.}(2015){Coley}, {Corbet}, \& {Krimm}}]{Coley2015}
{Coley}, J.~B., {Corbet}, R.~H.~D., \& {Krimm}, H.~A. 2015,
  \href{http://dx.doi.org/10.1088/0004-637X/808/2/140}{\JournalTitle{\apj},
  808, 140}

\bibitem[{{Corbet} {et~al.}(2007{\natexlab{a}}){Corbet}, {Markwardt},
  {Barbier}, {Barthelmy}, {Cummings}, {Gehrels}, {Krimm}, {Palmer}, {Sakamoto},
  {Sato}, {Tueller}, \& {Swift/Bat Survey Team}}]{Corbet2007a}
{Corbet}, R., {Markwardt}, C., {Barbier}, L., {et~al.} 2007{\natexlab{a}},
  \href{http://dx.doi.org/10.1143/PTPS.169.200}{\JournalTitle{Progress of
  Theoretical Physics Supplement}, 169, 200}

\bibitem[{{Corbet}(1984)}]{Corbet1984}
{Corbet}, R.~H.~D. 1984, \JournalTitle{\aap}, 141, 91

\bibitem[{{Corbet}(1986)}]{Corbet1986}
{Corbet}, R.~H.~D. 1986,
  \href{http://dx.doi.org/10.1093/mnras/220.4.1047}{\JournalTitle{\mnras}, 220,
  1047}

\bibitem[{{Corbet} {et~al.}(2010{\natexlab{a}}){Corbet}, {Barthelmy},
  {Baumgartner}, {Krimm}, {Markwardt}, {Skinner}, \& {Tueller}}]{Corbet2010a}
{Corbet}, R.~H.~D., {Barthelmy}, S.~D., {Baumgartner}, W.~H., {et~al.}
  2010{\natexlab{a}}, \JournalTitle{ATel}, 2599

\bibitem[{{Corbet} {et~al.}(2018){Corbet}, {Coley}, {Krimm}, \&
  {Pottschmidt}}]{Corbet2018}
{Corbet}, R.~H.~D., {Coley}, J.~B., {Krimm}, H.~A., \& {Pottschmidt}, K. 2018,
  \JournalTitle{ATel}, 11918

\bibitem[{{Corbet} \& {Krimm}(2013)}]{Corbet2013}
{Corbet}, R.~H.~D., \& {Krimm}, H.~A. 2013,
  \href{http://dx.doi.org/10.1088/0004-637X/778/1/45}{\JournalTitle{\apj}, 778,
  45}

\bibitem[{{Corbet} {et~al.}(2007{\natexlab{b}}){Corbet}, {Markwardt}, \&
  {Tueller}}]{Corbet2007b}
{Corbet}, R.~H.~D., {Markwardt}, C.~B., \& {Tueller}, J. 2007{\natexlab{b}},
  \href{http://dx.doi.org/10.1086/509319}{\JournalTitle{\apj}, 655, 458}

\bibitem[{{Corbet} {et~al.}(2010{\natexlab{b}}){Corbet}, {Pearlman}, \&
  {Pottschmidt}}]{Corbet2010b}
{Corbet}, R.~H.~D., {Pearlman}, A.~B., \& {Pottschmidt}, K. 2010{\natexlab{b}},
  \JournalTitle{ATel}, 2766

\bibitem[{{Cox}(2000)}]{Cox2000}
{Cox}, A.~N., ed. 2000, {Allen's Astrophysical Quantities}, 4th edn.
  ({Springer, AIP Press}), 389

\bibitem[{{Cranmer} \& {Owocki}(1996)}]{Cranmer1996}
{Cranmer}, S.~R., \& {Owocki}, S.~P. 1996,
  \href{http://dx.doi.org/10.1086/177166}{\JournalTitle{\apj}, 462, 469}

\bibitem[{{Cusumano} {et~al.}(2010){Cusumano}, {La Parola}, {Romano},
  {Segreto}, {Vercellone}, \& {Chincarini}}]{Cusumano2010}
{Cusumano}, G., {La Parola}, V., {Romano}, P., {et~al.} 2010,
  \href{http://dx.doi.org/10.1111/j.1745-3933.2010.00872.x}{\JournalTitle{\mnras},
  406, L16}

\bibitem[{{Cutri} {et~al.}(2003){Cutri}, {Skrutskie}, {van Dyk}, {Beichman},
  {Carpenter}, {Chester}, {Cambresy}, {Evans}, {Fowler}, {Gizis}, {Howard},
  {Huchra}, {Jarrett}, {Kopan}, {Kirkpatrick}, {Light}, {Marsh}, {McCallon},
  {Schneider}, {Stiening}, {Sykes}, {Weinberg}, {Wheaton}, {Wheelock}, \&
  {Zacarias}}]{Cutri2003}
{Cutri}, R.~M., {Skrutskie}, M.~F., {van Dyk}, S., {et~al.} 2003, {2MASS All
  Sky Catalog of point sources.}

\bibitem[{{D'A{\`i}} {et~al.}(2011){D'A{\`i}}, {Cusumano}, {La Parola},
  {Segreto}, {di Salvo}, {Iaria}, \& {Robba}}]{DAi2011}
{D'A{\`i}}, A., {Cusumano}, G., {La Parola}, V., {et~al.} 2011,
  \href{http://dx.doi.org/10.1051/0004-6361/201117035}{\JournalTitle{\aap},
  532, A73}

\bibitem[{{Demorest}(2007)}]{Demorest2007}
{Demorest}, P.~B. 2007, PhD thesis, University of California, Berkeley

\bibitem[{{Dickey} \& {Lockman}(1990)}]{Dickey1990}
{Dickey}, J.~M., \& {Lockman}, F.~J. 1990,
  \href{http://dx.doi.org/10.1146/annurev.aa.28.090190.001243}{\JournalTitle{\araa},
  28, 215}

\bibitem[{{Eastman} {et~al.}(2010){Eastman}, {Siverd}, \&
  {Gaudi}}]{Eastman2010}
{Eastman}, J., {Siverd}, R., \& {Gaudi}, B.~S. 2010,
  \href{http://dx.doi.org/10.1086/655938}{\JournalTitle{\pasp}, 122, 935}

\bibitem[{{Eggleton}(1983)}]{Eggleton1983}
{Eggleton}, P.~P. 1983,
  \href{http://dx.doi.org/10.1086/160960}{\JournalTitle{\apj}, 268, 368}

\bibitem[{{Falanga} {et~al.}(2015){Falanga}, {Bozzo}, {Lutovinov},
  {Bonnet-Bidaud}, {Fetisova}, \& {Puls}}]{Falanga2015}
{Falanga}, M., {Bozzo}, E., {Lutovinov}, A., {et~al.} 2015,
  \href{http://dx.doi.org/10.1051/0004-6361/201425191}{\JournalTitle{\aap},
  577, A130}

\bibitem[{{Farrell} {et~al.}(2006){Farrell}, {Sood}, \&
  {O'Neill}}]{Farrell2006}
{Farrell}, S.~A., {Sood}, R.~K., \& {O'Neill}, P.~M. 2006,
  \href{http://dx.doi.org/10.1111/j.1365-2966.2006.10150.x}{\JournalTitle{\mnras},
  367, 1457}

\bibitem[{{Farrell} {et~al.}(2008){Farrell}, {Sood}, {O'Neill}, \&
  {Dieters}}]{Farrell2008}
{Farrell}, S.~A., {Sood}, R.~K., {O'Neill}, P.~M., \& {Dieters}, S. 2008,
  \href{http://dx.doi.org/10.1111/j.1365-2966.2008.13588.x}{\JournalTitle{\mnras},
  389, 608}

\bibitem[{{Feldmeier} {et~al.}(1996){Feldmeier}, {Anzer}, {Boerner}, \&
  {Nagase}}]{Feldmeier1996}
{Feldmeier}, A., {Anzer}, U., {Boerner}, G., \& {Nagase}, F. 1996,
  \JournalTitle{\aap}, 311, 793

\bibitem[{{Foreman-Mackey} {et~al.}(2013){Foreman-Mackey}, {Hogg}, {Lang}, \&
  {Goodman}}]{ForemanMackey2013}
{Foreman-Mackey}, D., {Hogg}, D.~W., {Lang}, D., \& {Goodman}, J. 2013,
  \href{http://dx.doi.org/10.1086/670067}{\JournalTitle{\pasp}, 125, 306}

\bibitem[{{Fransson} \& {Fabian}(1980)}]{Fransson1980}
{Fransson}, C., \& {Fabian}, A.~C. 1980, \JournalTitle{\aap}, 87, 102

\bibitem[{{Gehrels} {et~al.}(2004){Gehrels}, {Chincarini}, {Giommi}, {Mason},
  {Nousek}, {Wells}, {White}, {Barthelmy}, {Burrows}, {Cominsky}, {Hurley},
  {Marshall}, {M{\'e}sz{\'a}ros}, {Roming}, {Angelini}, {Barbier}, {Belloni},
  {Campana}, {Caraveo}, {Chester}, {Citterio}, {Cline}, {Cropper}, {Cummings},
  {Dean}, {Feigelson}, {Fenimore}, {Frail}, {Fruchter}, {Garmire}, {Gendreau},
  {Ghisellini}, {Greiner}, {Hill}, {Hunsberger}, {Krimm}, {Kulkarni}, {Kumar},
  {Lebrun}, {Lloyd-Ronning}, {Markwardt}, {Mattson}, {Mushotzky}, {Norris},
  {Osborne}, {Paczynski}, {Palmer}, {Park}, {Parsons}, {Paul}, {Rees},
  {Reynolds}, {Rhoads}, {Sasseen}, {Schaefer}, {Short}, {Smale}, {Smith},
  {Stella}, {Tagliaferri}, {Takahashi}, {Tashiro}, {Townsley}, {Tueller},
  {Turner}, {Vietri}, {Voges}, {Ward}, {Willingale}, {Zerbi}, \&
  {Zhang}}]{Gehrels2004}
{Gehrels}, N., {Chincarini}, G., {Giommi}, P., {et~al.} 2004,
  \href{http://dx.doi.org/10.1086/422091}{\JournalTitle{\apj}, 611, 1005}

\bibitem[{{Goodman} \& {Weare}(2010)}]{Goodman2010}
{Goodman}, J., \& {Weare}, J. 2010,
  \href{http://dx.doi.org/10.2140/camcos.2010.5.65}{\JournalTitle{Comm. App.
  Math. Comp. Sci.}, 5, 65}

\bibitem[{{Goossens} {et~al.}(2013){Goossens}, {Bird}, {Drave}, {Bazzano},
  {Hill}, {McBride}, {Sguera}, \& {Sidoli}}]{Goossens2013}
{Goossens}, M.~E., {Bird}, A.~J., {Drave}, S.~P., {et~al.} 2013,
  \href{http://dx.doi.org/10.1093/mnras/stt1166}{\JournalTitle{\mnras}, 434,
  2182}

\bibitem[{{Grebenev} {et~al.}(2005){Grebenev}, {Bird}, {Molkov}, {Soldi},
  {Kretschmar}, {Diehl}, {Budz-Joergensen}, \& {McBreen}}]{Grebenev2005}
{Grebenev}, S.~A., {Bird}, A.~J., {Molkov}, S.~V., {et~al.} 2005,
  \JournalTitle{ATel}, 457

\bibitem[{{G{\"u}ver} \& {{\"O}zel}(2009)}]{Guver2009}
{G{\"u}ver}, T., \& {{\"O}zel}, F. 2009,
  \href{http://dx.doi.org/10.1111/j.1365-2966.2009.15598.x}{\JournalTitle{\mnras},
  400, 2050}

\bibitem[{{Hanson} {et~al.}(1996){Hanson}, {Conti}, \& {Rieke}}]{Hanson1996}
{Hanson}, M.~M., {Conti}, P.~S., \& {Rieke}, M.~J. 1996,
  \href{http://dx.doi.org/10.1086/192366}{\JournalTitle{\apjs}, 107, 281}

\bibitem[{{Hill} {et~al.}(2008){Hill}, {Dean}, {Landi}, {McBride}, {de Rosa},
  {Bird}, {Bazzano}, \& {Sguera}}]{Hill2008}
{Hill}, A.~B., {Dean}, A.~J., {Landi}, R., {et~al.} 2008,
  \href{http://dx.doi.org/10.1111/j.1365-2966.2008.12849.x}{\JournalTitle{\mnras},
  385, 423}

\bibitem[{{Horne} \& {Baliunas}(1986)}]{Horne1986}
{Horne}, J.~H., \& {Baliunas}, S.~L. 1986,
  \href{http://dx.doi.org/10.1086/164037}{\JournalTitle{\apj}, 302, 757}

\bibitem[{{Hu} {et~al.}(2017){Hu}, {Chou}, {Ng}, {Lin}, \& {Yen}}]{Hu2017}
{Hu}, C.-P., {Chou}, Y., {Ng}, C.-Y., {Lin}, L.~C.-C., \& {Yen}, D.~C.-C. 2017,
  \href{http://dx.doi.org/10.3847/1538-4357/aa79a3}{\JournalTitle{\apj}, 844,
  16}

\bibitem[{{Hung} {et~al.}(2010){Hung}, {Hickox}, {Boroson}, \&
  {Vrtilek}}]{Hung2010}
{Hung}, L.-W., {Hickox}, R.~C., {Boroson}, B.~S., \& {Vrtilek}, S.~D. 2010,
  \href{http://dx.doi.org/10.1088/0004-637X/720/2/1202}{\JournalTitle{\apj},
  720, 1202}

\bibitem[{{Jackson}(1975)}]{Jackson1975}
{Jackson}, J.~C. 1975,
  \href{http://dx.doi.org/10.1093/mnras/172.3.483}{\JournalTitle{\mnras}, 172,
  483}

\bibitem[{{Jahoda} {et~al.}(2006){Jahoda}, {Markwardt}, {Radeva}, {Rots},
  {Stark}, {Swank}, {Strohmayer}, \& {Zhang}}]{Jahoda2006}
{Jahoda}, K., {Markwardt}, C.~B., {Radeva}, Y., {et~al.} 2006,
  \href{http://dx.doi.org/10.1086/500659}{\JournalTitle{\apjs}, 163, 401}

\bibitem[{{Jahoda} {et~al.}(1996){Jahoda}, {Swank}, {Giles}, {Stark},
  {Strohmayer}, {Zhang}, \& {Morgan}}]{Jahoda1996}
{Jahoda}, K., {Swank}, J.~H., {Giles}, A.~B., {et~al.} 1996,
  \href{http://dx.doi.org/10.1117/12.256034}{\JournalTitle{\procspie}, 2808,
  59}

\bibitem[{{Jain} {et~al.}(2009){Jain}, {Paul}, \& {Dutta}}]{Jain2009}
{Jain}, C., {Paul}, B., \& {Dutta}, A. 2009,
  \href{http://dx.doi.org/10.1111/j.1745-3933.2009.00668.x}{\JournalTitle{\mnras},
  397, L11}

\bibitem[{{Jenke} {et~al.}(2012){Jenke}, {Finger}, {Wilson-Hodge}, \&
  {Camero-Arranz}}]{Jenke2012}
{Jenke}, P.~A., {Finger}, M.~H., {Wilson-Hodge}, C.~A., \& {Camero-Arranz}, A.
  2012,
  \href{http://dx.doi.org/10.1088/0004-637X/759/2/124}{\JournalTitle{\apj},
  759, 124}

\bibitem[{{Joss} \& {Rappaport}(1984)}]{Joss1984}
{Joss}, P.~C., \& {Rappaport}, S.~A. 1984,
  \href{http://dx.doi.org/10.1146/annurev.aa.22.090184.002541}{\JournalTitle{\araa},
  22, 537}

\bibitem[{{Kalberla} {et~al.}(2005){Kalberla}, {Burton}, {Hartmann}, {Arnal},
  {Bajaja}, {Morras}, \& {P{\"o}ppel}}]{Kalberla2005}
{Kalberla}, P.~M.~W., {Burton}, W.~B., {Hartmann}, D., {et~al.} 2005,
  \href{http://dx.doi.org/10.1051/0004-6361:20041864}{\JournalTitle{\aap}, 440,
  775}

\bibitem[{{Kaper} {et~al.}(2006){Kaper}, {van der Meer}, {van Kerkwijk}, \&
  {van den Heuvel}}]{Kaper2006}
{Kaper}, L., {van der Meer}, A., {van Kerkwijk}, M., \& {van den Heuvel}, E.
  2006, \JournalTitle{The Messenger}, 126, 27

\bibitem[{{Kelley} {et~al.}(1980){Kelley}, {Rappaport}, \&
  {Petre}}]{Kelley1980}
{Kelley}, R., {Rappaport}, S., \& {Petre}, R. 1980,
  \href{http://dx.doi.org/10.1086/158026}{\JournalTitle{\apj}, 238, 699}

\bibitem[{{Koenigsberger} {et~al.}(2006){Koenigsberger}, {Georgiev}, {Moreno},
  {Richer}, {Toledano}, {Canalizo}, \& {Arrieta}}]{Koenigsberger2006}
{Koenigsberger}, G., {Georgiev}, L., {Moreno}, E., {et~al.} 2006,
  \href{http://dx.doi.org/10.1051/0004-6361:20065305}{\JournalTitle{\aap}, 458,
  513}

\bibitem[{{Koenigsberger} {et~al.}(2003){Koenigsberger}, {Moreno}, \&
  {Cervantes}}]{Koenigsberger2003}
{Koenigsberger}, G., {Moreno}, E., \& {Cervantes}, F. 2003,
  \href{http://dx.doi.org/10.1017/S0074180900211698}{in IAU Symp. 212, A
  Massive Star Odyssey: From Main Sequence to Supernova, ed. K.~{van der
  Hucht}, A.~{Herrero}, \& C.~{Esteban}}, (San Francisco, CA: ASP), 101

\bibitem[{{Koh} {et~al.}(1997){Koh}, {Bildsten}, {Chakrabarty}, {Nelson},
  {Prince}, {Vaughan}, {Finger}, {Wilson}, \& {Rubin}}]{Koh1997}
{Koh}, D.~T., {Bildsten}, L., {Chakrabarty}, D., {et~al.} 1997,
  \href{http://dx.doi.org/10.1086/303929}{\JournalTitle{\apj}, 479, 933}

\bibitem[{{Krimm} {et~al.}(2013){Krimm}, {Holland}, {Corbet}, {Pearlman},
  {Romano}, {Kennea}, {Bloom}, {Barthelmy}, {Baumgartner}, {Cummings},
  {Gehrels}, {Lien}, {Markwardt}, {Palmer}, {Sakamoto}, {Stamatikos}, \&
  {Ukwatta}}]{Krimm2013}
{Krimm}, H.~A., {Holland}, S.~T., {Corbet}, R.~H.~D., {et~al.} 2013,
  \href{http://dx.doi.org/10.1088/0067-0049/209/1/14}{\JournalTitle{\apjs},
  209, 14}

\bibitem[{{Kuiper} {et~al.}(2005){Kuiper}, {Jonker}, {Hermsen}, \&
  {O'Brien}}]{Kuiper2005}
{Kuiper}, L., {Jonker}, P., {Hermsen}, W., \& {O'Brien}, K. 2005,
  \JournalTitle{ATel}, 654

\bibitem[{{Leahy} {et~al.}(1983){Leahy}, {Darbro}, {Elsner}, {Weisskopf},
  {Kahn}, {Sutherland}, \& {Grindlay}}]{Leahy1983}
{Leahy}, D.~A., {Darbro}, W., {Elsner}, R.~F., {et~al.} 1983,
  \href{http://dx.doi.org/10.1086/160766}{\JournalTitle{\apj}, 266, 160}

\bibitem[{{Lebrun} {et~al.}(2003){Lebrun}, {Leray}, {Lavocat}, {Cr{\'e}tolle},
  {Arqu{\`e}s}, {Blondel}, {Bonnin}, {Bou{\`e}re}, {Cara}, {Chaleil}, {Daly},
  {Desages}, {Dzitko}, {Horeau}, {Laurent}, {Limousin}, {Mathy}, {Mauguen},
  {Meignier}, {Molini{\'e}}, {Poindron}, {Rouger}, {Sauvageon}, \&
  {Tourrette}}]{Lebrun2003}
{Lebrun}, F., {Leray}, J.~P., {Lavocat}, P., {et~al.} 2003,
  \href{http://dx.doi.org/10.1051/0004-6361:20031367}{\JournalTitle{\aap}, 411,
  L141}

\bibitem[{{Lefever} {et~al.}(2007){Lefever}, {Puls}, \& {Aerts}}]{Lefever2007}
{Lefever}, K., {Puls}, J., \& {Aerts}, C. 2007,
  \href{http://dx.doi.org/10.1051/0004-6361:20066038}{\JournalTitle{\aap}, 463,
  1093}

\bibitem[{{Levine} {et~al.}(2004){Levine}, {Rappaport}, {Remillard}, \&
  {Savcheva}}]{Levine2004}
{Levine}, A.~M., {Rappaport}, S., {Remillard}, R., \& {Savcheva}, A. 2004,
  \href{http://dx.doi.org/10.1086/425567}{\JournalTitle{\apj}, 617, 1284}

\bibitem[{{Manousakis} \& {Walter}(2015)}]{Manousakis2015}
{Manousakis}, A., \& {Walter}, R. 2015,
  \href{http://dx.doi.org/10.1051/0004-6361/201526893}{\JournalTitle{\aap},
  584, A25}

\bibitem[{{Markwardt} {et~al.}(2005){Markwardt}, {Swank}, \&
  {Smith}}]{Markwardt2005a}
{Markwardt}, C.~B., {Swank}, J.~H., \& {Smith}, E. 2005, \JournalTitle{ATel},
  465

\bibitem[{{Masetti} {et~al.}(2006){Masetti}, {Orlandini}, {dal Fiume}, {del
  Sordo}, {Amati}, {Frontera}, {Palazzi}, \& {Santangelo}}]{Masetti2006}
{Masetti}, N., {Orlandini}, M., {dal Fiume}, D., {et~al.} 2006,
  \href{http://dx.doi.org/10.1051/0004-6361:20053654}{\JournalTitle{\aap}, 445,
  653}

\bibitem[{{Moreno} {et~al.}(2005){Moreno}, {Koenigsberger}, \&
  {Toledano}}]{Moreno2005}
{Moreno}, E., {Koenigsberger}, G., \& {Toledano}, O. 2005,
  \href{http://dx.doi.org/10.1051/0004-6361:20041995}{\JournalTitle{\aap}, 437,
  641}

\bibitem[{{Morris} {et~al.}(2009){Morris}, {Smith}, {Markwardt}, {Mushotzky},
  {Tueller}, {Kallman}, \& {Dhuga}}]{Morris2009}
{Morris}, D.~C., {Smith}, R.~K., {Markwardt}, C.~B., {et~al.} 2009,
  \href{http://dx.doi.org/10.1088/0004-637X/699/1/892}{\JournalTitle{\apj},
  699, 892}

\bibitem[{{Mullan}(1984)}]{Mullan1984}
{Mullan}, D.~J. 1984,
  \href{http://dx.doi.org/10.1086/162307}{\JournalTitle{\apj}, 283, 303}

\bibitem[{{Nespoli} {et~al.}(2008){Nespoli}, {Fabregat}, \&
  {Mennickent}}]{Nespoli2008}
{Nespoli}, E., {Fabregat}, J., \& {Mennickent}, R.~E. 2008,
  \JournalTitle{ATel}, 1396

\bibitem[{{Nespoli} {et~al.}(2010){Nespoli}, {Fabregat}, \&
  {Mennickent}}]{Nespoli2010}
{Nespoli}, E., {Fabregat}, J., \& {Mennickent}, R.~E. 2010,
  \href{http://dx.doi.org/10.1051/0004-6361/201014348}{\JournalTitle{\aap},
  516, A106}

\bibitem[{{Ogilvie} \& {Dubus}(2001)}]{Ogilvie2001}
{Ogilvie}, G.~I., \& {Dubus}, G. 2001,
  \href{http://dx.doi.org/10.1046/j.1365-8711.2001.04011.x}{\JournalTitle{\mnras},
  320, 485}

\bibitem[{{Oh} {et~al.}(2018){Oh}, {Koss}, {Markwardt}, {Schawinski},
  {Baumgartner}, {Barthelmy}, {Cenko}, {Gehrels}, {Mushotzky}, {Petulante},
  {Ricci}, {Lien}, \& {Trakhtenbrot}}]{Oh2018}
{Oh}, K., {Koss}, M., {Markwardt}, C.~B., {et~al.} 2018,
  \href{http://dx.doi.org/10.3847/1538-4365/aaa7fd}{\JournalTitle{\apjs}, 235,
  4}

\bibitem[{{Papadakis} \& {Lawrence}(1993)}]{Papadakis1993}
{Papadakis}, I.~E., \& {Lawrence}, A. 1993,
  \href{http://dx.doi.org/10.1093/mnras/261.3.612}{\JournalTitle{\mnras}, 261,
  612}

\bibitem[{{Petterson}(1975)}]{Petterson1975}
{Petterson}, J.~A. 1975,
  \href{http://dx.doi.org/10.1086/181942}{\JournalTitle{\apjl}, 201, L61}

\bibitem[{{Postnov} {et~al.}(2013){Postnov}, {Shakura}, {Staubert},
  {Kochetkova}, {Klochkov}, \& {Wilms}}]{Postnov2013}
{Postnov}, K., {Shakura}, N., {Staubert}, R., {et~al.} 2013,
  \href{http://dx.doi.org/10.1093/mnras/stt1363}{\JournalTitle{\mnras}, 435,
  1147}

\bibitem[{{Predehl} \& {Schmitt}(1995)}]{Predehl1995}
{Predehl}, P., \& {Schmitt}, J.~H.~M.~M. 1995, \JournalTitle{\aap}, 293, 889

\bibitem[{{Rieke} \& {Lebofsky}(1985)}]{Rieke1985}
{Rieke}, G.~H., \& {Lebofsky}, M.~J. 1985,
  \href{http://dx.doi.org/10.1086/162827}{\JournalTitle{\apj}, 288, 618}

\bibitem[{{Romano}(2015)}]{Romano2015}
{Romano}, P. 2015,
  \href{http://dx.doi.org/10.1016/j.jheap.2015.04.008}{\JournalTitle{Journal of
  High Energy Astrophysics}, 7, 126}

\bibitem[{{Scargle}(1982)}]{Scargle1982}
{Scargle}, J.~D. 1982,
  \href{http://dx.doi.org/10.1086/160554}{\JournalTitle{\apj}, 263, 835}

\bibitem[{{Scargle}(1989)}]{Scargle1989}
{Scargle}, J.~D. 1989,
  \href{http://dx.doi.org/10.1086/167757}{\JournalTitle{\apj}, 343, 874}

\bibitem[{{Schwarzenberg-Czerny}(1989)}]{Schwarzenberg-Czerny1989}
{Schwarzenberg-Czerny}, A. 1989,
  \href{http://dx.doi.org/10.1093/mnras/241.2.153}{\JournalTitle{\mnras}, 241,
  153}

\bibitem[{{Searle} {et~al.}(2008){Searle}, {Prinja}, {Massa}, \&
  {Ryans}}]{Searle2008}
{Searle}, S.~C., {Prinja}, R.~K., {Massa}, D., \& {Ryans}, R. 2008,
  \href{http://dx.doi.org/10.1051/0004-6361:20077125}{\JournalTitle{\aap}, 481,
  777}

\bibitem[{{Sepinsky} {et~al.}(2010){Sepinsky}, {Willems}, {Kalogera}, \&
  {Rasio}}]{Sepinksy2010}
{Sepinsky}, J.~F., {Willems}, B., {Kalogera}, V., \& {Rasio}, F.~A. 2010,
  \href{http://dx.doi.org/10.1088/0004-637X/724/1/546}{\JournalTitle{\apj},
  724, 546}

\bibitem[{{Standish}(1990)}]{Standish1990}
{Standish}, Jr., E.~M. 1990, \JournalTitle{\aap}, 233, 252

\bibitem[{{Swank} \& {Markwardt}(2001)}]{Swank2001}
{Swank}, J., \& {Markwardt}, C. 2001, in ASP Conf. Ser. 251, New Century of
  X-ray Astronomy, ed. H.~{Inoue} \& H.~{Kunieda}, (San Francisco, CA: ASP), 94

\bibitem[{{Taylor}(1992)}]{Taylor1992}
{Taylor}, J.~H. 1992,
  \href{http://dx.doi.org/10.1098/rsta.1992.0088}{\JournalTitle{Philosophical
  Transactions of the Royal Society of London Series A}, 341, 117}

\bibitem[{{Thompson} {et~al.}(2007){Thompson}, {Tomsick}, {in 't Zand},
  {Rothschild}, \& {Walter}}]{Thompson2007}
{Thompson}, T.~W.~J., {Tomsick}, J.~A., {in 't Zand}, J.~J.~M., {Rothschild},
  R.~E., \& {Walter}, R. 2007,
  \href{http://dx.doi.org/10.1086/513458}{\JournalTitle{\apj}, 661, 447}

\bibitem[{{Thompson} {et~al.}(2006){Thompson}, {Tomsick}, {Rothschild}, {in't
  Zand}, \& {Walter}}]{Thompson2006}
{Thompson}, T.~W.~J., {Tomsick}, J.~A., {Rothschild}, R.~E., {in't Zand},
  J.~J.~M., \& {Walter}, R. 2006,
  \href{http://dx.doi.org/10.1086/506251}{\JournalTitle{\apj}, 649, 373}

\bibitem[{{Townsend} {et~al.}(2011){Townsend}, {Coe}, {Corbet}, \&
  {Hill}}]{Townsend2011}
{Townsend}, L.~J., {Coe}, M.~J., {Corbet}, R.~H.~D., \& {Hill}, A.~B. 2011,
  \href{http://dx.doi.org/10.1111/j.1365-2966.2011.19153.x}{\JournalTitle{\mnras},
  416, 1556}

\bibitem[{{Ubertini} {et~al.}(2003){Ubertini}, {Lebrun}, {Di Cocco}, {Bazzano},
  {Bird}, {Broenstad}, {Goldwurm}, {La Rosa}, {Labanti}, {Laurent}, {Mirabel},
  {Quadrini}, {Ramsey}, {Reglero}, {Sabau}, {Sacco}, {Staubert}, {Vigroux},
  {Weisskopf}, \& {Zdziarski}}]{Ubertini2003}
{Ubertini}, P., {Lebrun}, F., {Di Cocco}, G., {et~al.} 2003,
  \href{http://dx.doi.org/10.1051/0004-6361:20031224}{\JournalTitle{\aap}, 411,
  L131}

\bibitem[{{van der Klis} \& {Bonnet-Bidaud}(1984)}]{vanDerKlis1984}
{van der Klis}, M., \& {Bonnet-Bidaud}, J.~M. 1984, \JournalTitle{\aap}, 135,
  155

\bibitem[{{van Kerkwijk} {et~al.}(1995){van Kerkwijk}, {van Paradijs}, \&
  {Zuiderwijk}}]{vanKerkwijk1995}
{van Kerkwijk}, M.~H., {van Paradijs}, J., \& {Zuiderwijk}, E.~J. 1995,
  \JournalTitle{\aap}, 303, 497

\bibitem[{{Vaughan}(2005)}]{Vaughan2005}
{Vaughan}, S. 2005,
  \href{http://dx.doi.org/10.1051/0004-6361:20041453}{\JournalTitle{\aap}, 431,
  391}

\bibitem[{{Wegner}(1994)}]{Wegner1994}
{Wegner}, W. 1994,
  \href{http://dx.doi.org/10.1093/mnras/270.2.229}{\JournalTitle{\mnras}, 270,
  229}

\bibitem[{{Willingale} {et~al.}(2013){Willingale}, {Starling}, {Beardmore},
  {Tanvir}, \& {O'Brien}}]{Willingale2013}
{Willingale}, R., {Starling}, R.~L.~C., {Beardmore}, A.~P., {Tanvir}, N.~R., \&
  {O'Brien}, P.~T. 2013,
  \href{http://dx.doi.org/10.1093/mnras/stt175}{\JournalTitle{\mnras}, 431,
  394}

\bibitem[{{Winkler} {et~al.}(2003){Winkler}, {Courvoisier}, {Di Cocco},
  {Gehrels}, {Gim{\'e}nez}, {Grebenev}, {Hermsen}, {Mas-Hesse}, {Lebrun},
  {Lund}, {Palumbo}, {Paul}, {Roques}, {Schnopper}, {Sch{\"o}nfelder},
  {Sunyaev}, {Teegarden}, {Ubertini}, {Vedrenne}, \& {Dean}}]{Winkler2003}
{Winkler}, C., {Courvoisier}, T.~J.-L., {Di Cocco}, G., {et~al.} 2003,
  \href{http://dx.doi.org/10.1051/0004-6361:20031288}{\JournalTitle{\aap}, 411,
  L1}

\bibitem[{{Wojdowski} {et~al.}(1998){Wojdowski}, {Clark}, {Levine}, {Woo}, \&
  {Zhang}}]{Wojdowski1998}
{Wojdowski}, P., {Clark}, G.~W., {Levine}, A.~M., {Woo}, J.~W., \& {Zhang},
  S.~N. 1998, \href{http://dx.doi.org/10.1086/305893}{\JournalTitle{\apj}, 502,
  253}

\end{thebibliography}


\end{document}